\documentclass[
12pt,
letterpaper,
reprint,
pra,
aps,
floatfix,
showkeys,
superscriptaddress,
longbibliography,
colorlinks
]{revtex4-2}

\usepackage[utf8]{inputenc}
\usepackage{amsmath,amsthm,amssymb,amsfonts}
\theoremstyle{remark}
\newtheorem{claim}{\textbf{Claim}} 
\newtheorem{theorem}{\textbf{Theorem}} 
\newtheorem{lemma}{\textbf{Lemma}} 
\newtheorem{example}{\textbf{Example}} 
\usepackage{mathtools}
\usepackage{graphicx}
\graphicspath{{./figures/}}
\usepackage{booktabs}
\usepackage{xcolor}
\usepackage{rotating}
\usepackage{orcidlink}
\usepackage{adjustbox}
\usepackage{array}
\usepackage[mathlines]{lineno}
\usepackage{appendix}
\usepackage{aligned-overset}
\usepackage{algorithm}
\usepackage{algpseudocode}
\usepackage{bbm}
\usepackage{ulem}
\usepackage{cancel}

\AtBeginDocument{%
  \heavyrulewidth=.08em
  \lightrulewidth=.05em
  \cmidrulewidth=.03em
  \belowrulesep=.65ex
  \belowbottomsep=0pt
  \aboverulesep=.4ex
  \abovetopsep=0pt
  \cmidrulesep=\doublerulesep
  \cmidrulekern=.5em
  \defaultaddspace=.5em
}
\usepackage{hyperref}

\newcommand{\abs}[1]{\left\lvert#1\right\rvert} 
\newcommand{\absSmall}[1]{\lvert#1\rvert} 
\newcommand{\norm}[1]{\left\|#1\right\|} 
\newcommand{\normSmall}[1]{\|#1\|} 
 
\newcommand{\bra}[1]{\langle#1\rvert} 
\newcommand{\ket}[1]{\lvert#1\rangle} 
\newcommand{\proj}[1]{\ket{#1}\bra{#1}} 
 
\newcommand{\braopket}[3]{\langle #1 | #2 | #3\rangle}

\newcommand{\tr}[1]{\text{Tr}\left[#1\right]}
\newcommand{\trSmall}[1]{\text{Tr}[#1]}  

\newcommand{\trSubSmall}[2]{\text{Tr}_{#1}[#2]}

\newcommand{\evTextSubSmall}[2]{\mathbb{E}_{#1}[#2]}

\newcommand{\varSubSmall}[2]{\text{Var}_{#1}[#2]}
 
\newcommand{\covSubSmall}[3]{\text{Cov}_{#1}[#2,#3]} 

\newcommand{\bmatrixByJames}[1]{\left[\;\begin{matrix}#1\end{matrix}\;\right]}
\newcommand{\pmatrixByJames}[1]{\left(\;\begin{matrix}#1\end{matrix}\;\right)}

\newcommand{\h}[1]{\hat{#1}}

\newcommand{\R}{\mathbb{R}} 
\newcommand{\C}{\mathbb{C}} 
\newcommand{\Z}{\mathbb{Z}}

\newcommand{\IQ}{\mathcal{I}_Q}
\newcommand{\IC}{\mathcal{I}_C}

\newcommand{\si}{\sigma}

\newcommand{\T}{\text{T}}

\newcommand{\diag}[1]{\text{diag}\left(#1\right)}

\newcommand{\order}[1]{\mathcal{O}(#1)}

\usepackage{outlines}

\begin{document}
\title{Lindblad estimation with fast and precise quantum control} 

\author{James~W.~Gardner\,\orcidlink{0000-0002-8592-1452}}
\email{james.gardner@anu.edu.au}
\affiliation{OzGrav-ANU, Centre for Gravitational Astrophysics, Research Schools of Physics, and of Astronomy and Astrophysics, The Australian National University, Canberra, ACT 2601, Australia}
\affiliation{Walter Burke Institute for Theoretical Physics, California Institute of Technology, Pasadena, California 91125, USA} 
\author{Simon A. Haine\,\orcidlink{0000-0003-1534-1492}}
\affiliation{Department of Quantum Science and Technology and Department of Fundamental and Theoretical Physics, Research School of Physics, The Australian National University, Canberra, ACT 0200, Australia}
\author{Joseph~J.~Hope\,\orcidlink{0000-0002-5260-1380}}
\affiliation{Department of Quantum Science and Technology and Department of Fundamental and Theoretical Physics, Research School of Physics, The Australian National University, Canberra, ACT 0200, Australia}
\author{Yanbei Chen\,\orcidlink{0000-0002-9730-9463}\,}
\affiliation{Walter Burke Institute for Theoretical Physics, California Institute of Technology, Pasadena, California 91125, USA}
\author{Tuvia Gefen\,\orcidlink{0000-0002-3235-4917}}
\email{getuvia@gmail.com}
\affiliation{Institute for Quantum Information and Matter, California Institute of Technology, Pasadena, California 91125, USA}
\affiliation{Racah Institute of Physics, The Hebrew University of Jerusalem, Jerusalem 91904, Givat Ram, Israel}

\date{\today}
\begin{abstract}
Enhancing precision sensors for stochastic signals using quantum techniques is a promising emerging field of physics. Estimating a weak stochastic waveform is the core task of many fundamental physics experiments including searches for stochastic gravitational waves, quantum gravity, and axionic dark matter. Simultaneously, noise spectroscopy and characterisation, e.g.\ estimation of various decay mechanisms in quantum devices, is relevant to a broad range of fundamental and technological applications. We consider the ultimate limit on the sensitivity of these devices for Lindblad estimation given any quantum state, fast and precise control sequence, and measurement scheme. We show that it is optimal to rapidly projectively measure and re-initialise the quantum state. We develop optimal protocols for a wide range of applications including stochastic waveform estimation, spectroscopy with qubits, and Lindblad estimation.
\end{abstract}
\maketitle
\allowdisplaybreaks

\begin{figure*}
    \centering
    \includegraphics[width=0.9\textwidth]{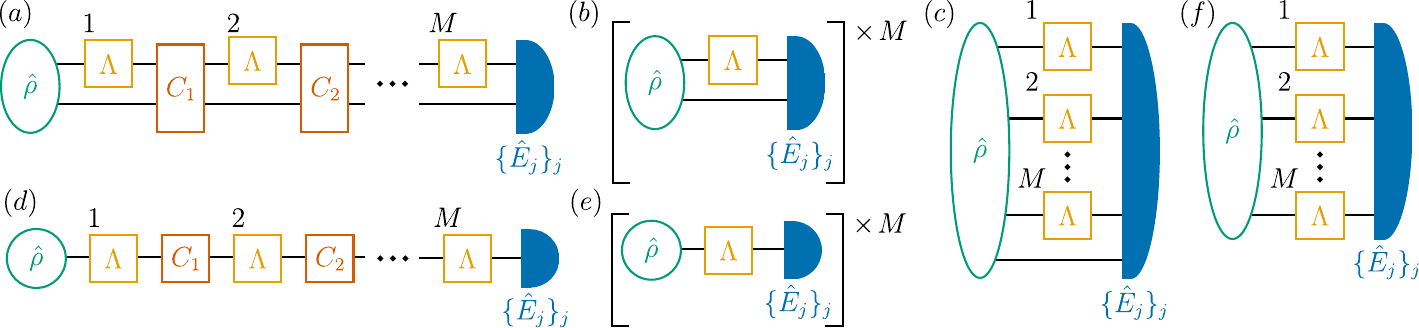}
    \caption{Block diagrams of different quantum metrological strategies. 
    \textbf{(a)}~Sequential strategy with fast and precise control where the encoding channel $\Lambda$ is queried $M$ times on a short timescale $t$ and between each query a control operation $C$ is performed. The final state after a total time of $T = M t$ is then measured. 
    \textbf{(b)}~Measure-and-reset strategy where the encoding channel is queried once, a measurement is performed, and then the state is re-initialised and the procedure is repeated a total of $M$ times. 
    \textbf{(c)}~Parallel strategy where $M$ devices are simultaneously queried and measured. 
    \textbf{(d--f)}~Unextended (i.e.\ ancilla-free) cases of the extended (i.e.\ noiseless ancilla--assisted) cases shown in panels (a--c), respectively.}
    \label{fig:diagram}
\end{figure*}

A wide variety of fundamental physics searches and emerging technologies involve sensing a weak stochastic signal: From stochastic waveform estimation~\cite{gardner2024stochastic} for searching for gravitational waves~\cite{AasiCQG15AdvancedLIGO,RomanoLRR17DetectionMethods,renzini2022stochastic}, quantum gravity~\cite{VerlindePLB21ObservationalSignatures,McCuller22SinglePhotonSignal,Vermeulen24PhotonCounting}, or axionic dark matter~\cite{rosenberg2000searches,graham2015experimental,AgrawalPRL24StimulatedEmission,ShinQI23UltimatePrecision}; to spectroscopy~\cite{mouradian2021quantum,rotem2019limits}, nano-scale nuclear magnetic resonance~\cite{gefen2019overcoming,cohen2020achieving}, and noise characterisation in quantum sensors and devices~\cite{norris2016qubit,shaw2024multi, stilck2024efficient,arshad2024real,PhysRevA.107.012611}. A common challenge across these different applications is how to differentiate between the weak stochastic signal and the noise in the system: The noise can completely mask the weak stochastic signal and significantly degrade the sensitivity. It is unclear whether this degraded sensitivity is fundamental to sensing a stochastic signal in the presence of noise, or whether it can be overcome with quantum-enhanced techniques. We address this problem by determining the optimal quantum metrological strategy to sense a weak stochastic signal in the presence of noise.

We now explain how we model this question as a Lindblad estimation problem. Under the Born-Markov approximation, the time evolution of the density matrix $\h\rho$ describing our quantum sensor is given by the following Lindblad master equation~\cite{alicki2007quantum,carmichael2009open,manzano2020short,campaioli2024quantum,albert2018lindbladians}:
\begin{align}
\label{eq:general Lindblad master equation}
    \dot{\h \rho} &= -i[\h H, \h\rho] + \gamma_1 \sum_{j=1}^{K} \mathcal{L}_{j,(s)}(\h \rho) + \sum_{k=K+1}^{K+N} \gamma_k \mathcal{L}_{k,(n)}(\h \rho)
\end{align}
where $\h H$ is the Hamiltonian with $\hbar=1$. The signal that we want to estimate is the common decay rate $\gamma_1$ of $K$ signal Lindblad jump operators $\{\h L_{j,(s)}\}_{j=1}^{K}$ each of which determines a Lindbladian superoperator as follows:
\begin{align*}
    \mathcal{L}_{j,(s)}(\h \rho) = \h L_{j,(s)} \h \rho \h L_{j,(s)}^\dag - \frac{1}{2} \{\h L_{j,(s)}^\dag \h L_{j,(s)},\h \rho\}.
\end{align*}
The signal $\gamma_1$, however, is much weaker than the noise described by the other $N$ decay rates $\{\gamma_k\}_{k=K+1}^{K+N}$ and noise Lindblad jump operators $\{\h L_{k,(n)}\}_{k=K+1}^{K+N}$ which similarly determine the noise Lindbladian superoperators $\{\mathcal{L}_{k,(n)}\}_{k=K+1}^{K+N}$ in Eq.~\ref{eq:general Lindblad master equation}. We want to determine how best to extract the information about the signal $\gamma_1$ encoded over time in the quantum state $\h\rho$ by preparing different initial states, applying different control operations during the time evolution, and performing different measurements at the end.

We want to determine the fundamental precision limit of this Lindblad estimation problem. This implies that we need to study the sequential metrological strategy with fast and precise control as shown in Fig.~\hyperref[fig:diagram]{\ref*{fig:diagram}a}. Here, the control operations that we apply to the quantum state can be arbitrary and continuously performed. This fast and precise control limit is hierarchically the most powerful metrological strategy as it can simulate all other strategies including, e.g., the measure-and-reset strategy shown in Fig.~\hyperref[fig:diagram]{\ref*{fig:diagram}b} or the parallel strategy shown in Fig.~\hyperref[fig:diagram]{\ref*{fig:diagram}c}~\cite{demkowicz2014using,sekatski2017quantum,zhou2021asymptotic}. The fast and precise control limit has been studied extensively for Hamiltonian estimation~\cite{sekatski2017quantum,demkowicz2017adaptive, ZhouNC18AchievingHeisenberg}, where the signal is encoded unitarily by $\h H$ in the presence of noise jump operators $\{\h L_{k,(n)}\}_{k=K+1}^{K+N}$ in Eq.~\ref{eq:general Lindblad master equation} and the root mean square error in estimating the signal falls either as $1/t$ or $1/\sqrt{t}$ over time $t$, depending on the geometry of the signal and noise operators. The fast and precise control limit has also been studied previously for Lindblad estimation~\cite{wan2022bounds,sekatski2022optimal,kurdzialek2023using,zhou2024achieving}, where the signal is encoded non-unitarily by signal jump operators and the noise corresponds to nuisance Hamiltonian $\h H$ and noise jump operators. In this case, the error in estimating the signal always falls as $1/\sqrt{t}$. But, the optimal initial state, control sequence, and measurement scheme for Lindblad estimation remain unknown.

In this paper, we consider the fast and precise control limit of Lindblad estimation. Firstly, in Sec.~\ref{sec:Review of quantum estimation}, we review the theory of quantum estimation. Then, in Sec.~\ref{sec:Estimating a weak decay rate}, we apply this formalism to find the optimal strategy for Lindblad estimation. In Sec.~\ref{sec:Classical limit}, we discuss the commuting Hermitian case. Finally, we apply our results to Lindblad estimation of a single qubit in Sec.~\ref{sec:Single qubit}, multiple qubits in Sec.~\ref{sec:Many qubits}, and a stochastic waveform in Sec.~\ref{sec:waveform estimation}.

\section{Review of quantum estimation} 
\label{sec:Review of quantum estimation}
We now review the quantum theory of asymptotic unbiased estimation using the Fisher information formalism. For an introduction to this topic, see Refs.~\cite{Wiseman09QuantumMeasurement,PezzeRMP18QuantumMetrology}.

Let us model an experiment as taking an initial quantum state described by a density matrix $\h \rho$ and transforming it to some final state $\h \rho'$ that encodes some real signal $\theta$ of interest but also introduces some noise. This transformation is described by a quantum channel $\Lambda$, i.e.\ a completely positive trace-preserving linear map between density matrices, such that $\h\rho' = \Lambda(\h\rho)$. The quantum channel $\Lambda$ has the following Kraus representation: 
\begin{align*}
    \h\rho' = \sum_j \h K_j \h\rho \h K_j^\dag, \qquad \sum_j \h K_j^\dag \h K_j = \h I
\end{align*}
where $\{\h K_j\}_j$ are the non-unique Kraus operators of $\Lambda$ and $\h I$ is the identity operator. We want to estimate $\theta$ from measuring $M$ independent and identical copies of the final state $\h \rho'$. This measurement is described by a positive operator-valued measure (POVM) with a set of effects $\{\h E_j\}_j$ such that the measurement outcome $j$ occurs with probability $\trSmall{\h\rho' \h E_j}$. (The effects satisfy $\sum_j \h E_j = \h I$ but are not necessarily a set of orthogonal projections.) We want to find the protocol that yields an unbiased estimate of $\theta$ with the minimum root mean square error (RMSE) $\Delta\theta$. This unextended measure-and-reset strategy is shown in Fig.~\hyperref[fig:diagram]{\ref*{fig:diagram}e}. We do not yet assume anything about the timescale of $\Lambda$.

Let the outcomes from a given measurement of $\h\rho'$ be described by a random variable $X$ with probability distribution $p(X=x|\theta)$. The classical Cram\'er-Rao bound on the RMSE is then
\begin{align*}
    \Delta\theta &\geq \frac{1}{\sqrt{M \IC(\theta)}}
\end{align*}
where the inequality is asymptotically tight as $M\rightarrow\infty$ using the maximum likelihood estimator, the factor of $1/\sqrt{M}$ comes from the Central Limit Theorem, and the classical Fisher information (CFI) about $\theta$ is defined as
\begin{align}
    \label{eq:CFI definition, discrete}
    \IC(\theta) &= \sum_j \frac{\left[\partial_\theta p(X=x_j|\theta)\right]^2}{p(X=x_j|\theta)}
\end{align}
where the sum excludes any $j$ such that $p(X=x_j|\theta)=0$. 

The CFI about $\theta$ is defined for a given measurement, but there are many possible measurements of $\h \rho'$ to perform. The quantum Fisher information (QFI) about $\theta$, $\IQ(\theta)$, is defined as the maximum possible value of the CFI about $\theta$ maximised over all possible measurements such that the quantum Cram\'er-Rao bound on the RMSE is
\begin{align*}
    \Delta\theta &\geq \frac{1}{\sqrt{M \IQ(\theta)}}
    .
\end{align*}
Here, the QFI is given by $\IQ(\theta) = \trSmall{\h\rho' \h S^2}$ where $\h S$ is the symmetric logarithmic derivative of $\h\rho'$ with respect to $\theta$ which satisfies $\partial_\theta\h\rho' = \frac{1}{2}\{\h \rho', \h S\}$ and is given, in terms of the spectral decomposition of the final state $\h\rho'=\sum_j p_j \ket{\phi_j}\bra{\phi_j}$, by
\begin{align*}
    \h S &= \sum_{j,k}\frac{2}{p_{j}+p_{k}}\braopket{\phi_{j}}{\partial_\theta \h\rho'}{\phi_{k}} \ket{\phi_{j}}\bra{\phi_{k}}
\end{align*}
where the sum excludes any $j, k$ such that $p_j + p_k = 0$. The QFI thus equals
\begin{align}
    \label{eq:QFI definition}
    \IQ(\theta) = \sum_{j,k}\frac{2}{p_j+p_k}\abs{\braopket{\phi_j}{\partial_\theta\h \rho'}{\phi_k}}^2
    .
\end{align}
An optimal measurement of $\h\rho'$ is to project onto the eigenbasis of $\h S$. Two useful properties are that the Fisher information from independent and identical repetitions sum and that the fractional RMSE is bounded by
\begin{align}\label{eq:fractional RMSE}
    \frac{\Delta\theta}{\theta} &\geq \frac{1}{\sqrt{M \theta^2 \IQ(\theta)}}
\end{align}
such that changing parameter from estimating $\theta$ to $\theta^2$ only scales the fractional RMSE by a factor of two since $\theta^2\IQ(\theta) = 4\theta^4\IQ(\theta^2)$ by the chain rule.

We are often also interested in sensing a weak parameter $\theta$, i.e.\ in studying the limit of vanishing signal: $\theta\rightarrow0$. If $\lim_{\theta\rightarrow0}\partial_\theta \h\rho'=0$ holds, then the optimal measurement in this limit is the projection $\h\Pi_0$ onto the eigenbasis of the signal-free final state $\lim_{\theta\rightarrow0}\h\rho'$ such that~\cite{gefen2019overcoming,gorecki2022quantum,haine2018using}
\begin{align}
    \label{eq:QFI in the vanishing signal limit}
    \lim_{\theta\rightarrow0}\partial_\theta \h\rho'=0 
    \implies
    \IQ(\theta=0) = \IC^{\h\Pi_0}(\theta=0).
\end{align}
This simplifies calculating the QFI in the vanishing signal limit into finding the CFI from the results of measuring $\h\Pi_0$.

While finding the optimal measurement is straightforward, finding the optimal initial state $\h \rho$ to maximise the QFI is more complicated. Due to the convexity of the space of density matrices and the QFI, the initial state can be assumed to be pure, i.e.\ $\h\rho = \proj\psi$ for some $\ket{\psi}$. This means that, without loss of generality, the state $\h \rho$ shown in Fig.~\ref{fig:diagram} is pure and the POVM is a projection-valued measure $\{\h \Pi_j\}_j$. The initial pure state $\ket{\psi}$ may include entanglement of the system with a noiseless ancilla such that we can then perform joint measurements of the final state of the system and the unchanged ancilla. (This is represented by extending the Kraus operators of $\Lambda$ to $\{\h K_j\otimes\h I\}_j$.) We will call states entangled with a noiseless ancilla ``extended states'' and ancilla-free states ``unextended states''. The extended measure-and-reset strategy is shown in Fig.~\hyperref[fig:diagram]{\ref*{fig:diagram}b}. Finding the optimal extended and unextended states for a particular channel $\Lambda$ will be the main focus of our work. The optimal extended state always performs as well as the optimal unextended state and, as we will see, sometimes the optimal performance can only be attained using entanglement. Numerically, the optimal extended state and its accompanying QFI and measurement scheme can be found efficiently using a semi-definite program (SDP)~\cite{demkowicz2012elusive,kolodynski2013efficient,zhou2021asymptotic}. 

Finally, we can consider improving the QFI by using fast and precise control. Given an infinitesimal channel $\Lambda$ which is repeated $M$ times with control as shown in Fig.~\hyperref[fig:diagram]{\ref*{fig:diagram}a}, the upper bound on the total QFI is~\cite{demkowicz2014using,sekatski2017quantum}
\begin{align}
\label{eq:ECQFI with fast and precise control}
\IQ(\theta)\leq4M\norm{\h \alpha}+4M(M-1)\normSmall{\h \beta}(\normSmall{\h \beta}+2\sqrt{\norm{\h \alpha}})
\end{align}
where $\h\alpha=\dot{\mathbf{K}}^\dag\dot{\mathbf{K}}$ and $\h\beta=i\dot{\mathbf{K}}^\dag\mathbf{K}$ given the vector $\mathbf{K}=(\h K_1, \mathellipsis, \h K_r)^\T$ from a minimum rank $r$ Kraus representation $\{\h K_j\}_{j=1}^r$ of $\Lambda$. (Here, $\dot{\mathbf{K}}$ is the derivative of $\mathbf{K}$ with respect to the parameter of interest $\theta$.) This chosen Kraus representation is not unique, however, as any $r$-by-$r$ unitary matrix $u$ forms an equally valid Kraus representation $u\mathbf{K}$. All minimum rank Kraus representations of $\Lambda$ can be reached this way. Under this gauge transformation, $\h\alpha$ and $\h\beta$ become 
 \begin{align}\label{eq:alpha and beta}
\h\alpha=(\dot{\mathbf{K}}-ih\mathbf{K})^{\dagger}(\dot{\mathbf{K}}-ih\mathbf{K}), 
\quad \h\beta=i(\dot{\mathbf{K}}-ih\mathbf{K})^{\dagger}\mathbf{K}
 \end{align}
where $h$ is any $r$-by-$r$ Hermitian matrix. In the relevant case that $\h \beta=0$, then Eq.~\ref{eq:ECQFI with fast and precise control} minimised over all possible Kraus representations becomes
\begin{align}\label{eq:ECQFI with fast and precise control, gauge form}
\IQ(\theta)\leq4M\min_h\norm{(\dot{\mathbf{K}}-ih\mathbf{K})^{\dagger}(\dot{\mathbf{K}}-ih\mathbf{K})}.
\end{align}
This upper bound is tight assuming access to a noiseless ancilla, i.e.\ the extended case shown in Fig.~\hyperref[fig:diagram]{\ref*{fig:diagram}a} rather than the unextended case shown in Fig.~\hyperref[fig:diagram]{\ref*{fig:diagram}d}. Since fast and precise control is hierarchically the most powerful metrological strategy, Eq.~\ref{eq:ECQFI with fast and precise control, gauge form} provides the ultimate sensitivity limit assuming that $\h \beta=0$. Eq.~\ref{eq:ECQFI with fast and precise control, gauge form} can also be efficiently calculated numerically via an SDP.

\section{Lindblad estimation}
\label{sec:Estimating a weak decay rate}
We now return to our Lindblad estimation problem in Eq.~\ref{eq:general Lindblad master equation}~\footnote{We use the terms ``Lindblad estimation'' and ``stochastic signal sensing'' interchangeably. If the Born-Markov approximation does not hold, however, then the stochastic signal may not be described by a Lindbladian and a different approach is necessary.}. Before applying the above quantum estimation formalism to this problem, we need to explain some of the simplifying assumptions that we make. We assume that there is no nuisance Hamiltonian $\h H=0$ and that the signal is much weaker than the noise such that the following condition holds for all $k=K+1,\mathellipsis,K+N$
\begin{align}\label{eq:vanishing signal condition}
    \gamma_1 \ll \frac{\gamma_k \normSmall{\mathcal{L}_{k,(n)}(\h \rho)}}{\normSmall{\sum_{j=1}^{K} \mathcal{L}_{j,(s)}(\h \rho)}}
\end{align}
where $\normSmall{\cdot}$ is the operator norm~\footnote{The operator norm $\normSmall{\cdot}$ is induced by the $\ell_2$ norm of the kets. For a given operator $\h A$, $\normSmall{\h A}$ is also called the spectral norm and equals the largest singular value of $\h A$.}. Informally, this condition means that the signal decay rate $\gamma_1$ is much slower than all of the noise decay rates $\gamma_k$~\footnote{In the noiseless case, there are no other natural scales to compare to the size of the signal. Instead, we first derive the Kraus operators in Eq.~\ref{eq:Kraus representation} assuming that the short time condition in Eq.~\ref{eq:short time} holds which implies that $\gamma_1t\ll1$. Then, we take the limit of $\gamma_1t\rightarrow0$ of the Kraus operators in Eq.~\ref{eq:Kraus representation} to reach Eq.~\ref{eq:K, Kdot}. This is what we mean by the vanishing signal limit in the noiseless case.}. 
We choose to estimate $\sqrt{\gamma_1}$ instead of $\gamma_1$ without loss of generality to avoid a divergence in the QFI in this vanishing signal limit in the noiseless case~\cite{gardner2024stochastic}. The fractional RMSEs with respect to $\sqrt{\gamma_1}$ and $\gamma_1$ are the same up to a factor of two by Eq.~\ref{eq:fractional RMSE} such that any sensitivity improvement that we demonstrate for $\sqrt{\gamma_1}$ is similar for $\gamma_1$. When estimating $\sqrt{\gamma_1}$, we assume that all of the jump operators $\{\h L_j\}_{j=1}^{K+N}$ are known. We also assume for now that we know the noise decay rates $\{\gamma_j\}_{j=K+1}^{K+N}$, but we will show later that we can remove this assumption. The jump operators $\h L_j$ are traceless, i.e.\ $\trSmall{\h L_j}=0$, without loss of generality since any component of $\h L_j$ along $\h I$ will not affect Eq.~\ref{eq:general Lindblad master equation}. We make no assumption about the dimension of the Hilbert space which may be infinite.

We want to find the optimal metrological strategy and ultimate precision limit of this Lindblad estimation problem. We thus study the sequential strategy with fast and precise control since it is hierarchically the most powerful metrological strategy. This means that we can perform arbitrary quantum control operations including the use of noiseless ancillae as shown in Fig.~\hyperref[fig:diagram]{\ref*{fig:diagram}a}. We need to first find the quantum channel $\Lambda$ that represents the time evolution under the master equation in Eq.~\ref{eq:general Lindblad master equation} after an infinitesimal time $t$. Given the initial state $\h\rho$, then the final state $\h\rho'=\Lambda(\h\rho)$ is
\begin{align}
\label{eq:general Lindblad master equation, short time}
    \h\rho' 
    &= \h\rho + t \sum_{j=1}^{K+N} \gamma_j \mathcal{L}_j(\h \rho) 
    \\&= \sum_{j=0}^{K+N} \h K_j \h\rho \h K_j^\dag\nonumber
\end{align}
where one possible Kraus representation of $\Lambda$ is
\begin{align}
\label{eq:Kraus representation}
    \h K_0 &= \h I - \frac{t}{2}\sum_{k=1}^{K+N} \gamma_k \h L_k^\dag\h L_k
    \\\h K_j &= \sqrt{\gamma_j t}\h L_j, \quad j=1,\mathellipsis,K+N\nonumber
    .
\end{align}
In Eq.~\ref{eq:general Lindblad master equation, short time} and henceforth, we drop all $\order{t^2}$ terms provided that the short evolution time $t$ satisfies 
\begin{align}
    \label{eq:short time}
    t \ll \frac{4\normSmall{\sum_{j = 1}^{K+N} \gamma_j \mathcal{L}_j(\h \rho)}}{\normSmall{\sum_{j,k = 1}^{K+N} \gamma_j\gamma_k \mathcal{L}_j(\mathcal{L}_k(\h \rho))}}.
\end{align}
Intuitively, the time $t$ must be faster than the shortest $\order{1/\gamma}$ timescale set by the largest decay rate $\gamma$. For example, for sensing a bosonic loss given by the annihilation operator $\h L_{1,(s)}=\h a$ with a Fock state $\ket{\h n = n}$ such that $\h n\ket{\h n = n} = n\ket{\h n = n}$ where $\h n = \h a^\dag \h a$ and $n\geq1$, then Eq.~\ref{eq:short time} implies that the time must satisfy $t\ll 4/[\gamma_1(2n-1)]$. 
We now want to find the optimal initial state, control sequence, and projective measurement for the sequential strategy with fast and precise control.

Let us briefly address the existing literature on Lindblad estimation. Estimating a weak decay rate while knowing the jump operators is different to the problem of estimating the whole Lindbladian with no prior information. In particular, the methods developed previously to estimate the whole Lindbladian~\cite{boulant2003robust,ben2020direct, stilck2024efficient} are not fine-tuned to isolate the noise-free component of a weak signal and will have vanishing QFI in this limit. While estimating the parameter of a Lindbladian with fast and precise control has been previously studied in Ref.~\cite{sekatski2022optimal}, the particular assumption made in that work about the jump operators does not hold in general, unlike our results.

\subsection{Fast and precise control}
Let us first calculate the ultimate sensitivity limit given by Eq.~\ref{eq:ECQFI with fast and precise control, gauge form} for sensing a stochastic signal or Lindblad estimation with fast and precise control as shown in Fig.~\hyperref[fig:diagram]{\ref*{fig:diagram}a}. In the vanishing signal limit of $\sqrt{\gamma_1}\rightarrow0$, the Kraus operators in Eq.~\ref{eq:Kraus representation} and their derivatives with respect to $\sqrt{\gamma_1}$ become
\begin{align}\label{eq:K, Kdot}
\mathbf{K} = \pmatrixByJames{\h I - \frac{t}{2}\sum_{k=K+1}^{K+N} \gamma_k \h L_k^\dag\h L_k \\ 0\\ \vdots\\ 0\\ \sqrt{\gamma_{K+1} t}\h L_{K+1,(n)}\\ \vdots\\ \sqrt{\gamma_{K+N} t}\h L_{K+N,(n)}}
, \quad 
\dot{\mathbf{K}} = \pmatrixByJames{0\\ \h L_{1,(s)}\\ \vdots\\ \h L_{K,(s)}\\ 0\\ \vdots\\ 0}\sqrt{t}
.
\end{align}
We choose the following ansatz for the gauge matrix: 
\begin{align}
    \label{eq:gauge}
    h &= \pmatrixByJames{
    0 & \mathbf{g}^\dag & \mathbf{0}_{1,N} \\
    \mathbf{g} & \mathbf{0}_{K,K} & \mathbf{G} \\
    \mathbf{0}_{N,1} & \mathbf{G}^\dag & \mathbf{0}_{N,N}\\
    }
\end{align}
where $\mathbf{0}_{m,n}$ is the $m$-by-$n$ zero matrix and $\vec{c}_j\in\C^{N+1}$ for $j=1,\mathellipsis,K$ such that $h$ is determined by
\begin{align*}
    \mathbf{g} &= -i\sqrt{t}\pmatrixByJames{(\vec{c}_1)_1\\ \vdots\\ (\vec{c}_K)_1}
    , \quad
    \mathbf{G} = -i\pmatrixByJames{
    \frac{(\vec{c}_1)_2}{\sqrt{\gamma_{K+1}}} & \cdots & \frac{(\vec{c}_1)_{N+1}}{\sqrt{\gamma_{K+N}}}\\
    \vdots & \ddots & \vdots \\ 
    \frac{(\vec{c}_K)_2}{\sqrt{\gamma_{K+1}}} & \cdots & \frac{(\vec{c}_K)_{N+1}}{\sqrt{\gamma_{K+N}}}\\
    }
\end{align*}
Using this gauge matrix $h$, the Kraus operators in Eq.~\ref{eq:K, Kdot} become
\begin{align}
    \label{eq:Kdot - i h K}
\dot{\mathbf{K}}-ih\mathbf{K}=
\pmatrixByJames{0\\ \h L_{1,(s)}-\vec{c}_1^\T\mathbf{L}\\ \vdots\\ \h L_{K,(s)}-\vec{c}_K^\T\mathbf{L}\\ 0\\ \vdots\\ 0}\sqrt{t}
\end{align}
where $\mathbf{L}=(\h I,\h L_{K+1,(n)},\mathellipsis,\h L_{K+N,(n)})^\T$. 
This implies that $\h\beta = 0$ in Eq.~\ref{eq:alpha and beta}. The upper bound on the QFI with fast and precise control in Eq.~\ref{eq:ECQFI with fast and precise control, gauge form} is thus:
\begin{align}
    \label{eq:ECQFI minimisation, Multiple sources of signal and noise}
    \IQ(\sqrt{\gamma_1}=0) \leq 4T \min_{\{\vec{c}_k\}_{k=1}^K} \biggl\|\sum_{k=1}^K \h A_k^\dag \h A_k\biggr\|
\end{align}
where $\h A_{k}=\h L_{k,(s)}-\vec{c}_k^\T\mathbf{L}$ and $T=Mt$ is the total time. Recalling that $\normSmall{X^\dag X}=\normSmall{X}^2$ for the operator norm, then Eq.~\ref{eq:ECQFI minimisation, Multiple sources of signal and noise} has the following geometric interpretation in the case of a single signal ($K=1$):
\begin{align}
    \label{eq:ECQFI_as_distance}
     \IQ(\sqrt{\gamma_1}=0) &\leq 4T \min_{\vec{c}} \normSmall{\h L_{1,(s)} - \vec{c}^\T\mathbf{L}}^2
     \\& = 4 T\, d(\h L_{1,(s)},\text{span}\{\mathbf{L}\})^2 \nonumber
\end{align}
where $d(\h L_{1,(s)},\text{span}\{ \mathbf{L}\})$ is the distance induced by the operator norm between the signal operator $\h L_{1,(s)}$ and the subspace of noise operators $\text{span}\{\mathbf{L}\}$. This distance vanishes if and only if $\h L_{1,(s)}\in\text{span}\{\mathbf{L}\}$ such that the QFI is nonzero only if $\h L_{1,(s)}\notin\text{span}\{\mathbf{L}\}$, i.e.\ some component of the signal operator lies outside of the subspace of noise operators.

We prove that the upper bound in Eq.~\ref{eq:ECQFI minimisation, Multiple sources of signal and noise} is tight and determine the optimal sensing strategy:
\begin{theorem}
\label{claim:optimal QFI}
Consider sensing the decay rate $\gamma_1$ common to $K$ signal operators $\{\h L_{k,(s)}\}_{k=1}^{K}$ in the presence of $N$ noise operators $\{\h L_{j,(n)}\}_{j=K+1}^{K+N}$. In the vanishing signal limit of $\gamma_{1} \rightarrow 0$, the optimal QFI with respect to $\sqrt{\gamma_1}$ is:
\begin{align*}
    \IQ(\sqrt{\gamma_1}=0) &= 4 T \max_{\ket{\psi}} \sum_{k=1}^{K} \braopket{\psi}{\h L_{k,(s)}^\dag (\h I - \h \Pi) \h L_{k,(s)}}{\psi}
\end{align*}
where $\h \Pi$ is the projection onto the noisy subspace $\text{span}\{\mathbf{L}\ket{\psi}\}$ of the images of the state $\ket{\psi}$ under the noise operators $\mathbf{L}=(\h I,\h L_{K+1,(n)},\mathellipsis,\h L_{K+N,(n)})^\T$. This QFI is achieved by a measure-and-reset strategy that continuously prepares the optimal state $\ket{\psi}$ above and projects onto the noisy subspace with $\h \Pi$.
\end{theorem}
This is the first main result of our work. In Appendix~\ref{app:ECQFI_bounds}, we show that the expression in Theorem~\ref{claim:optimal QFI} provides an upper bound on the optimal QFI by evaluating the minimisation over $\{\vec{c}_k\}_{k=1}^K$ in Eq.~\ref{eq:ECQFI minimisation, Multiple sources of signal and noise}. We now show that this upper bound is tight and can be saturated by a measure-and-reset strategy.

\subsection{Measure-and-reset strategy}
One possible control operation is to measure projectively and then re-initialise the state. If we repeatedly perform this control operation, then we reach the measure-and-reset strategy shown in Fig.~\hyperref[fig:diagram]{\ref*{fig:diagram}b}. (The same is true in the ancilla-free case as shown in Figs.~\hyperref[fig:diagram]{\ref*{fig:diagram}d--e}.) The measure-and-reset strategy is thus similar to a quantum jump experiment~\cite{nagourney1986shelved}.

In general, the optimal measure-and-reset strategy is outperformed by the optimal fast and precise control strategy~\cite{demkowicz2014using,sekatski2017quantum,zhou2021asymptotic}. For example, rapidly projectively measuring a state undergoing Hamiltonian evolution leads to the quantum Zeno effect~\cite{itano1990quantum} unlike the optimal fast and precise control strategy which leads to a coherent build-up of signal~\cite{ZhouNC18AchievingHeisenberg}. For sensing a stochastic signal, however, we will show that the optimal measure-and-reset strategy attains the ultimate sensitivity limit with fast and precise control. Intuitively, this is because the stochastic signal does not build up coherently over time.

Let us find the optimal measure-and-reset strategy and show that it saturates the ultimate sensitivity limit in Theorem~\ref{claim:optimal QFI}. We assume that we can rapidly projectively measure $\h\rho'$ in Eq.~\ref{eq:general Lindblad master equation, short time} and then re-initialise the state to $\h\rho=\proj{\psi}$ if we measure that a jump occurred. We repeat this process $M\gg1$ times over a long total time $T=Mt$. The total QFI is thus $M$ times the QFI from a single measurement after a time $t$. We now need to find the optimal initial pure state $\ket{\psi}$ and projective measurement to estimate $\gamma_1$. By Eq.~\ref{eq:general Lindblad master equation, short time}, $\partial_{\sqrt{\gamma_1}} \h\rho'=0$ holds in the vanishing signal limit of $\sqrt{\gamma_1}\rightarrow0$ such that the optimal measurement is to project onto the eigenbasis of the signal-free final state $\lim_{\sqrt{\gamma_1}\rightarrow0}\h\rho'$ by Eq.~\ref{eq:QFI in the vanishing signal limit}. To calculate this eigenbasis, let us first find an orthonormal basis to express the final state.

\begin{figure}
    \centering
    \includegraphics[width=0.9\columnwidth]{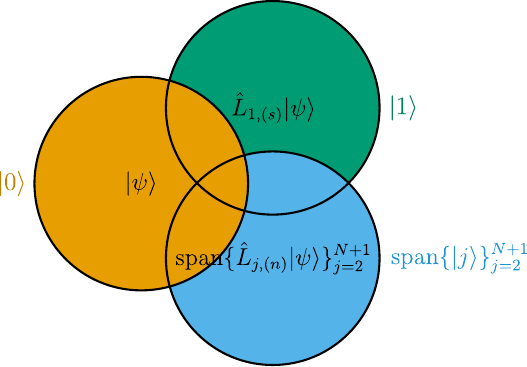}
    \caption{Visualisation of the Gram-Schmidt process for orthogonalising the images of the initial state $\ket{\psi}$ under $N$ noise operators $\text{span}\{\h L_{j,(n)}\ket{\psi}\}_{j=2}^{N+1}$ and one signal operator $\h L_{1,(s)}\ket{\psi}$. The component of the weak signal within the noisy subspace $\text{span}\{\ket{0},\ket{j}\}_{j=2}^{N+1}$ is lost leaving only the noise-free component $\ket{1}$.}
    \label{fig:venn_diagram}
\end{figure}
We perform the Gram-Schmidt process to the images of the initial state $\ket{\psi}$ under the noise and signal jump operators in the following order:
\begin{align}
    \label{eq:initial state and images}
    &\ket{\psi}, \h L_{K+1,(n)}\ket{\psi}, \h L_{K+2,(n)}\ket{\psi}, \mathellipsis, \h L_{K+N,(n)}\ket{\psi},\\&\hspace{0.5cm}\h L_{1,(s)}\ket{\psi}, \h L_{2,(s)}\ket{\psi}, \mathellipsis, \h L_{K,(s)}\ket{\psi}.\nonumber
\end{align}
That is, we orthogonalise first the initial state, then the images under the noise operators, and finally the images under the signal operators. This is shown for a single signal operator in Fig.~\ref{fig:venn_diagram}. This process defines the orthonormal basis $\{\ket{j}\}_{j=0}^{K+N}$ where $\ket{0}=\ket{\psi}$ such that
\begin{align*}
    \h L_{K+j,(n)} \ket{0} &= c_{K+j,0}\ket{0} + \sum_{k=1}^j c_{K+j,K+k} \ket{K+k} 
    \\\h L_{j,(s)} \ket{0} &= \sum_{k=1}^N c_{j,K+k} \ket{K+k} + \sum_{k=0}^{j} c_{j,k} \ket{k} 
    .
\end{align*}
We have thus split the space into the noisy subspace $\text{span}\{\ket{0},\ket{j}\}_{j=K+1}^{K+N}=\text{span}\{\mathbf{L}\ket{\psi}\}$ and the noise-free subspace $\text{span}\{\ket{j}\}_{j=1}^{K}$ which can only be reached if a signal jump occurs. Let $\mathbf{v}=(\ket{0},\ket{K+1},\mathellipsis,\ket{K+N})^\T$ such that $\mathbf{L} \ket{\psi} = \mathbf{M}\mathbf{v}$ where 
\begin{align*}
    \mathbf{M}&=
    \bmatrixByJames{\braopket{0}{\mathbf{L}}{0}&\braopket{K+1}{\mathbf{L}}{0}&\cdots&\braopket{K+N}{\mathbf{L}}{0}}
    .
\end{align*}
The projection onto the noisy subspace is thus
\begin{align}
    \h \Pi
    &= \proj{0} + \sum_{j=K+1}^{K+N} \proj{j}\nonumber
    \\&= \mathbf{v}^\T\mathbf{v}^*\nonumber
    \\&= \mathbf{L}^\T\ket{\psi}\braopket{\psi}{\mathbf{L}^*\mathbf{L}^\T}{\psi}^{-1}\bra{\psi}\mathbf{L}^*
    \label{eq:Pi}
\end{align}
where we used that $\mathbf{v} = \mathbf{M}^{-1} \mathbf{L} \ket{\psi}$ and $\h\Pi\mathbf{L}\ket{\psi}=\mathbf{L}\ket{\psi}$. Analytically, we thus need to invert the $N+1$ by $N+1$ matrix $\braopket{\psi}{\mathbf{L}^*\mathbf{L}^\T}{\psi}$ to find $\h\Pi$, whereas numerically the Gram-Schmidt process can be efficiently performed to find $\h \Pi$. 
Here, we have assumed that the images under the noises in Eq.~\ref{eq:initial state and images} are linearly independent such that the noisy subspace has dimension $N+1$. If the noisy subspace instead has dimension $N'+1<N+1$, then we can calculate $\h\Pi = \mathbf{v}^\T\mathbf{v}^*$ from $\mathbf{L} \ket{\psi} = \mathbf{M}\mathbf{v}$ and the Moore-Penrose pseudoinverse of the $N+1$ by $N'+1$ matrix $\mathbf{M}$.

We calculate the QFI of the measure-and-reset strategy with a given state $\ket{\psi}$ using Eq.~\ref{eq:QFI in the vanishing signal limit}. We determine the eigenbasis of the signal-free final state $\lim_{\sqrt{\gamma_1}\rightarrow0}\h\rho'$ from Eq.~\ref{eq:general Lindblad master equation, short time} in the orthonormal basis $\{\ket{j}\}_{j=0}^{K+N}$. We show that the QFI given a state $\ket{\psi}$ equals
\begin{align} 
    \IQ(\sqrt{\gamma_1}=0) &= 4 T \sum_{k=1}^{K} \braopket{\psi}{\h L_{k,(s)}^\dag (\h I - \h \Pi) \h L_{k,(s)}}{\psi}.
 \label{eq:QFI_with_measure_reset} 
\end{align}
This QFI for the optimal initial state $\ket{\psi}$ thus saturates the ultimate sensitivity limit in Theorem~\ref{claim:optimal QFI}. We prove these results in Appendix~\ref{app:QFI for multiple signal and noise operators}. We show that the eigenbasis is not exactly $\{\ket{j}\}_{j=0}^{K+N}$ but that the CFI from projecting onto $\{\ket{j}\}_{j=0}^{K+N}$ equals the QFI to leading order in time $t$. Moreover, it suffices to project onto the noisy subspace with $\h\Pi$ since any nonzero noise is louder than the vanishing signal such that the component of the signal within the noisy subspace is completely lost. It remains to find the optimal initial state $\ket{\psi}$ for a given set of jump operators. We derive the conditions on a given state being optimal in Appendix~\ref{app:optimal_states}. Although we have found the optimal strategy overall, we have not identified the optimal strategy when the initial state, control sequence, or final measurement is non-optimal and fixed.

\subsection{Properties of the optimal strategy}
\label{sec:properties}
Let us now discuss some properties of the optimal measure-and-reset strategy which saturates the ultimate sensitivity limit in Theorem~\ref{claim:optimal QFI}. 

The necessary and sufficient condition for the QFI to be nonzero in the limit of vanishing signal is the following:
\begin{align}
    \label{eq:QFI nonzero condition}
    \exists j\in\{1,\mathellipsis,K\}, \quad \h L_{j,(s)} \notin \text{span}\{\mathbf{L}\}.
\end{align}
We previously observed in Eq.~\ref{eq:ECQFI_as_distance} that this is a necessary condition in the case of a single signal operator. More generally, this is a necessary condition since any component of a signal operator parallel to a noise operator or the identity does not contribute to the QFI in Theorem~\ref{claim:optimal QFI}. If all of the signal operators lie in $\text{span}\{\mathbf{L}\}$, then the QFI is thus zero in the limit of vanishing signal. We may assume henceforth without loss of generality that each signal operator is orthogonal to each noise operator, and that the noise operators are pairwise orthogonal. But, the signal operators need not be pairwise orthogonal.

This phenomenon in noise sensing where the QFI vanishes as the parameter does is called the ``Rayleigh curse'' by analogy to quantum super-resolution~\cite{tsang2016quantum}. The noise-induced Rayleigh curse was recently studied in several works, e.g.\ see Refs.~\cite{gardner2024stochastic,gefen2019overcoming,oh2021quantum,shi2024quantum}. Here, we present the general criteria for avoiding it.

That Eq.~\ref{eq:QFI nonzero condition} is also a sufficient condition on the QFI not vanishing can be shown by direct calculation for a particular extended state. Consider the Bell state defined as $\ket{\psi}=\frac{1}{\sqrt{d}}\sum_{j=1}^d\ket{e_j}\otimes\ket{e_j}$ for a given basis $\{\ket{e_j}\}_{j=1}^d$. (Here, we consider a Hilbert space with finite dimension $d$.) The Bell state has the property that $\braopket{\psi}{\h A\otimes \h I}{\psi} = \frac{1}{d}\trSmall{\h A}$ for any $\h A$ such that the QFI in Theorem~\ref{claim:optimal QFI} becomes
\begin{align}
    \label{eq:QFI, Bell state}    
    \IQ(\sqrt{\gamma_1}=0) &= 4T\frac{1}{d} \sum_{j=1}^{K} \trSmall{\h L_{j,(s)}^\dag \h L_{j,(s)}}
\end{align}
where we replaced each signal operator $\h L_{j,(s)}$ with its component orthogonal to $\text{span}\{\mathbf{L}\}$. If Eq.~\ref{eq:QFI nonzero condition} holds, then at least one of these components is nonzero and the QFI in Eq.~\ref{eq:QFI, Bell state} is nonzero. The Bell state is not necessarily the optimal initial state, however, as we will see later.

The condition in Eq.~\ref{eq:QFI nonzero condition} states when it is possible to overcome the noise limitations in Lindblad estimation. Similar noise limitations were studied extensively in the deterministic (unitary) case. In the deterministic case, we want to sense some parameter $\theta$ from measurements of $\h\rho'=\h U\h\rho\h U^\dag$ where $\h U = \exp(-i\theta \h H T)$ for some Hermitian $\h H$. The QFI in this case is $\IQ(\theta) = 4 T^2 \varSubSmall{\ket{\psi}}{\h H}$. Here, the QFI is independent of the parameter $\theta$ and coherently increases as $T^2$ compared to the stochastic case which only scales as $T$. These rates are sometimes called ``Heisenberg scaling with time'' for $T^2$ and ``Standard Quantum Limit scaling with time'' for $T$, respectively. The fact that Lindblad estimation with fast and precise control cannot achieve Heisenberg scaling with time is well-known~\cite{wan2022bounds,sekatski2022optimal,kurdzialek2023using,zhou2024achieving}. For the deterministic case, the optimal strategy with fast and precise control is not to measure-and-reset since we want to coherently build up the signal as $T^2$, instead the optimal strategy is to perform quantum error correction. This strategy achieves Heisenberg scaling with time provided that the following necessary and sufficient condition holds:~\cite{ZhouNC18AchievingHeisenberg}
\begin{align*}
    \h H \notin \text{span}\{\h I, \h L_{j,(n)},\h L_{j,(n)}^\dag,\h L_{j,(n)}^\dag \h L_{k,(n)}\}_{j,k=K}^{K+N}.
\end{align*}
In comparison, in the stochastic case, the condition in Eq.~\ref{eq:QFI nonzero condition} is on the QFI not vanishing and can be met even if, e.g., $\h L_{1,(s)}=\h L_{2,(n)}^\dag$ or $\h L_{1,(s)}=\h L_{2,(n)}^\dag\h L_{2,(n)}$. Thus, the condition in the deterministic case to achieve $T^2$ scaling is not equivalent to the condition in the stochastic case to achieve nonzero QFI scaling as $T$.

We also observe in the vanishing signal limit that the measurement probabilities in Eq.~\ref{eq:probabilities 1,...,K} and thus the QFI of the optimal strategy do not depend on the noise decay rates $\{\gamma_j\}_{j=K+1}^{K+N}$ or the size of the noise jump operators $\{\normSmall{\h L_{j,(n)}}\}_{j=K+1}^{K+N}$ provided that they are nonzero. Instead, they only depend on the geometry of the noise operators which determines the noisy subspace projection $\h\Pi$ since any nonzero noise is louder than the vanishing signal. We can thus relax our assumption that we know the noise decay rates and can assume that the noise jump operators have unit norm without loss of generality. For finite signals, however, the QFI will depend on and require knowing the noise decay rates and jump operator norms~\cite{gardner2024stochastic}. 

Finally, we see that the QFI also only depends on the reduced density matrix of the initial state, $\h\rho_S = \trSubSmall{A}{\proj{\psi}}$, after tracing out any ancilla. This is because the QFI in Theorem~\ref{claim:optimal QFI} depends only on the expectation values of operators that act on the system. This means that the maximisation can be performed over the reduced density matrix $\h\rho_S$ instead of the full extended state $\ket{\psi}$~\cite{haine2015quantum}. The QFI here is different than if we prepare the system in the mixed state $\h\rho_S$, which is suboptimal, because to derive Theorem~\ref{claim:optimal QFI} we assumed that we can initialise and measure the ancilla too.

\subsection{Examples of the optimal QFI}
Let us now study the QFI of the optimal strategy from Theorem~\ref{claim:optimal QFI} for different numbers of noise operators.

\subsubsection{Noiseless case}
Consider the noiseless case with $K$ signal jump operators, where the optimal QFI is
\begin{align}
    \label{eq:QFI, any K, N=0}
    \IQ(\sqrt{\gamma_1}=0) &= 4 T \max_{\ket{\psi}} \sum_{j=1}^{K} \varSubSmall{\ket{\psi}}{\h L_{j,(s)}}
    .
\end{align}
Here, the variance of an operator $\h A$ is defined as $\varSubSmall{\ket{\psi}}{\h A}=\covSubSmall{\ket{\psi}}{\h A}{\h A}$ where the covariance of two operators $\h A$ and $\h B$ is defined as
\begin{align*}
    \covSubSmall{\ket{\psi}}{\h A}{\h B} = \braopket{\psi}{\h A^\dag \h B}{\psi} - \braopket{\psi}{\h A^\dag}{\psi} \braopket{\psi}{\h B}{\psi}
    .
\end{align*}
The following measure-and-reset strategy achieves the noiseless QFI in Eq.~\ref{eq:QFI, any K, N=0}: Prepare the optimal initial state $\ket\psi$ that maximises the sum of variances above and project onto it since $\h\Pi=\proj{\psi}$.

Let us consider the noiseless case with a single signal jump operator $\h L_{1,(s)}$. We need to find the state $\ket{\psi}$ that maximises the variance of $\h L_{1,(s)}$ since the QFI is
\begin{align}
    \label{eq:QFI, K=1, N=0}
    \IQ(\sqrt{\gamma_1}=0) &= 4 T \max_{\ket{\psi}} \varSubSmall{\ket{\psi}}{\h L_{1,(s)}}
    .
\end{align}
If $\h L_{1,(s)}$ is Hermitian, then this process corresponds to a random unitary channel with $\h L_{1,(s)}$ as the generator~\cite{gardner2024stochastic}. The optimal states are then equal superpositions of the eigenvectors of $\h L_{1,(s)}$ with the maximum $\lambda_+$ and minimum $\lambda_-$ eigenvalues, i.e.\ $\ket{\psi} = \frac{1}{\sqrt{2}}(\ket{\lambda_+} + e^{i\phi}\ket{\lambda_-})$ for any $\phi$, such that the QFI is $\IQ(\sqrt{\gamma_1}=0) = T (\lambda_+ - \lambda_-)^2$. Since this is the optimal state, there is no benefit from entangling with an ancilla and thus Fig.~\hyperref[fig:diagram]{\ref*{fig:diagram}b} and Fig.~\hyperref[fig:diagram]{\ref*{fig:diagram}e} are equivalent in this case.

If instead $\h L_{1,(s)}$ is non-Hermitian, then finding the optimal state is more difficult. The maximal variance and optimal state can be obtained numerically by using an SDP to compute $\min_{c\in\mathbb{C}}\bigl\|\h A^{\dag}\h A\bigr\|$ where $\h A=\h L_{1,(s)}-c\h I$. The optimal state $\ket{\psi}$ lies in the maximal eigenspace of $\h A^{\dag}\h A$ and satisfies $\braopket{\psi}{\h A}{\psi}=0$.

The optimality and QFI of a continuous measure-and-reset strategy in the noiseless case have previously been shown in Ref.~\cite{sekatski2022optimal} but only for a set of signal jump operators satisfying a particular assumption that we do not make in deriving the results above.

\subsubsection{One source of noise}
Let us now consider the case where there is a single noise jump operator $\h L_{K+1,(n)}$ such that the QFI in Eq.~\ref{eq:QFI_with_measure_reset} becomes
\begin{align}
\label{eq:QFI, any K, N=1} 
&\IQ(\sqrt{\gamma_1}=0)
    \\&= 4T 
    \sum_{j=1}^K
    \left(\varSubSmall{\ket{\psi}}{\h L_{j,(s)}} - \frac{\abs{\covSubSmall{\ket{\psi}}{\h L_{j,(s)}}{\h L_{K+1,(n)}}}^2}{\varSubSmall{\ket{\psi}}{\h L_{K+1,(n)}}}\right) \nonumber
\end{align}
where the optimal QFI in Theorem~\ref{claim:optimal QFI} is obtained by maximising this expression over all possible $\ket{\psi}$. Compared to the noiseless QFI in Eq.~\ref{eq:QFI, any K, N=0}, the noisy QFI loses the component of each signal within the noisy subspace resulting in a penalty related to the covariance between the signal and noise operators. This lost signal component within the noisy subspace is shown as the intersection of the two corresponding circles in Fig.~\ref{fig:venn_diagram} in the single signal case. This leads to a tradeoff when optimising Eq.~\ref{eq:QFI, any K, N=1} against $\ket{\psi}$ between maximising the sum of the variances and minimising the covariance penalty term. We will study some examples later where the noiseless QFI either can or cannot be recovered due to this tradeoff.

\subsubsection{More sources of noise}
In general, the expression for the QFI in terms of only the optimal initial state $\ket{\psi}$ and jump operators is given by expanding $\h\Pi$ from Eq.~\ref{eq:Pi} in Eq.~\ref{eq:QFI_with_measure_reset}. This results in an expression that depends only on the variances and covariances of the jump operators with respect to $\ket{\psi}$. For example, for two noise operators $N=2$, we provide the QFI explicitly in Appendix~\ref{app:Two sources of noise}. For additional noise operators $N>2$, inverting the $N+1$ by $N+1$ matrix $\braopket{\psi}{\mathbf{L}^*\mathbf{L}^\T}{\psi}$ in Eq.~\ref{eq:Pi} analytically is more complicated but can be done numerically instead.

\subsection{Impact of finite signal and time}
\label{sec:Impact of finite signal and time}
To derive Theorem~\ref{claim:optimal QFI}, we first assume in Eq.~\ref{eq:short time} that the interrogation time is infinitesimally short to reach the sequential limit with fast and precise control in Eq.~\ref{eq:ECQFI with fast and precise control} and then assume in Eq.~\ref{eq:vanishing signal condition} that the signal is vanishingly weak to derive the Kraus operators in Eq.~\ref{eq:K, Kdot}. We now address what happens to the QFI when the signal and time are instead finite with a couple of examples. 

For example, consider sensing decay $\h L_{1,(s)}=\h \si_-$ of a qubit prepared in $\ket{\uparrow}$ such that $\h \si_-\ket{\uparrow} = \ket{\downarrow}$. Let us examine a delayed measure-and-reset strategy that waits for a time $t$ before measuring the state given by the exact solution to the master equation in Eq.~\ref{eq:general Lindblad master equation}. The total QFI from $M=T/t$ repetitions of this strategy for a total time $T$ is
\begin{align*}
    \IQ(\sqrt{\gamma_1}) &= \frac{4 \gamma_1 t T}{e^{\gamma_1 t}-1} 
\end{align*}
which depends on both the finite signal $\gamma_1$ and time $t$. For a given finite signal $\gamma_1$, it is optimal to measure and reset as fast as possible since the QFI is maximised with value $4T$ in the short time limit of $t\ll2/\gamma_1$. This agrees with Eq.~\ref{eq:QFI, K=1, N=0} and is independent of the finite signal $\gamma_1$. 
We study the noiseless case with a finite signal further in Appendix~\ref{app:Non-vanishing signal}.

The fact that we first take the short time limit and then the vanishing signal limit matters. For example, consider sensing double decay $\h L_{1,(s)} = \h J_-^2$ of a three-level system prepared in $\ket{\h J_z=1}$ in the presence of single decay noise $\h L_{2,(n)} = \h J_-$ such that $\h J_-\ket{\h J_z=1}=\sqrt{2}\ket{\h J_z=0}$ and $\h J_-^2\ket{\h J_z=1}=2\ket{\h J_z=-1}$. The QFI of a single measurement in the short time limit is
\begin{align*}
    \IQ(\sqrt{\gamma_1}) 
    &= 16 t-8 t^2 \left(4 \gamma_1 + 2 \gamma_2+\frac{\gamma_2^2}{\gamma_1}\right)+\order{t^3} 
    .
\end{align*}
If we first take the short time limit, i.e.\ assume that $t\ll2\gamma_1/\gamma_2^2$, then this QFI equals $16 t$ independent of $\gamma_1$. This agrees with Eq.~\ref{eq:QFI, any K, N=1}. In comparison, if we instead first take the vanishing signal limit, i.e.\ assume that $t\gg2\gamma_1/\gamma_2^2$, then the QFI calculated from Eq.~\ref{eq:QFI definition} vanishes. We study the first case where $t\ll2\gamma_1/\gamma_2^2$ because it corresponds to finding the optimal metrological strategy with fast and precise control.

\section{Commuting Hermitian case}
\label{sec:Classical limit}
\begin{figure}
    \centering
    \includegraphics[width=\columnwidth]{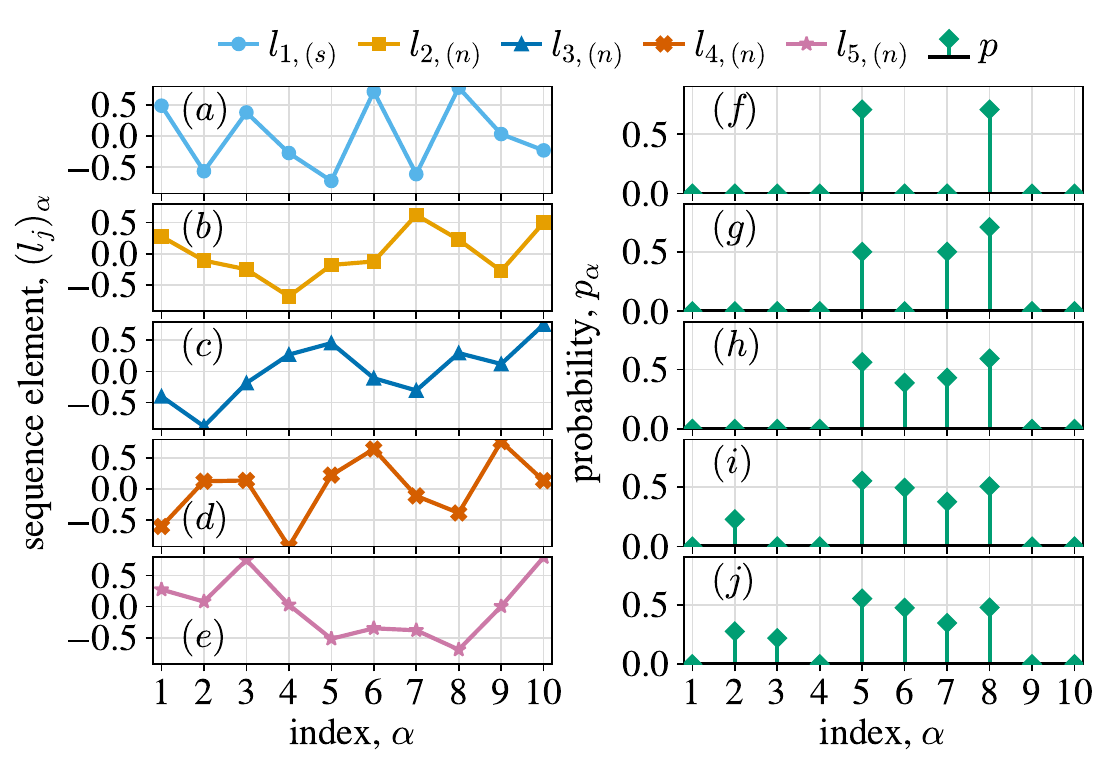}
    \caption{In the commuting Hermitian case, the jump operators $\h L_j$ become real vectors $l_j$ and the state $\ket{\psi}$ becomes a probability distribution $p$. \textbf{(a)} The signal vector $l_{1,(s)}$ and \textbf{(b--d)} noise vectors $l_{j,(n)}$ are randomly sampled from a uniform distribution. \textbf{(f)} Optimal probability distribution in the noiseless case.
    \textbf{(g--j)} Optimal distribution given the signal $l_{1,(s)}$ and first 1--4 noises, respectively. For example, panel (i) shows the distribution given the signal $l_{1,(s)}$ and three noises: $l_{2,(n)}$, $l_{3,(n)}$, and $l_{4,(n)}$.
    The support of the optimal distribution is length $N+2$ given $N$ noise operators.}
    \label{fig:classical_limit}
\end{figure}
In the following sections, we will use our results to find the optimal strategy with fast and precise control in a variety of applications. 

Let us first discuss the case where the jump operators are all Hermitian and pairwise commute. This implies that a simultaneous eigenbasis exists where $\h L_j = \diag{l_j}$ and $l_j \in \R^d$ for each $j = 1,\mathellipsis,K+N$. We can assume without loss of generality that the jump operators are traceless and pairwise orthonormal, i.e.\ $\sum_{\alpha=1}^d (l_j)_\alpha = 0$ and $\sum_{\alpha=1}^d (l_j)_\alpha (l_k)_\alpha = \delta_{jk}$ where the $\{l_j\}_{j=1}^{K+N}$ are thus zero-sum and pairwise orthonormal as real vectors. Here, we use Latin indices (e.g.\ $j,k,l\in \{1,\mathellipsis,K+N\}$) to label a particular vector and Greek indices (e.g.\ $\alpha,\beta,\gamma\in\{1,\mathellipsis,d\}$) to label the elements of that vector. Let us now consider the pure quantum state $\ket{\psi} = \sum_{\alpha,\beta=1}^d c_{\alpha\beta} \ket{e_\alpha}\otimes\ket{e_\beta}$ and its reduced density matrix $\h\rho_S$. The probability for the system to be in $\ket{e_\alpha}$ is $p_\alpha = \sum_{\beta=1}^d \abs{c_{\alpha\beta}}^2$ for $\alpha=1,\mathellipsis,d$. Let us define the expectation value of a vector $l\in\R^d$ with respect to a probability distribution $p \in \R^d$ where $\sum_{\alpha=1}^d p_\alpha = 1$ as $\evTextSubSmall{p}{l} = \sum_{\alpha=1}^d p_\alpha l_\alpha$. For any operator $\h A = \diag{l}$ where $l\in\R^d$ and $\sum_{\alpha=1}^d l_\alpha = 0$, then, since $\braopket{\psi}{\h A\otimes \h I}{\psi}=\trSmall{\h\rho_S \h A}$, the expectation value is
\begin{align}
    \label{eq:commuting case, ancilla not required}
    \braopket{\psi}{\h A\otimes \h I}{\psi}
    &= \evTextSubSmall{p}{l}
    .
\end{align}
The QFI in Theorem~\ref{claim:optimal QFI} only involves expectation values of operators $\{\h L_j\}_{j=1}^{K+N}$ or products of two such operators with respect to $\ket{\psi}$. By Eq.~\ref{eq:commuting case, ancilla not required}, this means that entanglement is not required and the QFI only depends on $\evTextSubSmall{p}{l_j}$ and $\evTextSubSmall{p}{l_j l_k}$. 

We can now restate the estimation problem for the case of a single source of noise. Given $l_j\in\R^d$ with $\sum_{\alpha=1}^d (l_j)_\alpha = 0$ and $\sum_{\alpha=1}^d (l_j)_\alpha (l_k)_\alpha = \delta_{jk}$ for $j,k\in\{1,2\}$, then the QFI in Eq.~\ref{eq:QFI, any K, N=1} is
\begin{align*}
    &\IQ(\sqrt{\gamma_1}=0) 
    \\&= 4T \max_p \left(\varSubSmall{p}{l_{1,(s)}}-\frac{\absSmall{\covSubSmall{p}{l_{1,(s)}}{l_{2,(n)}}}^2}{\varSubSmall{p}{l_{2,(n)}}}\right).
\end{align*}
We want to find the optimal probability distribution $p\in\R^d$ above. Similarly, for the case of $N$ noise sources, we replace all operator (co)variances in Theorem~\ref{claim:optimal QFI} for $K=1$ with their respective vector (co)variances. The uniform distribution $p_\alpha = 1/d$ makes the covariances vanish since the $\{l_j\}_{j=1}^{K+N}$ are pairwise orthogonal and leaves $\IQ(\sqrt{\gamma_1}=0) = 4T/d$ which is nonzero but not necessarily the optimal distribution, similarly to the Bell state in Eq.~\ref{eq:QFI, Bell state}. Since the QFI is now only expressed in terms of real vectors and probability distributions, the commuting Hermitian case can also be viewed as the classical limit of Lindblad estimation.

Let us now consider the commuting Hermitian case with one signal ($K=1$) and $N=d-2$ noises. Provided that a regularity condition of generalised non-degeneracy given in Appendix~\ref{app:classical_case} holds, then the optimal QFI is
\begin{align}
    \IQ(\sqrt{\gamma_1}=0) 
    = 4T\frac{||l_{1}||_{2}^{4}}{||l_{1}||_{1}^{2}}
\label{eq: ECQFI_classical_1}
\end{align}
where $\norm{l}_2^2 = \sum_{\alpha=1}^d|l_{\alpha}|^2$ and $\norm{l}_1 = \sum_{\alpha=1}^d|l_{\alpha}|$. The optimal distribution is $p_\alpha \propto \abs{l_{1,\alpha}}$ for $\alpha=1,\mathellipsis,d$, i.e.\ it mimics the magnitude of the signal $l_{1}$ up to normalisation. The regularity condition is unnecessary for $d=3$, i.e.\ for one signal and one noise in three dimensions. For $d=2$, i.e.\ the noiseless case in two dimensions, the QFI in Eq.~\ref{eq: ECQFI_classical_1} recovers the noiseless QFI of $2 T ||l_{1}||^{2}_{2}$. We prove these results in Appendix~\ref{app:classical_case}.

In the general case with one signal ($K=1$), $N$ noises, and any dimension $d$, we can determine the optimal QFI and initial state given a solution $c_{1,\text{min}},c_{i,\text{min}}$ to the following SDP:
\begin{align*}
    l_{1}=c_{1}\vec{1}+\sum_{i=2}^{N+1}c_{i}l_{i}+a_{\text{min}}
\end{align*}
where $\vec{1}=(1,\mathellipsis,1)^\T$. By Appendix~\ref{app:optimal_states}, the optimal initial state is supported only on indices $\alpha\in I$ that satisfy $|\left(a_{\text{min}}\right)_{\alpha}|=||a_{\text{min}}||_{\infty}$ where $\norm{l}_\infty = \max_{\alpha=1,\mathellipsis,d}|l_{\alpha}|$. We can thus restrict ourselves to the $m$-dimensional subspace spanned by the indices in $I$ and consider the corresponding sub-vectors $l_{j}^{\left(I\right)}$ and $a_{\text{min}}^{\left(I\right)}$. The optimal QFI is then:
\begin{align}
\IQ(\sqrt{\gamma_1}=0) = 4T\frac{||l_{1,\perp}^{\left(I\right)}||_{2}^{4}}{||l_{1,\perp}^{\left(I\right)}||_{1}^{2}}
\label{eq: ECQFI_classical_2}
\end{align}
where $l_{1,\perp}^{\left(I\right)}$ is the projection of $l_{1}^{\left(I\right)}$ onto the subspace orthogonal to $\text{span}\{\vec{1}, l_{2}^{\left(I\right)},...,l_{N+1}^{\left(I\right)}\}$. The optimal distribution is then $p_\alpha \propto |(l_{1,\perp}^{\left(I\right)})_\alpha|$ for $\alpha=1,\mathellipsis,m$ up to normalisation. We derive this optimal QFI and distribution in Appendix~\ref{app:classical_case}.

We can thus calculate the optimal QFI and initial state from the support of the optimal distribution. It remains to understand this support. We numerically observe the following phenomenon: the optimal distribution for $K=1$ in the presence of $N$ noises has a support of length $N+2$. (Our code for this result and the rest of our work is available online~\cite{repo}.) This is shown in Fig.~\ref{fig:classical_limit} for $d=10$ where we see that each additional noise adds one more dimension to the support of the optimal distribution. Analytically, we know that given $N$ noise operators and a generalised condition of non-degeneracy, the minimal length of the support is $N+2$ since otherwise the QFI must vanish by the dimension of the noisy subspace. This numerical phenomenon suggests that a length $N+2$ support is also sufficient, but we defer proving this result to future work. 




\section{A single qubit}
\label{sec:Single qubit}
Let us now consider a single qubit with one signal $\h L_{1,(s)}$ and one noise $\h L_{2,(n)}$. We want to find the optimal initial state and QFI for all possible choices of $\h L_{1,(s)}$ and $\h L_{2,(n)}$ to minimise the loss of precision compared to the noiseless QFI due to the noise. Furthermore, we want to determine the hierarchy of metrological strategies shown in Fig.~\ref{fig:diagram}: Whether an unextended measure-and-reset strategy (Fig.~\hyperref[fig:diagram]{\ref*{fig:diagram}e}) can saturate the optimal QFI or whether entanglement with other noisy probes (Fig.~\hyperref[fig:diagram]{\ref*{fig:diagram}f}) or a noiseless ancilla (Fig.~\hyperref[fig:diagram]{\ref*{fig:diagram}b}) is necessary. We will answer these questions by directly maximising the QFI in Eq.~\ref{eq:QFI, any K, N=1} over the initial state. As shown in Sec.~\ref{sec:properties}, we can assume without loss of generality that $\h L_{1,(s)}$ and $\h L_{2,(n)}$ are orthogonal, traceless, and normalised. This leads to the following result:
\begin{lemma}
\label{claim:unextended qubit}
For a single qubit, the only unextended states with nonzero QFI are the eigenstate/s of $\h L_{2,(n)}$. The optimal unextended QFI equals $4 T \varSubSmall{\ket{\psi}}{\h L_{1,(s)}}$ maximised over these eigenstates.
\end{lemma}
This result can be shown geometrically by seeing that if $\varSubSmall{\ket{\psi}}{\h L_{2,(n)}} > 0$, then any unextended state has vanishing QFI since the signal $\h L_{1,(s)}\ket{\psi}$ must be in the span of $\ket{\psi}$ and $\h L_{2,(n)}\ket{\psi}$ since the space is two-dimensional. The optimal unextended QFI then follows from Eq.~\ref{eq:QFI, any K, N=1}, by observing that if $\ket{\psi}$ is an eigenstate of $\h L_{2,(n)}$ then $\covSubSmall{\ket{\psi}}{\h L_{1,(s)}}{\h L_{2,(n)}}=0$.

We now proceed by case analysis on whether $\h L_{1,(s)}$ and $\h L_{2,(n)}$ are each Hermitian or non-Hermitian.

\subsection{$\h L_{1,(s)}, \h L_{2,(n)}$ are both Hermitian}
\label{sec:qubit, both Herm}
We first consider the case where $\h L_{1,(s)}$ and $\h L_{2,(n)}$ are Hermitian. A traceless Hermitian operator acting on a single qubit is a real linear combination of the Pauli operators $\h\si_x$, $\h\si_y$, and $\h\si_z$. Without loss of generality, we may assume that $\h L_{1,(s)} = \frac{1}{\sqrt{2}}\h\si_z$ and $\h L_{2,(n)} = \frac{1}{\sqrt{2}}(\cos(\theta)\h\si_x + \sin(\theta)\h\si_y)$. The maximal variance of $\h L_{1,(s)}$ is thus $\frac{1}{2}$ using any pure state on the equator of the Bloch sphere, i.e.\ $\ket{\psi}=\frac{1}{\sqrt2}(\ket{\uparrow}+e^{i\phi}\ket{\downarrow})$ where $\h\si_z\ket{\uparrow}=\ket{\uparrow}$ and $\h\si_z\ket{\downarrow}=-\ket{\downarrow}$, such that the optimal noiseless QFI in Eq.~\ref{eq:QFI, any K, N=0} is $\IQ(\sqrt{\gamma_1}=0) = 2 T$. 

In the noisy case, since $\ket{\psi}$ is an eigenstate of $\h L_{2,(n)}$ for $\phi=\theta$, we recover the optimal noiseless QFI using $\ket{\psi}$ as an input state. A noiseless ancilla qubit is therefore not needed in this case. For reference, the Bell state in Eq.~\ref{eq:QFI, Bell state} also recovers the optimal noiseless QFI since $\frac{1}{d} \trSmall{\h L_{1,(s)}^\dag \h L_{1,(s)}}=\frac{1}{2}$.

\subsection{$\h L_{1,(s)}$ is non-Hermitian and $\h L_{2,(n)}$ is Hermitian}
\label{sec:qubit, L1 non-Herm, L2 Herm}
We now consider the case where $\h L_{1,(s)}$ is non-Hermitian and $\h L_{2,(n)}$ is Hermitian. Without loss of generality, we can assume that $\h L_{1,(s)} = a \h \si_+ + b \h \si_-$ and $\h L_{2,(n)} = \frac{1}{\sqrt2} \h \si_z$ where $\h\si_\pm = (\h\si_x \pm i \h\si_y)/2$, $a\neq b^*$, and $\abs{a}^2+\abs{b}^2=1$. The noiseless QFI is $4T\max\{\abs{a}^2,\abs{b}^2\}$ since $\h L_{1,(s)}^\dag \h L_{1,(s)} =\text{diag}(\absSmall{b}^2,\absSmall{a}^2)$. The noiseless QFI is attained, e.g., by $\ket\downarrow$ if $\abs{a} \geq \abs{b}$ and $\ket{\uparrow}$ if $\abs{a} < \abs{b}$. The noiseless QFI is recovered in the noisy case with the same unextended states since they are eigenstates of $\h L_{2,(n)}$. 

\subsection{$\h L_{1,(s)}$ is Hermitian and $\h L_{2,(n)}$ is non-Hermitian}
\label{sec:L_1 Hermitian, L_2 non-Hermitian}
We now consider the case where $\h L_{1,(s)}$ is Hermitian and $\h L_{2,(n)}$ is non-Hermitian. Without loss of generality, let $\h L_{1,(s)} = \frac{1}{\sqrt{2}}\h\si_z$ and $\h L_{2,(n)} = a \h \si_+ + b \h \si_-$ where $a\neq b^*$ and $\abs{a}^2+\abs{b}^2=1$. The noiseless QFI remains $\IQ(\sqrt{\gamma_1}=0) = 2 T$. The Bell state in Eq.~\ref{eq:QFI, Bell state} recovers the optimal noiseless QFI for every $a,b$ since $\frac{1}{2}\trSmall{\h L_{1,(s)}^\dag \h L_{1,(s)}}= \max_{\ket{\psi}} \varSubSmall{\ket{\psi}}{\h L_{1,(s)}}$. Let us determine whether we can recover the optimal noiseless QFI without entanglement. The eigenstates of $\h L_{2,(n)}$ are
\begin{align*}
    \ket{\psi} = \frac{\sqrt{\abs{a}}\ket{\uparrow} \pm \sqrt{\abs{b}} e^{\frac{1}{2}i[\arg(b) - \arg(a)]}\ket{\downarrow}}{\sqrt{\abs{a}+\abs{b}}} 
    .
\end{align*}
By Lemma~\ref{claim:unextended qubit}, these are the only unextended states with nonzero QFI equal to 
\begin{align*}
    \IQ(\sqrt{\gamma_1}=0) = 4T\left(1 - \frac{1}{1 + 2 \abs{ab}}\right).
\end{align*}
There is therefore a gap between the optimal QFI obtained with a Bell state and the optimal unextended QFI obtained with an eigenstate of $\h L_{2,(n)}$. The size of the gap depends on the values of $a,b$ as follows:
\begin{enumerate}
    \item For $\abs{a}=\abs{b}=\frac{1}{\sqrt{2}}$, the gap vanishes and the unextended QFI is optimal.
    \item For any $0<\abs{ab}<\frac{1}{2}$, the optimal unextended QFI is suboptimal but nonzero.
    \item For $a=0$ or $b=0$, the optimal unextended QFI vanishes. 
\end{enumerate}

The case of $a=0$ and $b=1$ such that $\h L_{1,(s)} = \frac{1}{\sqrt{2}}\h\si_z$ and $\h L_{2,(n)} = \h \si_-$ is of particular practical interest. This case corresponds to estimating a weak dephasing rate ($T_{2}$)~\cite{arshad2024real} and to qubit spectroscopy~\cite{gefen2019overcoming, mouradian2021quantum} in the presence of strong amplitude-damping noise ($T_{1}$). These results imply that entanglement is necessary to achieve nonzero QFI for this sensing task. Hence, $T_{2}$ estimation in this regime can be significantly improved by using a Bell state compared to the standard unextended scheme. 
We remark that while this optimal protocol assumes a noiseless ancilla, we will show later in Sec.~\ref{sec:unextended_parallel_strategy} that this assumption can be removed: The same enhancement is achieved with a parallel strategy that entangles two noisy qubit probes. 

\subsection{$\h L_{1,(s)}, \h L_{2,(n)}$ are both non-Hermitian and in-plane}
\label{sec:qubit, both non-Herm, in-plane}
\begin{figure}
    \centering
    \includegraphics[width=\columnwidth]{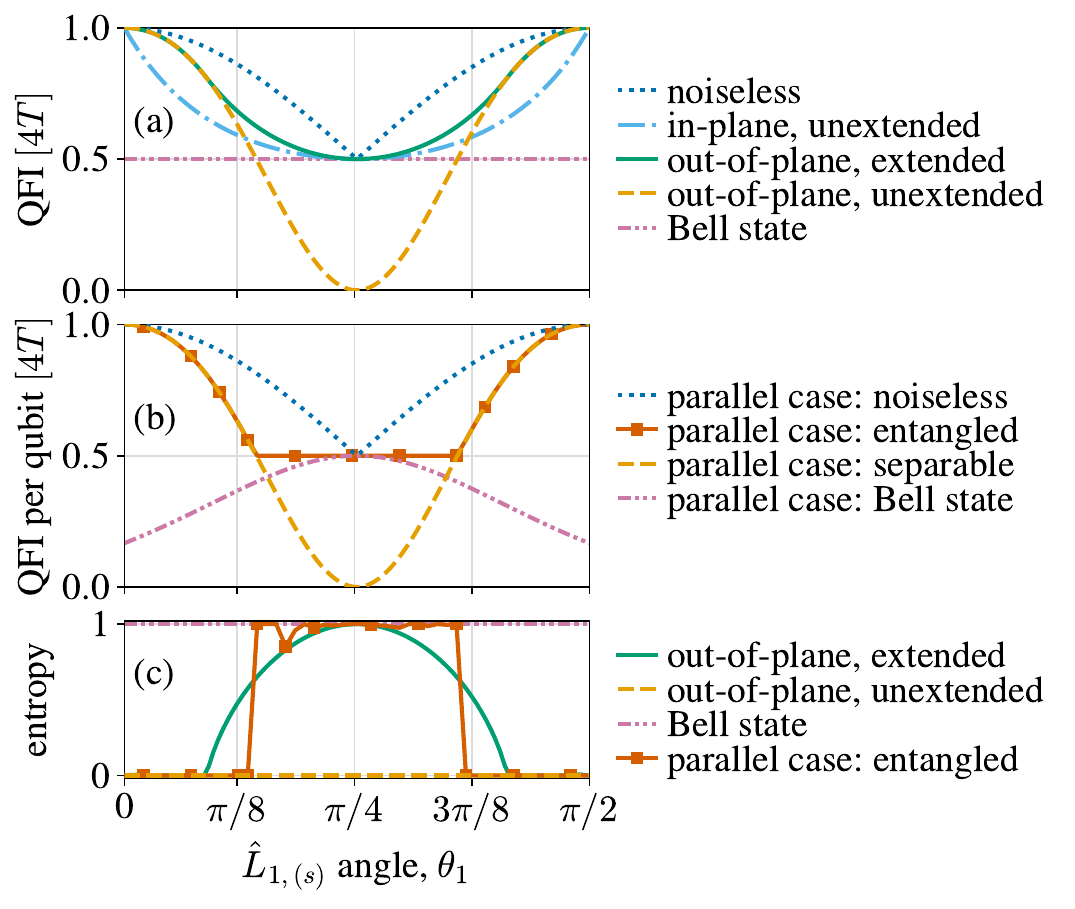} 
    \caption{\textbf{(a)} QFI versus $\theta_1$ for the case of $\h L_{1,(s)},\h L_{2,(n)}$ both non-Hermitian given in Eq.~\ref{eq:outofplane operators} where $\phi_1=\phi_2=0$ and $\theta_2=0$ for the $\h L_{2,(n)}$ in-plane case and $\theta_2=\pi/4$ in the $\h L_{2,(n)}$ out-of-plane case. In the in-plane case, an unextended state is optimal and a gap exists between the noisy and noiseless QFIs. In the out-of-plane case, there is a transition as $\theta_1$ increases below which an unextended state is optimal and above which a gap emerges between the optimal unextended and extended QFIs. 
    \textbf{(b)} For the unextended parallel strategy, the optimal state allowing for entanglement discontinuously changes from being separable to highly entangled when the QFI per qubit equals $2T$.
    \textbf{(c)} Entanglement entropy versus $\theta_1$ normalised to the entanglement entropy of a maximally entangled state.} 
    \label{fig:halfway_QFI}
\end{figure}
We now consider the case where $\h L_{1,(s)}$ and $\h L_{2,(n)}$ are non-Hermitian. We can further subdivide this case based on whether the real and imaginary parts of $\h L_{1,(s)}$ and $\h L_{2,(n)}$ all lie in the same plane. Let us first consider the in-plane case where, without loss of generality, 
\begin{align*}
    \h L_{1,(s)} &= \cos(\theta_1)\h \si_+ + \sin(\theta_1)e^{i\varphi_1}\h \si_-
    \\\h L_{2,(n)} &= \sin(\theta_1)e^{-i\varphi_1}\h \si_+ - \cos(\theta_1)\h \si_-
    .    
\end{align*}
The noiseless QFI is thus
\begin{align}\label{eq:QFI, single qubit, non-Hermitian, noiseless}
    \IQ(\sqrt{\gamma_1}=0) = 4T\max\{\cos^2(\theta_1),\sin^2(\theta_1)\}
\end{align}
such that the Bell state, which only achieves $2T$, is suboptimal for $\theta\neq(2k+1)\pi/4$ with $k\in\Z$. In comparison, the optimal QFI in the noisy case is
\begin{align}\label{eq:QFI, single qubit, non-Hermitian but in-plane, optimal}
    \IQ(\sqrt{\gamma_1}=0) = \frac{4T}{\max_\pm[\cos(\theta_1)\pm\sin(\theta_1)]^2}
\end{align}
which is attained by preparing either of the two eigenstates of $\h L_{2,(n)}$ denoted as $\ket{\lambda_j'}$ for $j=1,2$. This implies that any extended state of the form $\ket{\psi} = a_1 \ket{\lambda_1', e_1} + a_2 \ket{\lambda_2', e_2}$ with $\ket{e_1}$ and $\ket{e_2}$ orthogonal is also optimal. There is therefore no gap between the extended and unextended cases, however, there is a gap between the optimal noisy QFI in Eq.~\ref{eq:QFI, single qubit, non-Hermitian but in-plane, optimal} and the optimal noiseless QFI in Eq.~\ref{eq:QFI, single qubit, non-Hermitian, noiseless} for $\theta_1 \neq k \pi / 4$ with $k\in\Z$ as shown in Fig.~\hyperref[fig:halfway_QFI]{\ref*{fig:halfway_QFI}a}.

This result is obtained by direct optimisation of the QFI over the input state as follows. We parameterise the reduced density matrix of an arbitrary input state as:
\begin{align*}
    \h\rho_S = \pmatrixByJames{p & re^{i\phi} \\ re^{-i\phi} & 1 - p}
\end{align*}
where $0 \leq r^2 \leq p(1-p)$. The QFI in Eq.~\ref{eq:QFI, any K, N=1} for this input state then equals~\footnote{Here, the complex phase $\varphi_1$ in $\h L_{1,(s)}$ and $\h L_{2,(n)}$ has been compensated by setting $\phi\mapsto\phi + \varphi_1 / 2$ such that we may assume that $\varphi_1=0$ without loss of generality.}
\begin{align}\label{eq:QFI, single qubit, non-Hermitian but in-plane}
    &\IQ(\sqrt{\gamma_1}=0)
    \\&= \frac{8T \left[(1-p) p-r^2\right]}{1+(2 p-1) \cos (2 \theta_1 ) - 2 r^2[1 - \sin (2 \theta_1 ) \cos (2 \phi )]}. \nonumber
\end{align}
Maximising this QFI against the state parameters $p$, $r$, and $\phi$ leads to the optimal QFI in Eq.~\ref{eq:QFI, single qubit, non-Hermitian but in-plane, optimal}. This optimisation can be done analytically first against $r$ for fixed values of $p$ and $\phi$: the QFI is either constant in $r$ or its derivative with respect to $r^2$ is never zero such that it suffices to calculate the QFI on the boundary at $r=0$ and $r^2 = p(1-p)$. Taking $r=0$ and optimising Eq.~\ref{eq:QFI, single qubit, non-Hermitian but in-plane} over $p$ then yields Eq.~\ref{eq:QFI, single qubit, non-Hermitian but in-plane, optimal}.

\subsection{$\h L_{1,(s)}, \h L_{2,(n)}$ are both non-Hermitian and $\h L_{2,(n)}$~is~out-of-plane}
\label{sec:out of plane}
\begin{figure}
    \centering
    \includegraphics[width=\columnwidth]{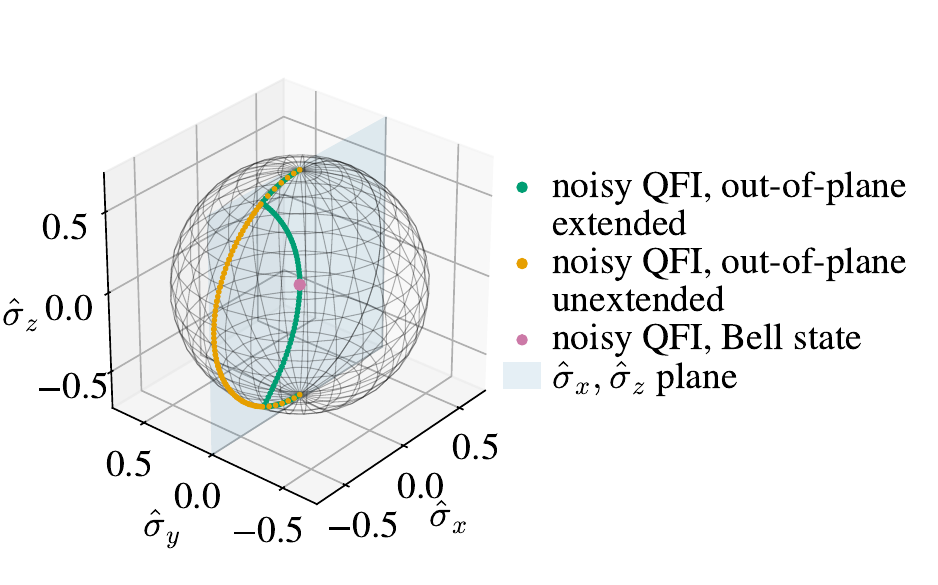}
    \caption{Bloch ball representation of the reduced density matrix of the optimal state for the case of $\h L_{1,(s)},\h L_{2,(n)}$ both non-Hermitian and $\h L_{2,(n)}$ out-of-plane. The families of optimal extended states and unextended states are shown for different values of $\theta_1\in(0,\pi/2)$ where $\theta_2=\pi/4$ and $\varphi_1=\varphi_2=0$. States on the surface of the Bloch ball are unextended, states inside the Bloch ball are extended, and states at the origin are maximally entangled such as the Bell state.
    Starting at $\theta_1=0$, the optimal state is $\ket{\uparrow}$ which is shown at the North pole of the Bloch ball. The optimal state remains unextended for increasing values of $\theta_1$ and rotates down the $\h\si_x,\h\si_z$ plane until a transition point determined by $\theta_2=\pi/4$ is reached after which the optimal state becomes extended and increasingly entangled until it reaches a maximally entangled state when $\theta_1=\pi/4$. The behaviour is then symmetric about $\theta_1=\pi/4$.} 
    \label{fig:bloch_sphere_qubit_out-of-plane}
\end{figure}
We now consider the case where $\h L_{1,(s)}$ and $\h L_{2,(n)}$ are non-Hermitian but the real and imaginary parts of $\h L_{1,(s)}$ and $\h L_{2,(n)}$ do not lie in the same plane. Without loss of generality,
\begin{align}
    \label{eq:outofplane operators}
    \h L_{1,(s)} &= \cos(\theta_1)\h \si_+ + \sin(\theta_1)e^{i\varphi_1}\h \si_-
    \\\h L_{2,(n)} &= \cos(\theta_2)[\sin(\theta_1)e^{-i\varphi_1}\h \si_+ - \cos(\theta_1)\h \si_-]\nonumber \\&+ \frac{1}{\sqrt2}\sin(\theta_2)e^{i\varphi_2}\h \si_z\nonumber
    .    
\end{align}
The noiseless QFI is still given in Eq.~\ref{eq:QFI, single qubit, non-Hermitian, noiseless} and the Bell state still achieves $2T$, but the remaining results from the previous case are different. 
As in the in-plane case, we want to optimise the QFI against the initial state parameters $(p,r,\phi)$ for any given values of $(\theta_1,\theta_2,\varphi_1,\varphi_2)$. The expression of the QFI in this case is too verbose to optimise analytically and thus we determine the optimal state numerically instead. As shown in Fig.~\hyperref[fig:halfway_QFI]{\ref*{fig:halfway_QFI}a}, a gap in the QFI exists in parts of the $(\theta_1,\theta_2,\varphi_1,\varphi_2)$ parameter space, not just between the noiseless and noisy cases as seen for $\theta_2=0$, but also between the unextended and extended cases. This means that entanglement is sometimes necessary to achieve the maximum QFI. The optimal state transitions from the unextended state to the Bell state as shown in Fig.~\hyperref[fig:halfway_QFI]{\ref*{fig:halfway_QFI}c} and Fig.~\ref{fig:bloch_sphere_qubit_out-of-plane}.

\subsection{Unextended parallel strategy}
\label{sec:unextended_parallel_strategy}
We now consider the unextended parallel strategy shown in Fig.~\hyperref[fig:diagram]{\ref*{fig:diagram}f} with $M=2$. Compared to the previous case shown in Fig.~\hyperref[fig:diagram]{\ref*{fig:diagram}b}, the noiseless ancilla is replaced by a copy of the system that is coupled independently to the same signal and noise. Experimentally, we have two copies of the same probe device which we can entangle with each other and perform joint measurements of. We model this as $K=2$ local signal operators and $N=2$ local noise operators and focus on the unextended states of this two-qubit system. Since we can always prepare the two noisy qubits in a separable state, the optimal QFI per qubit of this unextended parallel strategy (Fig.~\hyperref[fig:diagram]{\ref*{fig:diagram}f}) is always at least the optimal QFI of the unextended measure-and-reset strategy (Fig.~\hyperref[fig:diagram]{\ref*{fig:diagram}e}).

We previously required access to a noiseless ancilla to achieve the optimal QFI for the case of Hermitian $\h L_{1,(s)}$ and non-Hermitian $\h L_{2,(n)}$  (Sec.~\ref{sec:L_1 Hermitian, L_2 non-Hermitian}). We observed that for $\abs{a}\neq\abs{b}$, there was a gap between the optimal unextended QFI and the optimal QFI attained by the Bell state. We want to compare this now to the parallel strategy. Without loss of generality, the signal operators are $\h L_{j,(s)} = \frac{1}{\sqrt{2}}\h\si_z^{(j)}$ for $j=1,2$ where $\h \si_z^{(j)}$ acts on the $j$th qubit and the noise operators are $\h L_{j+2,(n)} = a \h \si_+^{(j)} + b \h \si_-^{(j)}$ for $j=1,2$ where $\abs{a}^2+\abs{b}^2=1$ and $\abs{a}\neq \abs{b}$. The optimal noiseless QFI is $4T$ by Eq.~\ref{eq:QFI, any K, N=0} which should be normalised to a QFI per qubit of $2T$. By Eq.~\ref{eq:QFI, any K, N=2}, this noiseless QFI per qubit is recovered by a Bell state since each covariance between a signal and a noise vanishes. This means that a Bell state of two noisy qubit probes is optimal in this case and access to noiseless ancilla is not required. 

Revisiting the practical problem with $a=0$ and $b=1$ of sensing a weak dephasing rate ($T_2$) in the presence of strong amplitude-damping noise ($T_1$), this result implies that we can overcome the vanishing QFI using a Bell state of two noisy qubit probes without requiring noiseless ancilla. Intuitively, the Bell state of the two probes allows quantum error detection of the amplitude-damping noises to differentiate them from the signals.

We also required access to a noiseless ancilla in Sec.~\ref{sec:out of plane} where both $\h L_{1,(s)}$ and $\h L_{2,(n)}$ were non-Hermitian but their real and imaginary parts did not lie in the same plane. In the parallel case, without loss of generality, the signal operators are
\begin{align*}
    \h L_{j,(s)} &= \cos(\theta_1)\h \si_+^{(j)} + \sin(\theta_1)e^{i\varphi_1}\h \si_-^{(j)}
    \intertext{where $j=1,2$ and the noise operators are}
    \h L_{j+2,(n)} &= \cos(\theta_2)[\sin(\theta_1)e^{-i\varphi_1}\h \si_+^{(j)} - \cos(\theta_1)\h \si_-^{(j)}] \\&+ \frac{1}{\sqrt2}\sin(\theta_2)e^{i\varphi_2}\h \si_z^{(j)}
    .
\end{align*}
In the noiseless case, by Eq.~\ref{eq:QFI, any K, N=0} and Eq.~\ref{eq:QFI, single qubit, non-Hermitian, noiseless}, the optimal QFI per qubit is
\begin{align*}
    \frac{1}{2}\IQ(\sqrt{\gamma_1}=0) = 4T\max\{\cos^2(\theta),\sin^2(\theta)\}
\end{align*}
and the Bell state only achieves a QFI per qubit of $2T$. We numerically optimise the noisy QFI per qubit from Eq.~\ref{eq:QFI, any K, N=2} over the initial state allowing for entanglement and compare this to the QFI per qubit of a product state of the eigenstates of $\h L_{3,(n)}$ and $\h L_{4,(n)}$. As shown in Fig.~\hyperref[fig:halfway_QFI]{\ref*{fig:halfway_QFI}b}, we observe a discontinuity in the optimal state versus $\theta_1$ as the optimal state is separable until the QFI per qubit equals $2T$ and the optimal state becomes highly entangled. These highly entangled states maintain a QFI per qubit of $2T$ and are not necessarily Bell states or the extended states from the previous case. The optimal QFI of the out-of-plane extended state in Fig.~\hyperref[fig:halfway_QFI]{\ref*{fig:halfway_QFI}a} is greater than the QFI per qubit of these entangled states although they outperform the separable states. This demonstrates the expected hierarchy of metrology strategies since the extended sequential strategy in Fig.~\hyperref[fig:diagram]{\ref*{fig:diagram}a} can simulate the unextended parallel strategy in Fig.~\hyperref[fig:diagram]{\ref*{fig:diagram}f} which can itself simulate the unextended measure-and-reset strategy in Fig.~\hyperref[fig:diagram]{\ref*{fig:diagram}e}, respectively.

\subsection{Hierarchy of metrological strategies}
Let us now summarise the hierarchy of metrological strategies for the single qubit case. Let denote the optimal noiseless QFI as $\IQ^{\text{noiseless}}$ and the optimal noisy QFI as $\IQ^{\text{optimal}}$, both corresponding to an extended sequential strategy with fast and precise control (Fig.~\hyperref[fig:diagram]{\ref*{fig:diagram}a}) which we have shown in Theorem~\ref{claim:optimal QFI} is attained by an extended measure-and-reset strategy (Fig.~\hyperref[fig:diagram]{\ref*{fig:diagram}b}). Let us also denote the optimal QFI with an unextended measure-and-reset strategy (Fig.~\hyperref[fig:diagram]{\ref*{fig:diagram}e}) as $\IQ^{\text{unextended}}$, where we have not determined whether this attains the QFI of the optimal unextended sequential strategy (Fig.~\hyperref[fig:diagram]{\ref*{fig:diagram}d}). Finally, let us denote the optimal QFI per qubit with an unextended parallel measure-and-reset strategy (Fig.~\hyperref[fig:diagram]{\ref*{fig:diagram}f}) as $\IQ^{\text{parallel}}$. The hierarchy of metrological strategies for a single qubit then depends on whether $\h L_{1,(s)}$ and $\h L_{2,(n)}$ are each Hermitian or non-Hermitian as follows:
\begin{enumerate}
    \item $\h L_{2,(n)}$ is Hermitian and $\h L_{1,(s)}$ is Hermitian or non-Hermitian (Sec.~\ref{sec:qubit, both Herm} or Sec.~\ref{sec:qubit, L1 non-Herm, L2 Herm}, respectively):
    \begin{align*}
        \IQ^{\text{unextended}}=\IQ^{\text{parallel}}=\IQ^{\text{optimal}}=\IQ^{\text{noiseless}}.
    \end{align*}
    \item $\h L_{2,(n)}$ is non-Hermitian and $\h L_{1,(s)}$ is Hermitian (Sec.~\ref{sec:L_1 Hermitian, L_2 non-Hermitian}):
    \begin{align*}
        \IQ^{\text{unextended}}\leq \IQ^{\text{parallel}} = \IQ^{\text{optimal}}=\IQ^{\text{noiseless}}.
    \end{align*}
    \item $\h L_{1,(s)}$, $\h L_{2,(n)}$ are non-Hermitian and in-plane (Sec.~\ref{sec:qubit, both non-Herm, in-plane}):
    \begin{align*}
        \IQ^{\text{unextended}}=\IQ^{\text{parallel}}=\IQ^{\text{optimal}}\leq\IQ^{\text{noiseless}}.
    \end{align*}
    \item $\h L_{1,(s)}, \h L_{2,(n)}$ are non-Hermitian and $\h L_{2,(n)}$ out-of-plane (Sec.~\ref{sec:out of plane}):
    \begin{align*}
        \IQ^{\text{unextended}}\leq\IQ^{\text{parallel}}\leq\IQ^{\text{optimal}}\leq\IQ^{\text{noiseless}}.
    \end{align*}
\end{enumerate}
For each of the above inequalities (i.e.\ $\leq$), we have examples where it becomes a strict inequality (i.e.\ $<$) or an equality (i.e.\ $=$). 

\section{Many qubits}
\label{sec:Many qubits}
We now consider some applications for $n$ qubits and discuss first the detection of correlated noise, focusing on correlated dephasing and decay, and then the general Pauli Lindblad estimation problem.

\subsection{Correlated dephasing and decay}
With correlated signals and noises, such as correlated dephasing and decay, we can restrict ourselves to the $n+1$ dimensional subspace of symmetric pure states in the unextended case. This is different from the unextended parallel case shown in Fig.~\hyperref[fig:diagram]{\ref*{fig:diagram}f} which represents uncorrelated local dephasing and decay.

Finding optimal estimation strategies of correlated dephasing and decay is relevant for various platforms and applications. Correlated dephasing appears due to laser noise~\cite{macieszczak2014bayesian, shaw2024multi, direkci2024heisenberg} and in nano-scale nuclear magnetic resonance and spectroscopy problems~\cite{gefen2019overcoming, mouradian2021quantum}. Correlated decay appears in various scenarios~\cite{mok2024universal} such as superradiant lasers~\cite{bohnet2012steady}, Dicke-driven phase transitions~\cite{ferioli2023non}, and chiral spin networks~\cite{pichler2015quantum}. For each of these applications, we know that the measure-and-reset strategy is optimal. We thus only need to determine the optimal initial state in each case.

For example, consider sensing correlated decay $\h L_{1,(s)} = \h J_-$ in the presence of correlated amplification $\h L_{2,(n)} = \h J_+$ and dephasing $\h L_{3,(n)} = \h J_z$ such that we want to optimise Eq.~\ref{eq:QFI, any K, N=2}. Here, the collective spin operators are defined by $\h J_j = \frac{1}{2}\sum_{k=1}^n \h \si_j^{(k)}$ for $j=x,y,z$ and $\h J_\pm = \h J_x \pm i \h J_y$. Since $(\h J_-)_{j, j - 1} = \sqrt{j (n + 1 - j)}$, then $(\h J_+\h J_-)_{j,k}=\delta_{jk}(j+1)(n - j)$ and the optimal state is thus the $j$th eigenstate of $\h J_z$ where $j=(n-1)/2$ and the QFI is $4T(n+1)^2$ if $n$ is odd or $j=n/2,n/2-1$ and the QFI equals $4Tn(n+2)$ if $n$ is even. The QFI grows quadratically in the number of correlated sensors $n$ but only linearly in the total time $T$.

In comparison, to sense correlated dephasing along $\h L_{1,(s)} = \h J_z$ in the presence of amplitude damping $\h L_{2,(n)} = \h J_-$ and amplification $\h L_{3,(n)} = \h J_+$ as well as dephasing along $\h J_x$ and $\h J_y$, the optimal state for $n>1$ is $(\ket{\h J_z = n/2}+\ket{\h J_z = -n/2})/\sqrt2$. This state maximises the variance of $\h J_z$ and its images under the noises are orthogonal such that it recovers the noiseless QFI of $Tn^2$.

In more complicated cases than these two examples, we can instead use an SDP to find the optimal state.

\subsection{Pauli Lindblad estimation}
\label{sec:Pauli_terms}
Let us now consider the full $2^n$ dimensional space of pure states of $n$ qubits. We assume that each jump operator $\h L$ is an $n$-qubit Pauli operator, i.e.\ a tensor product of $n$ single qubit Pauli operators including the identity, such that it is Hermitian ($\h L^\dag = \h L$) and involutory ($\h L^2 = \h I$). We also assume that all of the signal and noise jump operators, $\{\h L_{j,(s)}\}_{j=1}^{K}$ and $\{\h L_{j,(n)}\}_{j=K+1}^{K+N}$, correspond to different Pauli operators and thus are pairwise orthogonal. There are $4^n-1$ traceless orthogonal Pauli operators. The Pauli weight of a given Pauli operator is the number of non-identity Pauli operators it contains such that there are $3^n$ traceless orthogonal Pauli operators with a Pauli weight of $n$.

This Pauli structure simplifies the analysis and leads to the following result about the QFI in the extended case:
\begin{claim}
\label{claim:Pauli}
For any set of Pauli jump operators, the noiseless QFI can be recovered by the extended measure-and-reset strategy with a Bell state.
\end{claim}
This claim can be shown as follows. By Eq.~\ref{eq:QFI, any K, N=0}, the noiseless QFI is $4KT$. The noisy QFI of a Bell state given in Eq.~\ref{eq:QFI, Bell state} then recovers this value because $\trSmall{\h L_{j,(s)}^{\dag}\h L_{j,(s)}}=d$ for each signal Pauli operator. This can be understood similarly to superdense coding as the initial Bell state and its up to $4^n-1$ images under the different Pauli operators are all pairwise orthogonal. 

In the unextended case, however, Claim~\ref{claim:Pauli} does not hold. For example, for a single qubit and sensing $\h L_{1,(s)}=\h \sigma_{x}$ in the presence of $\h L_{2,(n)}=\h \sigma_{y}$ and $\h L_{3,(n)}=\h \sigma_{z}$, the unextended QFI vanishes. This is because the unextended initial state and its two images under the noises always span the two-dimensional space since the state cannot be a simultaneous eigenstate of $\h \sigma_{y}$ and $\h \sigma_{z}$.
While the general sufficient and necessary conditions for attaining the noiseless QFI or for the noisy QFI vanishing are unknown in the unextended case, we now provide a couple of examples where the performance can be evaluated:
\begin{example}
\label{example:QFI recovered}
If the Pauli weight of the signal operator is strictly greater than the maximum Pauli weight of the noise operators, then the noiseless QFI can be recovered by preparing an unextended product state.
\end{example}
We prove this result in Appendix~\ref{app:Pauli_terms}. For example, if sensing $\h L_{1,(s)}=\h \sigma_{x}^{(1)}\h \sigma_{x}^{(2)}$ in the presence of all six local noise operators of two qubits, then preparing an unextended product state $\ket{\uparrow}\otimes\ket{\uparrow}$ recovers the noiseless QFI. Here, we restrict to one signal operator ($K=1$) while Claim~\ref{claim:Pauli} applies for any $K\geq1$.
Alternatively, the noiseless QFI can always be recovered in the commuting case:
\begin{example}
\label{example:commuting Paulis}
For any commuting set of Pauli jump operators, the noiseless QFI can be recovered by the unextended measure-and-reset strategy.
\end{example}
This result can be shown as follows. Claim~\ref{claim:Pauli} implies that a Bell state recovers the noiseless QFI and Eq.~\ref{eq:commuting case, ancilla not required} implies that the ancilla is not required for the commuting Hermitian case. This means that a uniform superposition $\ket{\psi}=\frac{1}{\sqrt{d}}\sum_{j=1}^d\ket{e_j}$ in the simultaneous eigenbasis $\{\ket{e_j}\}_{j=1}^d$ recovers the noiseless QFI.

\section{Stochastic waveform estimation}
\label{sec:waveform estimation}
\begin{figure}
    \centering
    \includegraphics[width=\columnwidth]{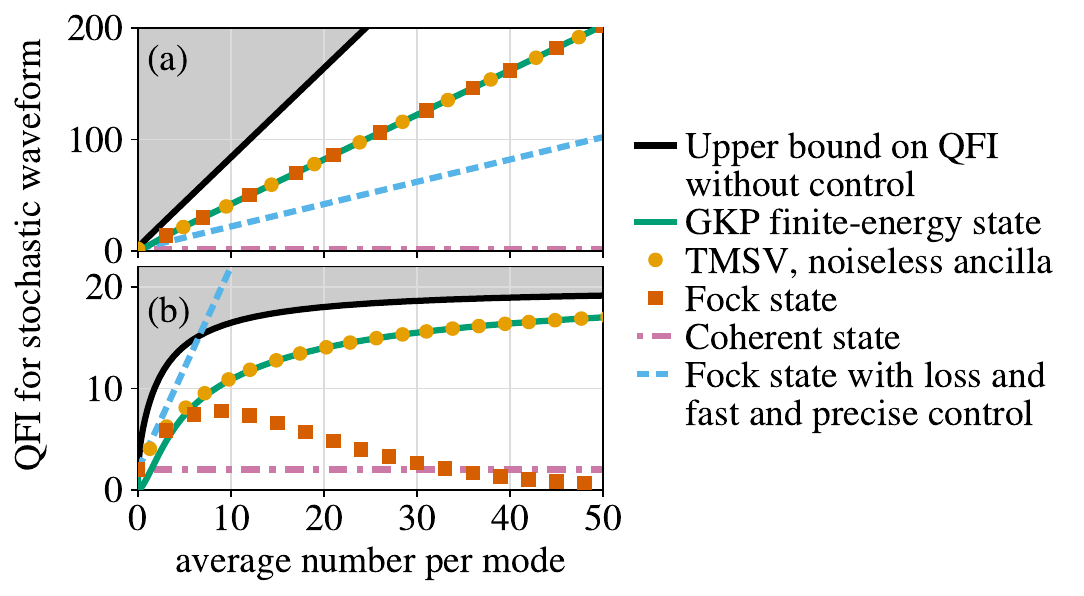}
    \caption{QFI for stochastic waveform estimation versus the average occupation number per mode of the initial state. \textbf{(a)}~In the noiseless case without control, the upper bound on the QFI for any initial state is attained by an SMSV state. \textbf{(b)}~For a loss of 10\%, the QFI without control is bounded at 20 and attained by a Gottesman-Kitaev-Preskill (GKP) state~\cite{GottesmanPRA01EncodingQubit}. In comparison, the QFI with fast and precise control is unbounded and attained by a Fock state. The noiseless QFI is not recovered, however, as shown in panel (a). We omit the factor of $T$ in the QFI with control. (Figure adapted from Fig.~4 of Ref.~\cite{gardner2024stochastic}.)}
    \label{fig:waveform_estimation}
\end{figure}
Let us consider stochastic waveform estimation using a linear quantum device, where we want to estimate the power spectral density of some continuous random variable $y(t)$ in time $t$ by measuring an outgoing bosonic mode from the device. This scenario describes several fundamental physics experiments including searches for quantum gravity, stochastic gravitational waves, and axionic dark matter. In Ref.~\cite{gardner2024stochastic}, the problem of stochastic waveform estimation in the vanishing signal limit was shown to be equivalent to the problem studied here of estimating $\sqrt{\gamma_1}$ for $\h L_{1,(s)} = \h x$ which is a random displacement channel along $\h p$. Here, given the annihilation operator $\h a$ of a single harmonic oscillator, the quadrature operators are $\h x = (\h a + \h a^\dag) / \sqrt{2}$ and $\h p = (-i \h a + i \h a^\dag) / \sqrt{2}$ such that a generalised quadrature at angle $\theta$ is given by $\h x_{\theta}=\cos(\theta) \h x+ \sin(\theta) \h p$. The signal and noise processes in Ref.~\cite{gardner2024stochastic}, however, were modelled in separate stages and fast and precise control of the device was not considered.
We now assume that fast and precise control is possible on timescales for which the signal remains stochastic.

Let us first consider the noiseless case. By Eq.~\ref{eq:QFI, K=1, N=0}, the noiseless QFI is unbounded as $\varSubSmall{\ket{\psi}}{\h x}$ can be made arbitrarily large by preparing, e.g., a single-mode squeezed vacuum (SMSV) state as shown in Fig.~\hyperref[fig:waveform_estimation]{\ref*{fig:waveform_estimation}a}. Given a constraint on the average number $\braopket{\psi}{\h n}{\psi} = \bar n$ where the number operator is $\h n = \h a^\dag \h a$, then an SMSV state has $\varSubSmall{\ket{\psi}}{\h x} = \bar n + \frac{1}{2} + \sqrt{\bar n (\bar n + 1)}$ and is optimal~\cite{gardner2024stochastic}. The noiseless QFI is thus $8 \bar n T$ in the high energy limit of large $\bar n \gg 1$. In comparison, if $\bar n \in\Z$, then the Fock state $\ket{\h n = \bar n}$ such that $\h n \ket{\h n = \bar n} = \bar n \ket{\h n = \bar n}$ has $\varSubSmall{\ket{\h n = \bar n}}{\h x} = \bar n + \frac{1}{2}$ and thus a noiseless QFI of $4 \bar n T$ in the high energy limit as shown in Fig.~\hyperref[fig:waveform_estimation]{\ref*{fig:waveform_estimation}a}.

Let us now consider the case of a single noise source. We prove the following claim:
\begin{claim}
\label{claim:x, a or x, adag}
The optimal QFI with fast and precise control for sensing a random displacement channel generated by a quadrature $\h x_{\theta}$ in the presence of loss $\h a$ or gain $\h a^\dag$ is $2 (\bar n + 1) T$ or $2 \bar n T$, respectively, given a constraint on the average number per mode of $\braopket{\psi}{\h n}{\psi} = \bar n$. This QFI is doubled if sensing isotropic random displacements generated by $\h x$ and $\h p$. The QFI is achieved by a measure-and-reset strategy for any initial state $\ket{\psi}$ that satisfies the following necessary and sufficient condition:
\begin{align}\label{eq:SWE optimality condition}
    \braopket{\psi}{\h a}{\psi}=\braopket{\psi}{\h a^2}{\psi}=0.  
\end{align}
Examples of optimal protocols include continuously preparing a Fock state (if $\bar n\in\Z$) or a finite superposition of Fock states or a binomial code state (if $\bar n\notin\Z$) and performing a number-resolving measurement. Alternatively, preparing a two-mode squeezed vacuum (TMSV) state and performing number-resolving measurements of the system and noiseless ancilla is also optimal.
\end{claim}
This claim can be proved as follows. We showed in Sec.~\ref{sec:properties} that only the component of $\h L_{1,(s)}$ orthogonal to $\h L_{2,(n)}$ contributes to the QFI. The component of $\h x_{\theta}$ orthogonal to $\h a$ is $\h a^\dag/\sqrt{2}$ up to a complex phase factor $e^{i\theta}$ which does not affect the QFI. The QFI in Eq.~\ref{eq:QFI, any K, N=1} for $\h L_{1,(s)}=\h a^\dag/\sqrt{2}$ and $\h L_{2,(n)}=\h a$ is
\begin{align}\label{eq:SWE QFI}
    &\IQ(\sqrt{\gamma_1}=0) \\&= 2T \max_{\ket{\psi}} \left( \bar n + 1 - \absSmall{\braopket{\psi}{\h a}{\psi}}^2 - \frac{\absSmall{\braopket{\psi}{\h a^2}{\psi} - \braopket{\psi}{\h a}{\psi}^2}^2}{\bar n - \absSmall{\braopket{\psi}{\h a}{\psi}}^2}\right).\nonumber
\end{align}
This equals $2(\bar n + 1)T$ for a fixed $\braopket{\psi}{\h n}{\psi} = \bar n$ and is attained by a given state $\ket{\psi}$ if and only if it satisfies the condition in Eq.~\ref{eq:SWE optimality condition}. If $\bar n\in\Z$, then a Fock state $\ket{\h n = \bar n}$ suffices. If $\bar n\notin\Z$, then, for example, we can instead prepare the following state
\begin{align*}
    \ket{\psi} = \sqrt{1 - \frac{\bar n}{n_\star}} \ket{\h n = 0} + \sqrt{\frac{\bar n}{n_\star}} \ket{\h n = n_\star}
\end{align*}
where $n_\star = \max\{3, \lceil \bar n \rceil\}$. Any binomial code state~\cite{michael2016new} $\ket{\psi} = \sum_{n=0}^\infty c_n \ket{\h n = q n + r}$ with $q,r\in\Z$ satisfying $q\geq 3$, $0\leq r<q$, and $\bar n = r + q \sum_{n=0}^\infty n \abs{c_n}^2$ is also optimal. In the extended case, a TMSV state given by
\begin{align*}
    \ket{\psi} = \frac{1}{\cosh(r)}\sum_{k=0}^{\infty}\left[- e^{i\phi} \tanh(r)\right]^{k}|\h n=k\rangle|\h n=k\rangle
\end{align*}
is optimal and requires continuously performing a number-resolving measurement of the system and ancilla. Similarly, the QFI in the presence of a gain $\h a^\dag$ is
\begin{align}\label{eq:SWE QFI gain}
    &\IQ(\sqrt{\gamma_1}=0) \\&= 2T \max_{\ket{\psi}}\left( \bar n - \absSmall{\braopket{\psi}{\h a}{\psi}}^2 - \frac{\absSmall{\braopket{\psi}{\h a^2}{\psi} - \braopket{\psi}{\h a}{\psi}^2}^2}{\bar n + 1 - \absSmall{\braopket{\psi}{\h a}{\psi}}^2}\right)\nonumber
\end{align}
which equals $2 \bar n T$ and Eq.~\ref{eq:SWE optimality condition} remains the optimality condition such that the same states above are optimal. Finally, for the $K=2, N=1$ isotropic case of $\h L_{1,(s)} = \h x$ and $\h L_{2,(s)} = \h p$ in the presence of $\h L_{3,(n)}=\h a$ or $\h a^\dag$, the total QFI in Eq.~\ref{eq:QFI, any K, N=1} sums the QFI from each signal operator. This proves Claim~\ref{claim:x, a or x, adag}.

With fast and precise control, the QFI in the presence of loss or gain is thus unbounded and goes as $2 \bar n T$ for $\bar n \gg 1$ but does not recover the noiseless QFI of $8 \bar n T$ as shown in Fig.~\ref{fig:waveform_estimation} for a loss. Without fast and precise control, however, the QFI for any state is bounded~\cite{gardner2024stochastic} as shown in Fig.~\hyperref[fig:waveform_estimation]{\ref*{fig:waveform_estimation}b}. Fast and precise control can therefore significantly increase the sensitivity of stochastic waveform estimation compared to the case without control. 

In comparison to the optimal states above, preparing an SMSV state yields vanishing QFI in the presence of loss or gain, even with fast and precise control. This result has been previously observed in the case without fast and precise control for a fixed nonzero loss in the weak signal limit~\cite{gardner2024stochastic}. With fast and precise control, the QFI in the presence of loss in Eq.~\ref{eq:SWE QFI} or gain in Eq.~\ref{eq:SWE QFI gain} vanishes since $\braopket{\psi}{\h a}{\psi}=0$ but $\absSmall{\braopket{\psi}{\h a^2}{\psi}}^2 = \bar n (\bar n + 1)$ for the SMSV state. 

Other noise sources are also relevant to experiments sensing, e.g., random displacements generated by $\h x$, such as parallel classical noise $\h x$ and perpendicular classical noise $\h p$. The orthogonal component of $\h x$ is zero for parallel classical noise and nonzero for perpendicular classical noise. This implies that the QFI in the presence of parallel classical noise vanishes, and that with fast and precise control the QFI in the presence of perpendicular classical noise recovers the noiseless QFI using an SMSV state since $\covSubSmall{\ket{\psi}}{\h x}{\h p} = 0$. Without fast and precise control, however, the QFI in the presence of perpendicular classical noise vanishes for an SMSV state with fixed energy in the weak signal limit~\cite{gardner2024stochastic}. This again shows how fast and precise control can improve stochastic waveform estimation. 

Let us now interrogate our assumption of fast and precise control for stochastic waveform estimation. This requirement is stringent: we must perform projective measurements and re-initialise the state of the device on the timescale of $t$ satisfying Eq.~\ref{eq:short time} before a second jump of the noise operators occurs. For example, if a single photon is lost from the differential mode of the arms of an optical interferometer, then we need to respond and re-initialise the state before a second photon is lost. This demand is beyond the present and projected capabilities of optical interferometry experiments such as searching for stochastic gravitational waves with the Laser Interferometric Gravitational-wave Observatory (LIGO)~\cite{AasiCQG15AdvancedLIGO,RomanoLRR17DetectionMethods,renzini2022stochastic} or quantum gravity with the Gravity from the Quantum Entanglement of Space-Time (GQuEST) future experiment~\cite{Vermeulen24PhotonCounting}. Our results in the fast and precise control limit establish the best possible performance of these interferometric devices but do not reflect their attainable near-future performance. We will discuss later possible future work to close this gap and determine the optimal near-future performance of these interferometers under more realistic assumptions.

A more promising application of our present results is to search for ultra-light dark matter such as axions using a superconducting microwave cavity in a static magnetic field~\cite{kim2010axions,choi2021recent,rosenberg2000searches,graham2015experimental,cameron1993search,du2018search, DixitPRL21SearchingDark,AgrawalPRL24StimulatedEmission}. In the fast and precise control limit, we consider operating the device on a timescale shorter than the noise processes but still longer than the coherence time of the axion such that the signal remains stochastic~\footnote{On these short timescales, whether the signal is entirely stochastic or instead needs to be modelled as a combination of a coherent part and an incoherent part is a topic of ongoing research~\cite{gardner2024stochastic,shi2024quantum}.}.
The stochastic signal appears isotropically in the $\h x$ and $\h p$ quadratures such that the QFI in the presence of loss $\h a$ is $4(\bar n +1)T$ and is attained, e.g., by continuously preparing a Fock state and performing a number-resolving measurement by Claim~\ref{claim:x, a or x, adag}. The ability to prepare highly non-Gaussian states in the microwave domain~\cite{deng2024quantum,AgrawalPRL24StimulatedEmission,eickbusch2022fast} is encouraging for eventually reaching the fast and precise control limit as technology further improves.

\section{Conclusions}
In this paper, we have investigated the fast and precise control limit of Lindblad estimation using a quantum device. This limit was previously well-understood for Hamiltonian estimation, i.e.\ sensing a deterministic signal, but not for probing a weak stochastic signal. Several diverse fundamental physics searches look for stochastic signals, from stochastic gravitational waves to axionic dark matter, and so understanding this limit is crucial to establishing the ultimate sensitivity that these devices may reach. We showed that the optimal strategy is to rapidly project the quantum state onto a basis formed from the initial state and orthogonalised images of the initial state under each Lindblad jump operator and then re-initialise the state if a jump is recorded. This is a significant departure from Hamiltonian estimation where the signal should be allowed to coherently build up rather than rapidly measured. We demonstrated this result for sensing a stochastic signal with a single qubit and found a gap between the sensitivity in the noiseless and noisy cases and another gap between the performance with and without entanglement depending on the geometry of the jump operators. We also investigated sensing a stochastic signal with many qubits and Lindblad estimation of Pauli operators. Finally, we discussed the feasibility and benefits of reaching the fast and precise control limit for stochastic waveform estimation.

There are several possible avenues for future work. On a fundamental level, investigating the multi-parameter estimation of different weak decay rates, the effect of a nuisance Hamiltonian term, and the $M>2$ extended parallel strategy shown in Fig.~\hyperref[fig:diagram]{\ref*{fig:diagram}c} for the qubit case may be interesting. Additionally, it would be useful to know whether the unextended measure-and-reset strategy shown in Fig.~\hyperref[fig:diagram]{\ref*{fig:diagram}e} is the optimal unextended fast and precise control strategy shown in Fig.~\hyperref[fig:diagram]{\ref*{fig:diagram}d} for Lindblad estimation as we have only proven this for the extended case. Meanwhile, on a practical level, it may be valuable to consider weaker strategies than fast and precise control. For example, in optical interferometers it is difficult to measure and reset the state as fast as the time taken to lose a single photon. However, it may be possible to control the state on longer timescales such that the noise may jump multiple times but the weak signal may only jump once. Similarly, it may be fruitful to investigate the optimal strategy when restricted to weaker controls than projectively measuring and resetting.

\begin{acknowledgements}
We thank the following people for their advice: Richard Allen, Wenjie Gong, Daniel Mark, Lee McCuller, Chris Pattinson, John Preskill, Alex Retzker, Thomas Schuster, and Sisi Zhou. We also thank the Caltech Chen Quantum Group and the ANU CGA Squeezer Group.
This research is supported by the Australian Research Council Centre of Excellence for Gravitational Wave Discovery (Project No. CE170100004 and CE230100016). J.W.G. and this research are supported by an Australian Government Research Training Program Scholarship and also partially supported by the US NSF grant PHY-2011968. In addition, Y.C. acknowledges the support by the Simons Foundation (Award Number 568762). T.G. acknowledges funding provided by the Institute for Quantum Information and Matter (a US NSF Physics Frontiers Center) and the Quantum Science and Technology Scholarship of the Israel Council for Higher Education. S.A.H. acknowledges support through an Australian Research Council Future Fellowship grant FT210100809. 
This paper has been assigned LIGO Document No.\ P2400542. 
\end{acknowledgements}

\appendix
\section{Proof that Theorem~\ref{claim:optimal QFI} is an upper~bound}
\label{app:ECQFI_bounds}
We want to show the upper bound on the QFI with fast and precise control given in Theorem~\ref{claim:optimal QFI} by calculating the minimum in Eq.~\ref{eq:ECQFI minimisation, Multiple sources of signal and noise}. We may use the minimax theorem since the argument is concave in $\h\rho$ and convex in $\{\vec{c}_j\}_{j=1}^K$~\cite{v1928theorie}. This implies that Eq.~\ref{eq:ECQFI minimisation, Multiple sources of signal and noise} becomes
\begin{align}
\begin{split}
    &\IQ(\sqrt{\gamma_1}=0)
    \\&\leq 4T\max_{\h\rho}\min_{\{\vec{c}_j\}_{j=1}^K}\trSmall{\h\rho\sum_{j=1}^K \h A_j^\dag \h A_j}
    \\&= 4T\max_{\ket{\psi}}\sum_{j=1}^K \min_{\vec{c}_j} \braopket{\psi}{(\h L_{j,(s)}-\vec{c}_j^\T\mathbf{L})^\dag (\h L_{j,(s)}-\vec{c}_j^\T\mathbf{L})}{\psi}
\label{eq:ECQFI_claim_1_app}
\end{split}
\end{align}
where we replaced the density matrix $\h\rho$ by the extended pure state $\ket{\psi}$ that purifies it such that $\h\rho=\trSubSmall{A}{\proj{\psi}}$.

We now want to show that Eq.~\ref{eq:ECQFI_claim_1_app} is equivalent to Theorem~\ref{claim:optimal QFI}. This proof can be done in two different ways. The first way is to differentiate Eq.~\ref{eq:ECQFI_claim_1_app} with respect to $\vec{c}_j^\dag$, which yields that the minimum is achieved at 
\begin{align}\label{eq:optimal c_j}
    \vec{c}_j &= \braopket{\psi}{\mathbf{L}^*\mathbf{L}^\T}{\psi}^{-1}\braopket{\psi}{\mathbf{L}^*\h L_{j,(s)}}{\psi}.
\end{align}
For example, in the noiseless case this means that $\vec{c}_j = \braopket{\psi}{\h L_{j,(s)}}{\psi}$. Substituting this value for $\vec{c}_j$ in Eq.~\ref{eq:ECQFI minimisation, Multiple sources of signal and noise} and using Eq.~\ref{eq:Pi} then shows that the QFI is upper bounded as given in Theorem~\ref{claim:optimal QFI}.
The second way is to rewrite Eq.~\ref{eq:ECQFI_claim_1_app} as follows:
\begin{align*}
\IQ(\sqrt{\gamma_1}=0) \leq
4T\max_{\ket{\psi}}\underset{j=1}{\overset{K}{\sum}}\min_{\vec{c}_{j}}||(\h L_{j,(s)}-\vec{c}_{j}^{\T}\mathbf{L})\ket{\psi}||^{2}
\end{align*}
such that for every $\ket{\psi}$ and $j$, we want to find
$\min_{\vec{c}_{j}}||(\h L_{j,(s)}-\vec{c}_{j}^{\T}\mathbf{L})\ket{\psi}||^{2}$.   
By splitting the identity into projections onto orthogonal subspaces with $\h I = \h\Pi + (\h I - \h\Pi)$, we can show that 
\begin{align*}
&\min_{\vec{c}_{j}}||(\h L_{j,(s)}-\vec{c}_{j}^{\T}\mathbf{L})\ket{\psi}||^{2}\\
&=\min_{\vec{c}_{j}}||(\h\Pi\h L_{j,(s)}-\vec{c}_{j}^{\T}\mathbf{L})\ket{\psi}||^{2}+||(\h I-\h\Pi)\h L_{j,(s)}\ket{\psi}||^{2}\\
&=\bra{\psi}\h L_{j,(s)}^{\dagger}(\h I-\h\Pi)\h L_{j,(s)}\ket{\psi} 
\end{align*}
where the last equality is obtained by choosing $\vec{c}_{j}$ such that $\vec{c}_{j}^{\T}\mathbf{L}\ket{\psi}=\h\Pi\h L_{j,(s)}\ket{\psi}$. This again proves that Theorem~\ref{claim:optimal QFI} is an upper bound. 

\section{Proof that Theorem~\ref{claim:optimal QFI} can be achieved}
\label{app:QFI for multiple signal and noise operators}
We want to show that the QFI for the measure-and-reset strategy with multiple signal and noise sources achieves the ultimate sensitivity limit in Theorem~\ref{claim:optimal QFI}. By Eq.~\ref{eq:general Lindblad master equation, short time}, the signal-free density matrix is 
\begin{align*}
    \lim_{\sqrt{\gamma_1}\rightarrow0}\h\rho' \approx \h\rho + \sum_{j=K+1}^{K+N} \gamma_j t \mathcal{L}_{j,(n)}(\h\rho)
\end{align*} 
such that its matrix elements in the $\{\ket{j}\}_{j=0}^{K+N}$ basis are
\begin{align}
    \label{eq:signal-free DM matrix elements}
    \begin{split}
    \braopket{0}{\lim_{\sqrt{\gamma_1}\rightarrow0}\h\rho'}{0} &= 1 - \sum_{j=K+1}^{K+N} \gamma_{j} t \sum_{k=K+1}^{j} \abs{c_{j,k}}^2
    \\\braopket{j}{\lim_{\sqrt{\gamma_1}\rightarrow0}\h\rho'}{k} &= 0 
    \\\braopket{j}{\lim_{\sqrt{\gamma_1}\rightarrow0}\h\rho'}{0} &=  - \frac{1}{2} \sum_{k=K+1}^{K+N} \gamma_k t \bra{j}\h L_{k,(n)}^\dag \h L_{k,(n)}\ket{0} 
    \\\braopket{j'}{\lim_{\sqrt{\gamma_1}\rightarrow0}\h\rho'}{k'} &= \sum_{l=\max\{j',k'\}}^{K+N} \gamma_{l} t c_{l,j'} c_{l,k'}^*  
    \\\braopket{j'}{\lim_{\sqrt{\gamma_1}\rightarrow0}\h\rho'}{k} &= 0 
    \\\braopket{j'}{\lim_{\sqrt{\gamma_1}\rightarrow0}\h\rho'}{0} &=  \sum_{k=K+1}^{K+N} \gamma_k t \bigg(c_{k,0}^* \bra{j'}\h L_{k,(n)} \ket{0} \\&
    \hspace{1.5cm}
    - \frac{\bra{j'}\h L_{k,(n)}^\dag \h L_{k,(n)}\ket{0}}{2} \bigg) 
    \end{split}
\end{align}
where $1\leq j,k \leq K$ and $K+1\leq j',k' \leq K+N$ for the free indices above. 
Let us find the eigenvectors $\{\ket{\lambda_j}\}_{j=0}^{K+N}$ of this matrix since they form the optimal measurement basis by Eq.~\ref{eq:QFI in the vanishing signal limit}. Firstly, let us try the following ansatz
\begin{align*}
    \ket{\lambda_0} = \ket{0} + t \sum_{j=1}^{K+N} a_{0,j} \ket{j}, \quad \lambda_0 = \braopket{0}{\lim_{\sqrt{\gamma_1}\rightarrow0}\h\rho'}{0}
\end{align*}
such that the eigenvector equation is
\begin{align*}
    \left(\lim_{\sqrt{\gamma_1}\rightarrow0}\h\rho' - \lambda_0 \h I\right)\ket{\lambda_0}
    = \sum_{j=1}^{K+N} \left(\braopket{j}{\lim_{\sqrt{\gamma_1}\rightarrow0}\h\rho'}{0} - a_{0,j}t\right) \ket{j}
\end{align*}
which is zero if $a_{0,j} = \braopket{j}{\lim_{\sqrt{\gamma_1}\rightarrow0}\h\rho'}{0} / t$ for $j=1,\mathellipsis,K+N$.
Secondly, let us try the following ansatz for $j=1,\mathellipsis,K$
\begin{align*}
    \ket{\lambda_j}=\sum_{k=1}^{K} a'_{j,k} \ket{k} + t \sum_{k=0}^{K+N} a_{j,k} \ket{k}, \quad \lambda_j = 0
\end{align*}
such that the eigenvector equation is
\begin{align*}
    \lim_{\sqrt{\gamma_1}\rightarrow0}\h\rho'\ket{\lambda_j}
    &= \left(a_{j,0} t + \sum_{k=1}^{K} a'_{j,k} \braopket{0}{\lim_{\sqrt{\gamma_1}\rightarrow0}\h\rho'}{k}\right)\ket{0}
\end{align*}
which is zero if
\begin{align*}
    a_{j,0} = - \sum_{k=1}^K a'_{j, k} \frac{\braopket{0}{\lim_{\sqrt{\gamma_1}\rightarrow0}\h\rho'}{k}}{t} 
    .
\end{align*}
This means that there is a degenerate $K$-dimensional subspace of eigenvectors with eigenvalue zero for which we may freely choose which to project onto. Without loss of generality, we choose the eigenvectors which are $\order{t}$ perturbations of the $\{\ket{j}\}_{j=1}^{K}$ states such that 
\begin{align*}
    \ket{\lambda_j}=\ket{j} + t \sum_{k=0}^{K+N} a_{j,k} \ket{k}, \quad \lambda_j = 0
\end{align*}
where $a_{j,0} = -\braopket{0}{\lim_{\sqrt{\gamma_1}\rightarrow0}\h\rho'}{j}/t$ for $j=1,\mathellipsis,K$.
The remaining $N$ eigenvectors, corresponding to the noisy subspace, are orthogonal to $\{\ket{\lambda_j}\}_{j=0}^{K}$ such that they are of the following form for $j=K+1,\mathellipsis,K+N$
\begin{align*}
    \ket{\lambda_j}=\sum_{k=K+1}^{K+N} a'_{j,k} \ket{k} + t \sum_{k=0}^{K+N} a_{j,k} \ket{k}
    .
\end{align*}
Unlike the first $K+1$ eigenvectors which are $\order{t}$ perturbations of the initial state and signal eigenbasis $\{\ket{j}\}_{j=1}^{K}$, these last $N$ eigenvectors are not $\order{t}$ perturbations of $\{\ket{j}\}_{j=K+1}^{K+N}$ and are not necessarily degenerate.

We now calculate the probabilities $p_j = \tr{\proj{\lambda_j}\h\rho'}$ of projecting $\h\rho'$ onto this optimal measurement basis. The first $K+1$ probabilities are equal to the corresponding diagonal elements of $\h\rho'$, i.e.\ $p_j = \braopket{j}{\h\rho'}{j}$ for $j=0,1,\mathellipsis,K$. Since $p_0$ has a constant term, its contribution to the CFI in Eq.~\ref{eq:CFI definition, discrete} vanishes. Meanwhile, the probabilities for $j=1,\mathellipsis,K$ are
\begin{align}
    \label{eq:probabilities 1,...,K}
    p_j &= \gamma_1 t \sum_{k=j}^{K} \abs{c_{k,j}}^2
    \\&= \gamma_1 t \braopket{\psi}{\h L_{j,(s)}^\dag (\h I - \h \Pi) \h L_{j,(s)}}{\psi}\nonumber
\end{align}
such that the sum of their contributions to the CFI in Eq.~\ref{eq:CFI definition, discrete} equals the QFI in Theorem~\ref{claim:optimal QFI} after multiplying by the number of measurements $M=T/t$. Moreover, all of the information about $\sqrt{\gamma_1}$ comes from projecting onto the combined noise-free signal subspace with $\h I - \h \Pi = \sum_{j=1}^K \proj{j}$, i.e.\ we do not need to differentiate the different signals since the decay rate is common.

We now need to show that the remaining probabilities $p_j$ for $j=K+1,\mathellipsis,K+N$ from the noisy $N$-dimensional subspace contribute no information in the limit of $\sqrt{\gamma_1}\rightarrow0$. These probabilities only depend on the constant coefficients $a'_{j,k}$ of $\ket{\lambda_j}$ and thus on the $\braopket{j'}{\h\rho'}{k'}$ matrix elements where $K+1\leq j',k' \leq K+N$, such that $p_j$ for $j=K+1,\mathellipsis,K+N$ can be shown to equal
\begin{align*}
    \gamma_{1} t \sum_{m=1}^{K} \abs{\sum_{k=K+1}^{K+N} c_{m,k}^*a'_{j,k}}^2 + \sum_{m=K+1}^{K+N} \gamma_{m} t \abs{\sum_{k=K+1}^{m} c_{m,k}^*a'_{j,k}}^2
\end{align*}
The contribution to the CFI in Eq.~\ref{eq:CFI definition, discrete} corresponding to $p_j$ vanishes provided that the second term above is nonzero. That is, if $\absSmall{\sum_{k=K+1}^{m} c_{m,k}^*a'_{j,k}}^2$ is nonzero for at least one $m=K+1,\mathellipsis,K+N$. Let us show that this must be the case since otherwise we will find a contradiction if we try to set this term to zero for all $m=K+1,\mathellipsis,K+N$. Firstly, consider $m=K+1$ such that $\absSmall{c_{K+1,K+1}^*a'_{j,K+1}}^2=0$ if and only if $a_{j,K+1}'=0$ since we assume that $c_{j,j}\neq0$ without loss of generality. Then, consider $m=K+2$ such that the term is $\absSmall{c_{K+2,K+2}^*a_{j,K+2}'}^2$ given $a_{j,K+1}'=0$, which then demands that $a_{j,K+2}'=0$ too. This process continues similarly with the $m=K+3$ term and so on up to the $m=K+N$ term with $a_{j,m}'=0$ required at each stage. Since $\ket{\lambda_j}$ is normalised, however, $a_{j,m}'\neq0$ for at least one $m=K+1,\mathellipsis,K+N$ such that this process must fail and thus the contribution to the CFI from $p_j$ vanishes. 

Projecting onto $\{\ket{j}\}_{j=0}^{K+N}$ instead of $\{\ket{\lambda_j}\}_{j=0}^{K+N}$ yields the same CFI since the only differences between the two bases, i.e.\ $\order{t}$ perturbations and a rotation of the noisy subspace, do not affect the CFI to leading order. This completes the proof of Theorem~\ref{claim:optimal QFI}.

\section{Optimal initial states}
\label{app:optimal_states}
We establish here the sufficient and necessary conditions for a state $\h \rho$ to be optimal. Let the solution of the SDP in Eq.~\ref{eq:ECQFI minimisation, Multiple sources of signal and noise} be $\vec{c}_{j,\text{min}}$ such that $\h A_{j,\text{min}}=\h L_{j}-\vec{c}_{j,\text{min}}^\T\mathbf{L}$ for $1\leq j\leq K$. A state $\h \rho$ is then optimal if and only if it satisfies the following conditions:
\begin{enumerate}
    \item[(I)] The support of $\h \rho$ lies within the eigenspace of the maximal eigenvalue of $\sum_{j=1}^{K}\h A_{j,\text{min}}^{\dagger}\h A_{j,\text{min}}$.
    \item[(II)] For any signal index $1\leq j\leq K$ and noise index $K+1\leq k\leq K+N$, then $\trSmall{\h\rho \h A_{j,\text{min}}}=0$ and $\trSmall{\h\rho \h L_{k,(n)}^{\dagger}\h A_{j,\text{min}}}=0$ hold. 
\end{enumerate}

These conditions can be proven as a special case of the conditions found in Ref.~\cite{zhou2021asymptotic}, however, we now present a self-contained derivation for our problem.
$\h\rho$ is an optimal initial state if and only if it satisfies:
\begin{align}
\tr{\h\rho\sum_{j=1}^{K}\h A_{j,\text{min}}^{\dagger}\h A_{j,\text{min}}}&=\max_{\h\varrho} \min_{\vec{c}} \tr{\h\varrho\sum_{j=1}^{K}\h A_{j}^{\dagger}\h A_{j}}\label{eq:opt state line1}
\\
&=\min_{\vec{c}} \max_{\h\varrho}
\tr{\h\varrho\sum_{j=1}^{K}\h A_{j}^{\dagger}\h A_{j}}\label{eq:opt state line2}
\end{align}
Condition~(I) guarantees that $\h\rho$ and $\vec{c}_{\text{min}}$ saturate Eq.~\ref{eq:opt state line2}. We have shown that $\h\rho$ and $\vec{c}_{\text{min}}$ saturate Eq.~\ref{eq:opt state line1} if and only if $A_{j,\text{min}}\ket{\psi}$ is orthogonal to $\ket{\psi}$ and $\h L_{k,(n)}\ket{\psi}$ for every $K+1\leq k\leq K+N$ where $\ket{\psi}$ is a purification of $\h\rho$. This is exactly Condition~(II).

\section{QFI with two sources of noise}
\label{app:Two sources of noise}
In the case where there are $K$ signals and two noises $\h L_{K+1,(n)}$ and $\h L_{K+2,(n)}$, the QFI in Theorem~\ref{claim:optimal QFI} is
\begin{align}
    \label{eq:QFI, any K, N=2}
    \IQ(\sqrt{\gamma_1}=0) &= 4 T \max_{\ket{\psi}} \sum_{j=1}^K f_{\ket{\psi}}(\h L_{j,(s)}, \h L_{K+1,(n)}, \h L_{K+2,(n)}).
\end{align}
Here, the function $f_{\ket{\psi}}$ is defined as
\begin{align*}
    &f_{\ket{\psi}}(\h L_{j,(s)}, \h L_{K+1,(n)}, \h L_{K+2,(n)}) \\&= \varSubSmall{\ket{\psi}}{\h L_{j,(s)}} -\frac{\abs{\covSubSmall{\ket{\psi}}{\h L_{K+1,(n)}}{\h L_{j,(s)}}}^2}{\varSubSmall{\ket{\psi}}{\h L_{K+1,(n)}}}
    \\&\hspace{0.5cm}-\frac{\abs{\covSubSmall{\ket{\psi}}{\h L_{K+2,(n)}}{\h L_{j,(s)}}-\zeta}^2}{\varSubSmall{\ket{\psi}}{\h L_{K+2,(n)}}-\xi}
\end{align*}
where $\zeta$ and $\xi$ are given by
\begin{align*}
    \zeta &= \frac{\covSubSmall{\ket{\psi}}{\h L_{K+2,(n)}}{\h L_{K+1,(n)}}\covSubSmall{\ket{\psi}}{\h L_{K+1,(n)}}{\h L_{j,(s)}}}{\varSubSmall{\ket{\psi}}{\h L_{K+1,(n)}}}
    \\\xi &= \frac{\abs{\covSubSmall{\ket{\psi}}{\h L_{K+1,(n)}}{\h L_{K+2,(n)}}}^2}{\varSubSmall{\ket{\psi}}{\h L_{K+1,(n)}}}
\end{align*}
The second term in $f_{\ket{\psi}}$ removes the $\h L_{K+1,(n)}$ component of the signal and is the same as the case of a single noise operator in Eq.~\ref{eq:QFI, any K, N=1}. The third term in $f_{\ket{\psi}}$ then removes the remaining $\h L_{K+2,(n)}$ component of the signal that is orthogonal to the $\h L_{K+1,(n)}$ component already removed.

\section{Noiseless case with finite signal}
\label{app:Non-vanishing signal}
We now discuss the noiseless case of sensing a finite signal as discussed in Sec.~\ref{sec:Impact of finite signal and time}. We want to show that the QFI for a finite signal is the same as in the vanishing signal limit, provided that all of the signal jump operators are Hermitian. The proof is analogous to the vanishing signal case. Compared to Eq.~\ref{eq:K, Kdot}, the Kraus operators from Eq.~\ref{eq:Kraus representation} are now
\begin{align*}
    \mathbf{K} &= \pmatrixByJames{\h I - \frac{t}{2} \gamma_1 \sum_{k=1}^{K} \h L_{k,(s)}^\dag\h L_{k,(s)} \\ \sqrt{\gamma_{1} t}\h L_{1,(s)}\\ \vdots\\ \sqrt{\gamma_{1} t}\h L_{K,(s)}}
    \\
    \dot{\mathbf{K}} &= \pmatrixByJames{- t\sqrt{\gamma_1}\sum_{k=1}^{K} \h L_{k,(s)}^\dag\h L_{k,(s)}\\ \h L_{1,(s)}\sqrt{t}\\ \vdots\\ \h L_{K,(s)}\sqrt{t}}
    .
\end{align*}
We choose the same gauge as the vanishing signal limit in Eq.~\ref{eq:gauge}:
\begin{align*}
    h &= \pmatrixByJames{
    0 & \mathbf{g}^\dag \\
    \mathbf{g} & \mathbf{0}_{K,K}\\
    }, \quad 
    \mathbf{g} = -i\sqrt{t}\pmatrixByJames{c_1\\ \vdots\\ c_K}
\end{align*}
such that $\dot{\mathbf{K}}-ih\mathbf{K}$ now equals Eq.~\ref{eq:Kdot - i h K} plus higher order corrections in time $t$ as follows:
\begin{align*}
    \dot{\mathbf{K}}-ih\mathbf{K}&=
    \pmatrixByJames{0\\ \h L_{1,(s)}-c_1\h I\\ \vdots\\ \h L_{K,(s)}-c_K\h I}\sqrt{t} 
    \\&+ \pmatrixByJames{t\sqrt{\gamma_1}\sum_{k=1}^{K} c_k^* \h L_{k,(s)}-\h L_{k,(s)}^\dag\h L_{k,(s)}\\ c_1 \frac{1}{2} t^{3/2} \gamma_1 \sum_{k=1}^{K} \h L_{k,(s)}^\dag\h L_{k,(s)} \\ \vdots\\ c_K \frac{1}{2} t^{3/2} \gamma_1 \sum_{k=1}^{K} \h L_{k,(s)}^\dag\h L_{k,(s)}}
    .
\end{align*}
This means that $\h\alpha$ and $\h\beta$ in Eq.~\ref{eq:alpha and beta} are now
\begin{align*}
    \h\alpha &= t \sum_{k=1}^K \h A_k^\dag \h A_k + \order{t^2}
    \\\h\beta &= i\sqrt{\gamma_{1}} t \sum_{k=1}^K \left(c_k \h L_{k,(s)}^\dag - c_k^* \h L_{k,(s)}\right) + \order{t^2}
\end{align*}
where $\h A_{k}=\h L_{k,(s)}-c_k\h I$ here. We now apply our assumption that all of the signal jump operators are Hermitian to imply that $\h\beta = \order{t^2}$ since the coefficients $\{c_k\}_{k=1}^K$ are then real by Eq.~\ref{eq:optimal c_j}. Since $\h\alpha=\order{t}$ and $M=T/t$ with $T$ fixed, the second term in Eq.~\ref{eq:ECQFI with fast and precise control} can then be dropped since it is $\order{T^2\sqrt{t}}$ compared to the first term which is $\order{T}$. This establishes that the noiseless QFI for a non-vanishing signal $\sqrt{\gamma_1}>0$ is upper bounded by the vanishing signal limit in Eq.~\ref{eq:QFI, any K, N=0}, provided that the signal jump operators are Hermitian.

We now show that the CFI of the same measure-and-reset strategy as the vanishing signal limit, i.e.\ preparing and projecting onto the optimal state $\ket\psi$, attains this upper bound on the QFI. The probability $p=\braopket{\psi}{\h\rho'}{\psi}$ of projecting the density matrix $\h\rho'$, expressed in the $\{\ket{j}\}_{j=0}^K$ basis, onto the initial state $\ket\psi$ is
\begin{align*}
    p &= 1 - \gamma_1 t \sum_{j=1}^{K} \sum_{k=1}^{j} \abs{c_{jk}}^2.
\end{align*}
The CFI in Eq.~\ref{eq:CFI definition, discrete} of $p$ and $1-p$ with respect to $\sqrt{\gamma_1}$ therefore equals the upper bound to leading order in time, once multiplied by the number of measurements $M=T/t$. This is because the probability $p$ is the same in the vanishing signal limit in Eq.~\ref{eq:probabilities 1,...,K}.

This completes the proof, assuming that the signal jump operators are Hermitian. In the non-Hermitian case, the measure-and-reset CFI above is the same since we did not use the assumption of Hermitiancy in deriving it. We defer to future work showing that the same upper bound holds in the non-Hermitian case such that the optimal QFI is independent of $\gamma_1$.

\section{Proofs for the commuting Hermitian~case}
\label{app:classical_case}
We derive here the results for the commuting Hermitian case in Sec.~\ref{sec:Classical limit}. Let us first prove for $d=3$ and $N=1$ that the QFI is given in Eq.~\ref{eq: ECQFI_classical_1} and that the optimal distribution is $p_{i}\propto|l_{1,i}|$. We derive this using the SDP formulation of Eq.~\ref{eq:ECQFI minimisation, Multiple sources of signal and noise}. The solution of this SDP is given in the common eigenbasis as:
\begin{align}
    l_{1}=c_{1,\text{min}}\vec{1}+c_{2,\text{min}}l_{2}+a_{\text{min}}
    \label{eq:classical_sdp_expression}
\end{align}
where $a_{\text{min}}$ is the diagonal of $\h A_{\text{min}}$ given $c_{1,\text{min}}$ and $c_{2,\text{min}}$. The optimal QFI is then given by $4 T ||a_{\text{min}}||_{\infty}^{2}$. Eq.~\ref{eq:classical_sdp_expression} implies that $a_{\text{min}}\notin\text{span}\{l_2,\vec{1} \}$ such that $a_{\text{min}}$ is not orthogonal to $l_{1}$ and
\begin{align}
    |l_{1}|^{2}=l_{1}\cdot a_{\text{min}}.
    \label{eq:overlap_a_l1}
\end{align}
If $\left(l_{2}\right)_{1}=\left(l_{2}\right)_{3}$ then the optimal QFI is equal to the noiseless one: $4 T\left(l_{1,\text{max}}-l_{1,\text{min}}\right)^{2} $ with the optimal state being $(1/\sqrt{2}, 0, 1/\sqrt{2})^\T$.

Let us now split into cases regarding the size of the support of the optimal state which is either length two or three. If the support is length two, then $l_1=(1,0,-1)$ and $l_2=(1,-2,1)$ up to constants and permutations, and the optimal QFI equals the two-dimensional noiseless QFI. If the support is instead length three, then $a_{\text{min}}=x v_{i}$ where $ v_{1}=\left(1,-1,-1\right)$, $ v_{2}=\left(1,1,-1\right)$, or $ v_{3}=\left(1,-1,1\right)$ and $x=|l_{1}|^{2}/\sum_{i}|l_{1,i}|$ by Eq.~\ref{eq:overlap_a_l1} since $\max_i \left(l_{1}\cdot v_{i}\right)=\sum_{i} |l_i|$. This also implies that $\left( a_{\text{min}} \right)_i=x \; \text{sign}(l_{1,i})$. The optimal QFI is therefore $4 Tx^{2}$ which is Eq.~\ref{eq: ECQFI_classical_1}.

Let us now derive the optimal probability distribution $\left(p_{i}\right)_{i=1}^{3}$ for the initial state. By Condition~(II) in Appendix~\ref{app:optimal_states} and using $\left(a_{\text{min}}\right)_{i}\propto\text{sign}(l_{1,i})$, the necessary and sufficient conditions on the optimal distribution are the following:
\begin{align*}
\sum_{i=1}^3 p_{i}\,\text{sign}(l_{1,i})=0,\quad
\sum_{i=1}^3 p_{i}\,\text{sign}(l_{1,i})l_{2,i}=0.    
\end{align*}
Since $l_1\cdot\vec{1}=l_1\cdot\, l_2=0$, then the distribution $p_{i}\propto|l_{1,i}|$ is thus optimal. This completes the proof of Eq.~\ref{eq: ECQFI_classical_1} for $d=3$ and $N=1$.

This derivation can be generalised for any $d=N+2$. We introduce the following regularity or generalised non-degeneracy condition: The vectors $\{\vec{1}, l_2,\mathellipsis,l_{N+1}\}$ are regular if any $k$-dimensional sub-vectors of them span $\mathbb{R}^{k}$. (For any set of indices $J=\{j_{1},\mathellipsis,j_{k}\}$, the sub-vectors of $\{\vec{1}, l_2,\mathellipsis,l_{N+1}\}$ that correspond to these indices span $\mathbb{R}^{k}$.) This regularity condition does not always hold, e.g.\ it is not satisfied for commuting Pauli operators, but it does typically hold for uniformly randomly sampled commuting operators. 

Given this regularity condition, the optimal QFI is given in Eq.~\ref{eq: ECQFI_classical_1} and the optimal state corresponds to a distribution $p_{i}\propto|l_{1,i}|$. The proof of this is similar to the $d=3$ case. For example, Eq.~\ref{eq:classical_sdp_expression} becomes:
\begin{align}\label{eq:classical_sdp_expression 2}
    l_{1}=c_{1,\text{min}}\vec{1}+\underset{j=2}{\overset{N+1}{\sum}}c_{j,\text{min}}l_{j}+a_{\text{min}}.    
\end{align}
The regularity condition guarantees that the optimal state is supported by all $d$ eigenstates, which implies that $a_{\text{min}}$ takes the form of $a_{\text{min}}=x v$ where $v\in\{1,-1\}^d$ such that
\begin{align}\label{eq:classical x}
    x=\frac{|l_{1}|^{2}}{\text{max}_{v}(l_{1}\cdot v)}=\frac{|l_{1}|^{2}}{\sum_{i}|l_{1,i}|}=\frac{||l_{1}||_{2}^{2}}{||l_{1}||_{1}}.    
\end{align}
The derivation of the optimal input is then analogous to the $d=3$ case.



We can now also derive the more general optimal QFI expression of Eq.~\ref{eq: ECQFI_classical_2}. Given a solution of the SDP in Eq.~\ref{eq:classical_sdp_expression 2}, we can restrict ourselves to the support of the optimal state, i.e\ the eigenspace of the maximal eigenvalue of $\h A_{\text{min}}^{\dagger}\h A_{\text{min}}$. Eq.~\ref{eq:classical_sdp_expression 2} then becomes
\begin{align}\label{eq:classical_sdp_expression 3}
    l_{1}^{(I)}=c_{1,\text{min}}\vec{1}+\underset{j=2}{\overset{N+1}{\sum}}c_{j,\text{min}}l_{j}^{(I)}+xv
\end{align}
where $a_{\text{min}}^{\left(I\right)}=xv$ and $v\in\{1,-1\}^{|I|}$. We now cannot use Eq.~\ref{eq:classical x} to find $x$ since $l_{1}^{\left(I\right)}$ is not necessarily orthogonal to $\text{span}\{\vec{1}$, $l_{2}^{\left(I\right)},\mathellipsis,l_{N+1}^{\left(I\right)}\}$. Instead, we can replace $l_{1}^{\left(I\right)}$ with $l_{1,\perp}^{\left(I\right)}$ as this will only modify the coefficients $\left\{ c_{i}\right\} _{i=1}^{N+1}$ such that Eq.~\ref{eq:classical_sdp_expression 3} becomes
\begin{align*}
    l_{1,\perp}^{\left(I\right)}=c'_{1}\vec{1}+\sum_{j=2}^{N+1}c'_{j}l_{j}^{\left(I\right)}+xv.
\end{align*}
By orthogonality, we then obtain that $x=||l_{1,\perp}^{\left(I\right)}||_{2}^{2}/||l_{1,\perp}^{\left(I\right)}||_{1}$ which yields the optimal QFI expression. The proof for the optimal distribution is analogous to the previous case. 

\section{Proof of Example~\ref{example:QFI recovered}}
\label{app:Pauli_terms}


We want to prove Example~\ref{example:QFI recovered} concerning the QFI of the unextended measure-and-reset strategy for Pauli Lindblad estimation. We can assume that the signal operator is $\hat{L}_{1,\left(s\right)}=\prod_{j=1}^n\h \sigma_{x}^{(j)}$ without loss of generality. Then, the noiseless QFI of $4T$ is recovered by preparing $\ket{\uparrow}^{\otimes n}$ since it is a maximal variance state of $\hat{L}_{1,\left(s\right)}$ and, for any noise operator $\h L_{j,\left(n\right)}$ with Pauli weight less than $n$, there exists at least one qubit for which the state remains $\ket{\uparrow}$. Thus, the image of the initial state under the signal, $\ket{\downarrow}^{\otimes n}$, is orthogonal to the initial state and the image under each noise operator such that the noiseless QFI is achieved.

\nocite{apsrev42Control}
\bibliographystyle{bibliography/style.bst}

\renewcommand{\selectlanguage}[1]{}
\bibliography{bibliography/bib}

\begin{thebibliography}{70}%
\makeatletter
\providecommand \@ifxundefined [1]{%
 \@ifx{#1\undefined}
}%
\providecommand \@ifnum [1]{%
 \ifnum #1\expandafter \@firstoftwo
 \else \expandafter \@secondoftwo
 \fi
}%
\providecommand \@ifx [1]{%
 \ifx #1\expandafter \@firstoftwo
 \else \expandafter \@secondoftwo
 \fi
}%
\providecommand \natexlab [1]{#1}%
\providecommand \enquote  [1]{``#1''}%
\providecommand \bibnamefont  [1]{#1}%
\providecommand \bibfnamefont [1]{#1}%
\providecommand \citenamefont [1]{#1}%
\providecommand \href@noop [0]{\@secondoftwo}%
\providecommand \href [0]{\begingroup \@sanitize@url \@href}%
\providecommand \@href[1]{\@@startlink{#1}\@@href}%
\providecommand \@@href[1]{\endgroup#1\@@endlink}%
\providecommand \@sanitize@url [0]{\catcode `\\12\catcode `\$12\catcode `\&12\catcode `\#12\catcode `\^12\catcode `\_12\catcode `\%12\relax}%
\providecommand \@@startlink[1]{}%
\providecommand \@@endlink[0]{}%
\providecommand \url  [0]{\begingroup\@sanitize@url \@url }%
\providecommand \@url [1]{\endgroup\@href {#1}{\urlprefix }}%
\providecommand \urlprefix  [0]{URL }%
\providecommand \Eprint [0]{\href }%
\providecommand \doibase [0]{https://doi.org/}%
\providecommand \selectlanguage [0]{\@gobble}%
\providecommand \bibinfo  [0]{\@secondoftwo}%
\providecommand \bibfield  [0]{\@secondoftwo}%
\providecommand \translation [1]{[#1]}%
\providecommand \BibitemOpen [0]{}%
\providecommand \bibitemStop [0]{}%
\providecommand \bibitemNoStop [0]{.\EOS\space}%
\providecommand \EOS [0]{\spacefactor3000\relax}%
\providecommand \BibitemShut  [1]{\csname bibitem#1\endcsname}%
\let\auto@bib@innerbib\@empty
\bibitem [{\citenamefont {Gardner}\ \emph {et~al.}(2024)\citenamefont {Gardner}, \citenamefont {Gefen}, \citenamefont {Haine}, \citenamefont {Hope}, \citenamefont {Preskill}, \citenamefont {Chen},\ and\ \citenamefont {McCuller}}]{gardner2024stochastic}%
  \BibitemOpen
  \bibfield  {author} {\bibinfo {author} {\bibfnamefont {J.~W.}\ \bibnamefont {Gardner}}, \bibinfo {author} {\bibfnamefont {T.}~\bibnamefont {Gefen}}, \bibinfo {author} {\bibfnamefont {S.~A.}\ \bibnamefont {Haine}}, \bibinfo {author} {\bibfnamefont {J.~J.}\ \bibnamefont {Hope}}, \bibinfo {author} {\bibfnamefont {J.}~\bibnamefont {Preskill}}, \bibinfo {author} {\bibfnamefont {Y.}~\bibnamefont {Chen}},\ \bibnamefont {and}\ \bibinfo {author} {\bibfnamefont {L.}~\bibnamefont {McCuller}},\ }\bibfield  {title} {\bibinfo {title} {Stochastic waveform estimation at the fundamental quantum limit},\ }\href@noop {} {\bibfield  {journal} {\bibinfo  {journal} {arXiv preprint arXiv:2404.13867}\ } (\bibinfo {year} {2024})}\BibitemShut {NoStop}%
\bibitem [{\citenamefont {Aasi}\ \emph {et~al.}(2015)\citenamefont {Aasi}, \citenamefont {Abbott}, \citenamefont {Abbott}, \citenamefont {Abbott}, \citenamefont {Abernathy}, \citenamefont {Ackley}, \citenamefont {Adams}, \citenamefont {Adams}, \citenamefont {Addesso}, \citenamefont {Adhikari}, \citenamefont {Adya}, \citenamefont {Affeldt}, \citenamefont {Aggarwal}, \citenamefont {Aguiar}, \citenamefont {Ain}, \citenamefont {Ajith}, \citenamefont {Alemic}, \citenamefont {Allen}, \citenamefont {Amariutei}, \citenamefont {Anderson}, \citenamefont {Anderson}, \citenamefont {Arai}, \citenamefont {Araya}, \citenamefont {Arceneaux}, \citenamefont {Areeda}, \citenamefont {Ashton}, \citenamefont {Ast}, \citenamefont {Aston}, \citenamefont {Aufmuth}, \citenamefont {Aulbert}, \citenamefont {Aylott}, \citenamefont {Babak}, \citenamefont {Baker}, \citenamefont {Ballmer}, \citenamefont {Barayoga}, \citenamefont {Barbet}, \citenamefont {Barclay}, \citenamefont {Barish}, \citenamefont {Barker}, \citenamefont {Barr},
  \citenamefont {Barsotti}, \citenamefont {Bartlett}, \citenamefont {Barton}, \citenamefont {Bartos}, \citenamefont {Bassiri}, \citenamefont {Batch}, \citenamefont {Baune}, \citenamefont {Behnke}, \citenamefont {Bell}, \citenamefont {Bell}, \citenamefont {Benacquista}, \citenamefont {Bergman}, \citenamefont {Bergmann}, \citenamefont {Berry}, \citenamefont {Betzwieser}, \citenamefont {Bhagwat}, \citenamefont {Bhandare}, \citenamefont {Bilenko}, \citenamefont {Billingsley}, \citenamefont {Birch}, \citenamefont {Biscans}, \citenamefont {Biwer}, \citenamefont {Blackburn}, \citenamefont {Blackburn}, \citenamefont {Blair}, \citenamefont {Blair}, \citenamefont {Bock}, \citenamefont {Bodiya}, \citenamefont {Bojtos}, \citenamefont {Bond}, \citenamefont {Bork}, \citenamefont {Born}, \citenamefont {Bose}, \citenamefont {Brady}, \citenamefont {Braginsky}, \citenamefont {Brau}, \citenamefont {Bridges}, \citenamefont {Brinkmann}, \citenamefont {Brooks}, \citenamefont {Brown}, \citenamefont {Brown}, \citenamefont {Brown},
  \citenamefont {Buchman}, \citenamefont {Buikema}, \citenamefont {Buonanno}, \citenamefont {Cadonati}, \citenamefont {Bustillo}, \citenamefont {Camp}, \citenamefont {Cannon}, \citenamefont {Cao}, \citenamefont {Capano}, \citenamefont {Caride}, \citenamefont {Caudill}, \citenamefont {Cavagli{\`a}}, \citenamefont {Cepeda}, \citenamefont {Chakraborty}, \citenamefont {Chalermsongsak}, \citenamefont {Chamberlin}, \citenamefont {Chao}, \citenamefont {Charlton}, \citenamefont {Chen}, \citenamefont {Cho}, \citenamefont {Cho}, \citenamefont {Chow}, \citenamefont {Christensen}, \citenamefont {Chu}, \citenamefont {Chung}, \citenamefont {Ciani}, \citenamefont {Clara}, \citenamefont {Clark}, \citenamefont {Collette}, \citenamefont {Cominsky}, \citenamefont {Constancio}, \citenamefont {Cook}, \citenamefont {Corbitt}, \citenamefont {Cornish}, \citenamefont {Corsi}, \citenamefont {Costa}, \citenamefont {Coughlin}, \citenamefont {Countryman}, \citenamefont {Couvares}, \citenamefont {Coward}, \citenamefont {Cowart},
  \citenamefont {Coyne}, \citenamefont {Coyne}, \citenamefont {Craig}, \citenamefont {Creighton}, \citenamefont {Creighton}, \citenamefont {Cripe}, \citenamefont {Crowder}, \citenamefont {Cumming}, \citenamefont {Cunningham}, \citenamefont {Cutler}, \citenamefont {Dahl}, \citenamefont {Canton}, \citenamefont {Damjanic}, \citenamefont {Danilishin}, \citenamefont {Danzmann}, \citenamefont {Dartez}, \citenamefont {Dave}, \citenamefont {Daveloza}, \citenamefont {Davies}, \citenamefont {Daw}, \citenamefont {DeBra}, \citenamefont {Pozzo}, \citenamefont {Denker}, \citenamefont {Dent}, \citenamefont {Dergachev}, \citenamefont {DeRosa}, \citenamefont {DeSalvo}, \citenamefont {Dhurandhar}, \citenamefont {D{\textasciiacute}{\i}az}, \citenamefont {Palma}, \citenamefont {Dojcinoski}, \citenamefont {Dominguez}, \citenamefont {Donovan}, \citenamefont {Dooley}, \citenamefont {Doravari}, \citenamefont {Douglas}, \citenamefont {Downes}, \citenamefont {Driggers}, \citenamefont {Du}, \citenamefont {Dwyer}, \citenamefont
  {Eberle}, \citenamefont {Edo}, \citenamefont {Edwards}, \citenamefont {Edwards}, \citenamefont {Effler}, \citenamefont {Eggenstein}, \citenamefont {Ehrens}, \citenamefont {Eichholz}, \citenamefont {Eikenberry}, \citenamefont {Essick}, \citenamefont {Etzel}, \citenamefont {Evans}, \citenamefont {Evans}, \citenamefont {Factourovich}, \citenamefont {Fairhurst}, \citenamefont {Fan}, \citenamefont {Fang}, \citenamefont {Farr}, \citenamefont {Farr}, \citenamefont {Favata}, \citenamefont {Fays}, \citenamefont {Fehrmann}, \citenamefont {Fejer}, \citenamefont {Feldbaum}, \citenamefont {Ferreira}, \citenamefont {Fisher}, \citenamefont {Frei}, \citenamefont {Freise}, \citenamefont {Frey}, \citenamefont {Fricke}, \citenamefont {Fritschel}, \citenamefont {Frolov}, \citenamefont {{Fuentes-Tapia}}, \citenamefont {Fulda}, \citenamefont {Fyffe}, \citenamefont {Gair}, \citenamefont {Gaonkar}, \citenamefont {Gehrels}, \citenamefont {Gergely{\textasciiacute}}, \citenamefont {Giaime}, \citenamefont {Giardina}, \citenamefont
  {Gleason}, \citenamefont {Goetz}, \citenamefont {Goetz}, \citenamefont {Gondan}, \citenamefont {Gonz{\'a}lez}, \citenamefont {Gordon}, \citenamefont {Gorodetsky}, \citenamefont {Gossan}, \citenamefont {Go{\ss}ler}, \citenamefont {Gr{\"a}f}, \citenamefont {Graff}, \citenamefont {Grant}, \citenamefont {Gras}, \citenamefont {Gray}, \citenamefont {Greenhalgh}, \citenamefont {Gretarsson}, \citenamefont {Grote}, \citenamefont {Grunewald}, \citenamefont {Guido}, \citenamefont {Guo}, \citenamefont {Gushwa}, \citenamefont {Gustafson}, \citenamefont {Gustafson}, \citenamefont {Hacker}, \citenamefont {Hall}, \citenamefont {Hammond}, \citenamefont {Hanke}, \citenamefont {Hanks}, \citenamefont {Hanna}, \citenamefont {Hannam}, \citenamefont {Hanson}, \citenamefont {Hardwick}, \citenamefont {Harry}, \citenamefont {Harry}, \citenamefont {Hart}, \citenamefont {Hartman}, \citenamefont {Haster}, \citenamefont {Haughian}, \citenamefont {Hee}, \citenamefont {Heintze}, \citenamefont {Heinzel}, \citenamefont {Hendry},
  \citenamefont {Heng}, \citenamefont {Heptonstall}, \citenamefont {Heurs}, \citenamefont {Hewitson}, \citenamefont {Hild}, \citenamefont {Hoak}, \citenamefont {Hodge}, \citenamefont {Hollitt}, \citenamefont {Holt}, \citenamefont {Hopkins}, \citenamefont {Hosken}, \citenamefont {Hough}, \citenamefont {Houston}, \citenamefont {Howell}, \citenamefont {Hu}, \citenamefont {Huerta}, \citenamefont {Hughey}, \citenamefont {Husa}, \citenamefont {Huttner}, \citenamefont {Huynh}, \citenamefont {{Huynh-Dinh}}, \citenamefont {Idrisy}, \citenamefont {Indik}, \citenamefont {Ingram}, \citenamefont {Inta}, \citenamefont {Islas}, \citenamefont {Isler}, \citenamefont {Isogai}, \citenamefont {Iyer}, \citenamefont {Izumi}, \citenamefont {Jacobson}, \citenamefont {Jang}, \citenamefont {Jawahar}, \citenamefont {Ji}, \citenamefont {{Jim{\'e}nez-Forteza}}, \citenamefont {Johnson}, \citenamefont {Jones}, \citenamefont {Jones}, \citenamefont {Ju}, \citenamefont {Haris}, \citenamefont {Kalogera}, \citenamefont {Kandhasamy},
  \citenamefont {Kang}, \citenamefont {Kanner}, \citenamefont {Katsavounidis}, \citenamefont {Katzman}, \citenamefont {Kaufer}, \citenamefont {Kaufer}, \citenamefont {Kaur}, \citenamefont {Kawabe}, \citenamefont {Kawazoe}, \citenamefont {Keiser}, \citenamefont {Keitel}, \citenamefont {Kelley}, \citenamefont {Kells}, \citenamefont {Keppel}, \citenamefont {Key}, \citenamefont {Khalaidovski}, \citenamefont {Khalili}, \citenamefont {Khazanov}, \citenamefont {Kim}, \citenamefont {Kim}, \citenamefont {Kim}, \citenamefont {Kim}, \citenamefont {Kim}, \citenamefont {King}, \citenamefont {King}, \citenamefont {Kinzel}, \citenamefont {Kissel}, \citenamefont {Klimenko}, \citenamefont {Kline}, \citenamefont {Koehlenbeck}, \citenamefont {Kokeyama}, \citenamefont {Kondrashov}, \citenamefont {Korobko}, \citenamefont {Korth}, \citenamefont {Kozak}, \citenamefont {Kringel}, \citenamefont {Krishnan}, \citenamefont {Krueger}, \citenamefont {Kuehn}, \citenamefont {Kumar}, \citenamefont {Kumar}, \citenamefont {Kuo}, \citenamefont
  {Landry}, \citenamefont {Lantz}, \citenamefont {Larson}, \citenamefont {Lasky}, \citenamefont {Lazzarini}, \citenamefont {Lazzaro}, \citenamefont {Le}, \citenamefont {Leaci}, \citenamefont {Leavey}, \citenamefont {Lebigot}, \citenamefont {Lee}, \citenamefont {Lee}, \citenamefont {Lee}, \citenamefont {Leong}, \citenamefont {Levin}, \citenamefont {Levine}, \citenamefont {Lewis}, \citenamefont {Li}, \citenamefont {Libbrecht}, \citenamefont {Libson}, \citenamefont {Lin}, \citenamefont {Littenberg}, \citenamefont {Lockerbie}, \citenamefont {Lockett}, \citenamefont {Logue}, \citenamefont {Lombardi}, \citenamefont {Lormand}, \citenamefont {Lough}, \citenamefont {Lubinski}, \citenamefont {L{\"u}ck}, \citenamefont {Lundgren}, \citenamefont {Lynch}, \citenamefont {Ma}, \citenamefont {Macarthur}, \citenamefont {MacDonald}, \citenamefont {Machenschalk}, \citenamefont {MacInnis}, \citenamefont {Macleod}, \citenamefont {{Maga{\~n}a-Sandoval}}, \citenamefont {Magee}, \citenamefont {Mageswaran}, \citenamefont {Maglione},
  \citenamefont {Mailand}, \citenamefont {Mandel}, \citenamefont {Mandic}, \citenamefont {Mangano}, \citenamefont {Mansell}, \citenamefont {M{\'a}rka}, \citenamefont {M{\'a}rka}, \citenamefont {Markosyan}, \citenamefont {Maros}, \citenamefont {Martin}, \citenamefont {Martin}, \citenamefont {Martynov}, \citenamefont {Marx}, \citenamefont {Mason}, \citenamefont {Massinger}, \citenamefont {Matichard}, \citenamefont {Matone}, \citenamefont {Mavalvala}, \citenamefont {Mazumder}, \citenamefont {Mazzolo}, \citenamefont {McCarthy}, \citenamefont {McClelland}, \citenamefont {McCormick}, \citenamefont {McGuire}, \citenamefont {McIntyre}, \citenamefont {McIver}, \citenamefont {McLin}, \citenamefont {McWilliams}, \citenamefont {Meadors}, \citenamefont {Meinders}, \citenamefont {Melatos}, \citenamefont {Mendell}, \citenamefont {Mercer}, \citenamefont {Meshkov}, \citenamefont {Messenger}, \citenamefont {Meyers}, \citenamefont {Miao}, \citenamefont {Middleton}, \citenamefont {Mikhailov}, \citenamefont {Miller},
  \citenamefont {Miller}, \citenamefont {Millhouse}, \citenamefont {Ming}, \citenamefont {Mirshekari}, \citenamefont {Mishra}, \citenamefont {Mitra}, \citenamefont {Mitrofanov}, \citenamefont {Mitselmakher}, \citenamefont {Mittleman}, \citenamefont {Moe}, \citenamefont {Mohanty}, \citenamefont {Mohapatra}, \citenamefont {Moore}, \citenamefont {Moraru}, \citenamefont {Moreno}, \citenamefont {Morriss}, \citenamefont {Mossavi}, \citenamefont {{Mow-Lowry}}, \citenamefont {Mueller}, \citenamefont {Mueller}, \citenamefont {Mukherjee}, \citenamefont {Mullavey}, \citenamefont {Munch}, \citenamefont {Murphy}, \citenamefont {Murray}, \citenamefont {Mytidis}, \citenamefont {Nash}, \citenamefont {Nayak}, \citenamefont {Necula}, \citenamefont {Nedkova}, \citenamefont {Newton}, \citenamefont {Nguyen}, \citenamefont {Nielsen}, \citenamefont {Nissanke}, \citenamefont {Nitz}, \citenamefont {Nolting}, \citenamefont {Normandin}, \citenamefont {Nuttall}, \citenamefont {Ochsner}, \citenamefont {O'Dell}, \citenamefont {Oelker},
  \citenamefont {Ogin}, \citenamefont {Oh}, \citenamefont {Oh}, \citenamefont {Ohme}, \citenamefont {Oppermann}, \citenamefont {Oram}, \citenamefont {O'Reilly}, \citenamefont {Ortega}, \citenamefont {O'Shaughnessy}, \citenamefont {Osthelder}, \citenamefont {Ott}, \citenamefont {Ottaway}, \citenamefont {Ottens}, \citenamefont {Overmier}, \citenamefont {Owen}, \citenamefont {Padilla}, \citenamefont {Pai}, \citenamefont {Pai}, \citenamefont {Palashov}, \citenamefont {{Pal-Singh}}, \citenamefont {Pan}, \citenamefont {Pankow}, \citenamefont {Pannarale}, \citenamefont {Pant}, \citenamefont {Papa}, \citenamefont {Paris}, \citenamefont {Patrick}, \citenamefont {Pedraza}, \citenamefont {Pekowsky}, \citenamefont {Pele}, \citenamefont {Penn}, \citenamefont {Perreca}, \citenamefont {Phelps}, \citenamefont {Pierro}, \citenamefont {Pinto}, \citenamefont {Pitkin}, \citenamefont {Poeld}, \citenamefont {Post}, \citenamefont {Poteomkin}, \citenamefont {Powell}, \citenamefont {Prasad}, \citenamefont {Predoi}, \citenamefont
  {Premachandra}, \citenamefont {Prestegard}, \citenamefont {Price}, \citenamefont {Principe}, \citenamefont {Privitera}, \citenamefont {Prix}, \citenamefont {Prokhorov}, \citenamefont {Puncken}, \citenamefont {P{\"u}rrer}, \citenamefont {Qin}, \citenamefont {Quetschke}, \citenamefont {Quintero}, \citenamefont {Quiroga}, \citenamefont {{Quitzow-James}}, \citenamefont {Raab}, \citenamefont {Rabeling}, \citenamefont {Radkins}, \citenamefont {Raffai}, \citenamefont {Raja}, \citenamefont {Rajalakshmi}, \citenamefont {Rakhmanov}, \citenamefont {Ramirez}, \citenamefont {Raymond}, \citenamefont {Reed}, \citenamefont {Reid}, \citenamefont {Reitze}, \citenamefont {Reula}, \citenamefont {Riles}, \citenamefont {Robertson}, \citenamefont {Robie}, \citenamefont {Rollins}, \citenamefont {Roma}, \citenamefont {Romano}, \citenamefont {Romanov}, \citenamefont {Romie}, \citenamefont {Rowan}, \citenamefont {R{\"u}diger}, \citenamefont {Ryan}, \citenamefont {Sachdev}, \citenamefont {Sadecki}, \citenamefont {Sadeghian},
  \citenamefont {Saleem}, \citenamefont {Salemi}, \citenamefont {Sammut}, \citenamefont {Sandberg}, \citenamefont {Sanders}, \citenamefont {Sannibale}, \citenamefont {{Santiago-Prieto}}, \citenamefont {Sathyaprakash}, \citenamefont {Saulson}, \citenamefont {Savage}, \citenamefont {Sawadsky}, \citenamefont {Scheuer}, \citenamefont {Schilling}, \citenamefont {Schmidt}, \citenamefont {Schnabel}, \citenamefont {Schofield}, \citenamefont {Schreiber}, \citenamefont {Schuette}, \citenamefont {Schutz}, \citenamefont {Scott}, \citenamefont {Scott}, \citenamefont {Sellers}, \citenamefont {Sengupta}, \citenamefont {Sergeev}, \citenamefont {Serna}, \citenamefont {Sevigny}, \citenamefont {Shaddock}, \citenamefont {Shahriar}, \citenamefont {Shaltev}, \citenamefont {Shao}, \citenamefont {Shapiro}, \citenamefont {Shawhan}, \citenamefont {Shoemaker}, \citenamefont {Sidery}, \citenamefont {Siemens}, \citenamefont {Sigg}, \citenamefont {Silva}, \citenamefont {Simakov}, \citenamefont {Singer}, \citenamefont {Singer},
  \citenamefont {Singh}, \citenamefont {Sintes}, \citenamefont {Slagmolen}, \citenamefont {Smith}, \citenamefont {Smith}, \citenamefont {Smith}, \citenamefont {{Smith-Lefebvre}}, \citenamefont {Son}, \citenamefont {Sorazu}, \citenamefont {Souradeep}, \citenamefont {Staley}, \citenamefont {Stebbins}, \citenamefont {Steinke}, \citenamefont {Steinlechner}, \citenamefont {Steinlechner}, \citenamefont {Steinmeyer}, \citenamefont {Stephens}, \citenamefont {Steplewski}, \citenamefont {Stevenson}, \citenamefont {Stone}, \citenamefont {Strain}, \citenamefont {Strigin}, \citenamefont {Sturani}, \citenamefont {Stuver}, \citenamefont {Summerscales}, \citenamefont {Sutton}, \citenamefont {Szczepanczyk}, \citenamefont {Szeifert}, \citenamefont {Talukder}, \citenamefont {Tanner}, \citenamefont {T{\'a}pai}, \citenamefont {Tarabrin}, \citenamefont {Taracchini}, \citenamefont {Taylor}, \citenamefont {Tellez}, \citenamefont {Theeg}, \citenamefont {Thirugnanasambandam}, \citenamefont {Thomas}, \citenamefont {Thomas},
  \citenamefont {Thorne}, \citenamefont {Thorne}, \citenamefont {Thrane}, \citenamefont {Tiwari}, \citenamefont {Tomlinson}, \citenamefont {Torres}, \citenamefont {Torrie}, \citenamefont {Traylor}, \citenamefont {Tse}, \citenamefont {Tshilumba}, \citenamefont {Ugolini}, \citenamefont {Unnikrishnan}, \citenamefont {Urban}, \citenamefont {Usman}, \citenamefont {Vahlbruch}, \citenamefont {Vajente}, \citenamefont {Valdes}, \citenamefont {Vallisneri}, \citenamefont {van Veggel}, \citenamefont {Vass}, \citenamefont {Vaulin}, \citenamefont {Vecchio}, \citenamefont {Veitch}, \citenamefont {Veitch}, \citenamefont {Venkateswara}, \citenamefont {{Vincent-Finley}}, \citenamefont {Vitale}, \citenamefont {Vo}, \citenamefont {Vorvick}, \citenamefont {Vousden}, \citenamefont {Vyatchanin}, \citenamefont {Wade}, \citenamefont {Wade}, \citenamefont {Wade}, \citenamefont {Walker}, \citenamefont {Wallace}, \citenamefont {Walsh}, \citenamefont {Wang}, \citenamefont {Wang}, \citenamefont {Wang}, \citenamefont {Ward}, \citenamefont
  {Warner}, \citenamefont {Was}, \citenamefont {Weaver}, \citenamefont {Weinert}, \citenamefont {Weinstein}, \citenamefont {Weiss}, \citenamefont {Welborn}, \citenamefont {Wen}, \citenamefont {Wessels}, \citenamefont {Westphal}, \citenamefont {Wette}, \citenamefont {Whelan}, \citenamefont {Whitcomb}, \citenamefont {White}, \citenamefont {Whiting}, \citenamefont {Wilkinson}, \citenamefont {Williams}, \citenamefont {Williams}, \citenamefont {Williamson}, \citenamefont {Willis}, \citenamefont {Willke}, \citenamefont {Wimmer}, \citenamefont {Winkler}, \citenamefont {Wipf}, \citenamefont {Wittel}, \citenamefont {Woan}, \citenamefont {Worden}, \citenamefont {Xie}, \citenamefont {Yablon}, \citenamefont {Yakushin}, \citenamefont {Yam}, \citenamefont {Yamamoto}, \citenamefont {Yancey}, \citenamefont {Yang}, \citenamefont {Zanolin}, \citenamefont {Zhang}, \citenamefont {Zhang}, \citenamefont {Zhang}, \citenamefont {Zhang}, \citenamefont {Zhao}, \citenamefont {Zhou}, \citenamefont {Zhu}, \citenamefont {Zucker},
  \citenamefont {Zuraw},\ and\ \citenamefont {Zweizig}}]{AasiCQG15AdvancedLIGO}%
  \BibitemOpen
  \bibfield  {author} {\bibinfo {author} {\bibfnamefont {J.}~\bibnamefont {Aasi}}, \bibinfo {author} {\bibfnamefont {B.~P.}\ \bibnamefont {Abbott}}, \bibinfo {author} {\bibfnamefont {R.}~\bibnamefont {Abbott}}, \bibnamefont {et~al.},\ }\bibfield  {title} {\bibinfo {title} {Advanced {{LIGO}}},\ }\href {https://doi.org/10.1088/0264-9381/32/7/074001} {\bibfield  {journal} {\bibinfo  {journal} {Class. Quantum Grav.}\ }\textbf {\bibinfo {volume} {32}},\ \bibinfo {pages} {074001} (\bibinfo {year} {2015})}\BibitemShut {NoStop}%
\bibitem [{\citenamefont {Romano}\ and\ \citenamefont {Cornish}(2017)}]{RomanoLRR17DetectionMethods}%
  \BibitemOpen
  \bibfield  {author} {\bibinfo {author} {\bibfnamefont {J.~D.}\ \bibnamefont {Romano}}\ \bibnamefont {and}\ \bibinfo {author} {\bibfnamefont {{\relax Neil}.~J.}\ \bibnamefont {Cornish}},\ }\bibfield  {title} {\bibinfo {title} {Detection methods for stochastic gravitational-wave backgrounds: A unified treatment},\ }\href {https://doi.org/10.1007/s41114-017-0004-1} {\bibfield  {journal} {\bibinfo  {journal} {Living. Rev. Relativ.}\ }\textbf {\bibinfo {volume} {20}},\ \bibinfo {pages} {2} (\bibinfo {year} {2017})}\BibitemShut {NoStop}%
\bibitem [{\citenamefont {Renzini}\ \emph {et~al.}(2022)\citenamefont {Renzini}, \citenamefont {Goncharov}, \citenamefont {Jenkins},\ and\ \citenamefont {Meyers}}]{renzini2022stochastic}%
  \BibitemOpen
  \bibfield  {author} {\bibinfo {author} {\bibfnamefont {A.~I.}\ \bibnamefont {Renzini}}, \bibinfo {author} {\bibfnamefont {B.}~\bibnamefont {Goncharov}}, \bibinfo {author} {\bibfnamefont {A.~C.}\ \bibnamefont {Jenkins}},\ \bibnamefont {and}\ \bibinfo {author} {\bibfnamefont {P.~M.}\ \bibnamefont {Meyers}},\ }\bibfield  {title} {\bibinfo {title} {Stochastic gravitational-wave backgrounds: Current detection efforts and future prospects},\ }\href {https://doi.org/10.3390/galaxies10010034} {\bibfield  {journal} {\bibinfo  {journal} {Galaxies}\ }\textbf {\bibinfo {volume} {10}},\ \bibinfo {pages} {34} (\bibinfo {year} {2022})}\BibitemShut {NoStop}%
\bibitem [{\citenamefont {Verlinde}\ and\ \citenamefont {Zurek}(2021)}]{VerlindePLB21ObservationalSignatures}%
  \BibitemOpen
  \bibfield  {author} {\bibinfo {author} {\bibfnamefont {E.~P.}\ \bibnamefont {Verlinde}}\ \bibnamefont {and}\ \bibinfo {author} {\bibfnamefont {K.~M.}\ \bibnamefont {Zurek}},\ }\bibfield  {title} {\bibinfo {title} {Observational {{Signatures}} of {{Quantum Gravity}} in {{Interferometers}}},\ }\href {https://doi.org/10.1016/j.physletb.2021.136663} {\bibfield  {journal} {\bibinfo  {journal} {Phys. Lett. B}\ }\textbf {\bibinfo {volume} {822}},\ \bibinfo {pages} {136663} (\bibinfo {year} {2021})}\BibitemShut {NoStop}%
\bibitem [{\citenamefont {{L. McCuller, Single-{{Photon Signal Sideband Detection}} for {{High-Power Michelson Interferometers}} (2022)}}()}]{McCuller22SinglePhotonSignal}%
  \BibitemOpen
  \bibfield  {author} {\bibinfo {author} {\bibnamefont {{L. McCuller, Single-{{Photon Signal Sideband Detection}} for {{High-Power Michelson Interferometers}} (2022)}}},\ }\href {https://doi.org/10.48550/arXiv.2211.04016} {\bibinfo {title} {arxiv:2211.04016 [hep-ex, physics:physics, physics:quant-ph]}}\BibitemShut {NoStop}%
\bibitem [{\citenamefont {{S. M. Vermeulen, T. Cullen, D. Grass, I. A. O. MacMillan, A. J. Ramirez, J. Wack, B. Ko- rzh, V. S. H. Lee, K. M. Zurek, C. Stoughton, and L. McCuller, Photon Counting Interferome- try to Detect Geontropic Space-Time Fluctuations with GQuEST (2024)}}()}]{Vermeulen24PhotonCounting}%
  \BibitemOpen
  \bibfield  {author} {\bibinfo {author} {\bibnamefont {{S. M. Vermeulen, T. Cullen, D. Grass, I. A. O. MacMillan, A. J. Ramirez, J. Wack, B. Ko- rzh, V. S. H. Lee, K. M. Zurek, C. Stoughton, and L. McCuller, Photon Counting Interferome- try to Detect Geontropic Space-Time Fluctuations with GQuEST (2024)}}},\ }\href {https://doi.org/10.48550/arXiv.2404.07524} {\bibinfo {title} {arxiv:2404.07524 [astro-ph, physics:gr-qc, physics:physics, physics:quant-ph]}}\BibitemShut {NoStop}%
\bibitem [{\citenamefont {Rosenberg}\ and\ \citenamefont {Van~Bibber}(2000)}]{rosenberg2000searches}%
  \BibitemOpen
  \bibfield  {author} {\bibinfo {author} {\bibfnamefont {L.~J.}\ \bibnamefont {Rosenberg}}\ \bibnamefont {and}\ \bibinfo {author} {\bibfnamefont {K.~A.}\ \bibnamefont {Van~Bibber}},\ }\bibfield  {title} {\bibinfo {title} {Searches for invisible axions},\ }\href {https://doi.org/10.1016/S0370-1573(99)00045-9} {\bibfield  {journal} {\bibinfo  {journal} {Phys. Rep.}\ }\textbf {\bibinfo {volume} {325}},\ \bibinfo {pages} {1} (\bibinfo {year} {2000})}\BibitemShut {NoStop}%
\bibitem [{\citenamefont {Graham}\ \emph {et~al.}(2015)\citenamefont {Graham}, \citenamefont {Irastorza}, \citenamefont {Lamoreaux}, \citenamefont {Lindner},\ and\ \citenamefont {van Bibber}}]{graham2015experimental}%
  \BibitemOpen
  \bibfield  {author} {\bibinfo {author} {\bibfnamefont {P.~W.}\ \bibnamefont {Graham}}, \bibinfo {author} {\bibfnamefont {I.~G.}\ \bibnamefont {Irastorza}}, \bibinfo {author} {\bibfnamefont {S.~K.}\ \bibnamefont {Lamoreaux}}, \bibinfo {author} {\bibfnamefont {A.}~\bibnamefont {Lindner}},\ \bibnamefont {and}\ \bibinfo {author} {\bibfnamefont {K.~A.}\ \bibnamefont {van Bibber}},\ }\bibfield  {title} {\bibinfo {title} {Experimental searches for the axion and axion-like particles},\ }\href {https://doi.org/10.1146/annurev-nucl-102014-022120} {\bibfield  {journal} {\bibinfo  {journal} {Annu. Rev. Nucl. Part. Sci.}\ }\textbf {\bibinfo {volume} {65}},\ \bibinfo {pages} {485} (\bibinfo {year} {2015})}\BibitemShut {NoStop}%
\bibitem [{\citenamefont {Agrawal}\ \emph {et~al.}(2024)\citenamefont {Agrawal}, \citenamefont {Dixit}, \citenamefont {Roy}, \citenamefont {Chakram}, \citenamefont {He}, \citenamefont {Naik}, \citenamefont {Schuster},\ and\ \citenamefont {Chou}}]{AgrawalPRL24StimulatedEmission}%
  \BibitemOpen
  \bibfield  {author} {\bibinfo {author} {\bibfnamefont {A.}~\bibnamefont {Agrawal}}, \bibinfo {author} {\bibfnamefont {A.~V.}\ \bibnamefont {Dixit}}, \bibinfo {author} {\bibfnamefont {T.}~\bibnamefont {Roy}}, \bibinfo {author} {\bibfnamefont {S.}~\bibnamefont {Chakram}}, \bibinfo {author} {\bibfnamefont {K.}~\bibnamefont {He}}, \bibinfo {author} {\bibfnamefont {R.~K.}\ \bibnamefont {Naik}}, \bibinfo {author} {\bibfnamefont {D.~I.}\ \bibnamefont {Schuster}},\ \bibnamefont {and}\ \bibinfo {author} {\bibfnamefont {A.}~\bibnamefont {Chou}},\ }\bibfield  {title} {\bibinfo {title} {Stimulated {{Emission}} of {{Signal Photons}} from {{Dark Matter Waves}}},\ }\href {https://doi.org/10.1103/PhysRevLett.132.140801} {\bibfield  {journal} {\bibinfo  {journal} {Phys. Rev. Lett.}\ }\textbf {\bibinfo {volume} {132}},\ \bibinfo {pages} {140801} (\bibinfo {year} {2024})}\BibitemShut {NoStop}%
\bibitem [{\citenamefont {Shi}\ and\ \citenamefont {Zhuang}(2023)}]{ShinQI23UltimatePrecision}%
  \BibitemOpen
  \bibfield  {author} {\bibinfo {author} {\bibfnamefont {H.}~\bibnamefont {Shi}}\ \bibnamefont {and}\ \bibinfo {author} {\bibfnamefont {Q.}~\bibnamefont {Zhuang}},\ }\bibfield  {title} {\bibinfo {title} {Ultimate precision limit of noise sensing and dark matter search},\ }\href {https://doi.org/10.1038/s41534-023-00693-w} {\bibfield  {journal} {\bibinfo  {journal} {npj Quantum Inf}\ }\textbf {\bibinfo {volume} {9}},\ \bibinfo {pages} {1} (\bibinfo {year} {2023})}\BibitemShut {NoStop}%
\bibitem [{\citenamefont {Mouradian}\ \emph {et~al.}(2021)\citenamefont {Mouradian}, \citenamefont {Glikin}, \citenamefont {Megidish}, \citenamefont {Ellers},\ and\ \citenamefont {Haeffner}}]{mouradian2021quantum}%
  \BibitemOpen
  \bibfield  {author} {\bibinfo {author} {\bibfnamefont {S.~L.}\ \bibnamefont {Mouradian}}, \bibinfo {author} {\bibfnamefont {N.}~\bibnamefont {Glikin}}, \bibinfo {author} {\bibfnamefont {E.}~\bibnamefont {Megidish}}, \bibinfo {author} {\bibfnamefont {K.-I.}\ \bibnamefont {Ellers}},\ \bibnamefont {and}\ \bibinfo {author} {\bibfnamefont {H.}~\bibnamefont {Haeffner}},\ }\bibfield  {title} {\bibinfo {title} {Quantum sensing of intermittent stochastic signals},\ }\href {https://doi.org/10.1103/PhysRevA.103.032419} {\bibfield  {journal} {\bibinfo  {journal} {Phys. Rev. A}\ }\textbf {\bibinfo {volume} {103}},\ \bibinfo {pages} {032419} (\bibinfo {year} {2021})}\BibitemShut {NoStop}%
\bibitem [{\citenamefont {Rotem}\ \emph {et~al.}(2019)\citenamefont {Rotem}, \citenamefont {Gefen}, \citenamefont {Oviedo-Casado}, \citenamefont {Prior}, \citenamefont {Schmitt}, \citenamefont {Burak}, \citenamefont {McGuiness}, \citenamefont {Jelezko},\ and\ \citenamefont {Retzker}}]{rotem2019limits}%
  \BibitemOpen
  \bibfield  {author} {\bibinfo {author} {\bibfnamefont {A.}~\bibnamefont {Rotem}}, \bibinfo {author} {\bibfnamefont {T.}~\bibnamefont {Gefen}}, \bibinfo {author} {\bibfnamefont {S.}~\bibnamefont {Oviedo-Casado}}, \bibinfo {author} {\bibfnamefont {J.}~\bibnamefont {Prior}}, \bibinfo {author} {\bibfnamefont {S.}~\bibnamefont {Schmitt}}, \bibinfo {author} {\bibfnamefont {Y.}~\bibnamefont {Burak}}, \bibinfo {author} {\bibfnamefont {L.}~\bibnamefont {McGuiness}}, \bibinfo {author} {\bibfnamefont {F.}~\bibnamefont {Jelezko}},\ \bibnamefont {and}\ \bibinfo {author} {\bibfnamefont {A.}~\bibnamefont {Retzker}},\ }\bibfield  {title} {\bibinfo {title} {Limits on spectral resolution measurements by quantum probes},\ }\href@noop {} {\bibfield  {journal} {\bibinfo  {journal} {Physical review letters}\ }\textbf {\bibinfo {volume} {122}},\ \bibinfo {pages} {060503} (\bibinfo {year} {2019})}\BibitemShut {NoStop}%
\bibitem [{\citenamefont {Gefen}\ \emph {et~al.}(2019)\citenamefont {Gefen}, \citenamefont {Rotem},\ and\ \citenamefont {Retzker}}]{gefen2019overcoming}%
  \BibitemOpen
  \bibfield  {author} {\bibinfo {author} {\bibfnamefont {T.}~\bibnamefont {Gefen}}, \bibinfo {author} {\bibfnamefont {A.}~\bibnamefont {Rotem}},\ \bibnamefont {and}\ \bibinfo {author} {\bibfnamefont {A.}~\bibnamefont {Retzker}},\ }\bibfield  {title} {\bibinfo {title} {Overcoming resolution limits with quantum sensing},\ }\href {https://doi.org/10.1038/s41467-019-12817-y} {\bibfield  {journal} {\bibinfo  {journal} {Nat. Comm.}\ }\textbf {\bibinfo {volume} {10}},\ \bibinfo {pages} {4992} (\bibinfo {year} {2019})}\BibitemShut {NoStop}%
\bibitem [{\citenamefont {Cohen}\ \emph {et~al.}(2020)\citenamefont {Cohen}, \citenamefont {Gefen}, \citenamefont {Ortiz},\ and\ \citenamefont {Retzker}}]{cohen2020achieving}%
  \BibitemOpen
  \bibfield  {author} {\bibinfo {author} {\bibfnamefont {D.}~\bibnamefont {Cohen}}, \bibinfo {author} {\bibfnamefont {T.}~\bibnamefont {Gefen}}, \bibinfo {author} {\bibfnamefont {L.}~\bibnamefont {Ortiz}},\ \bibnamefont {and}\ \bibinfo {author} {\bibfnamefont {A.}~\bibnamefont {Retzker}},\ }\bibfield  {title} {\bibinfo {title} {Achieving the ultimate precision limit with a weakly interacting quantum probe},\ }\href@noop {} {\bibfield  {journal} {\bibinfo  {journal} {npj Quantum Information}\ }\textbf {\bibinfo {volume} {6}},\ \bibinfo {pages} {83} (\bibinfo {year} {2020})}\BibitemShut {NoStop}%
\bibitem [{\citenamefont {Norris}\ \emph {et~al.}(2016)\citenamefont {Norris}, \citenamefont {Paz-Silva},\ and\ \citenamefont {Viola}}]{norris2016qubit}%
  \BibitemOpen
  \bibfield  {author} {\bibinfo {author} {\bibfnamefont {L.~M.}\ \bibnamefont {Norris}}, \bibinfo {author} {\bibfnamefont {G.~A.}\ \bibnamefont {Paz-Silva}},\ \bibnamefont {and}\ \bibinfo {author} {\bibfnamefont {L.}~\bibnamefont {Viola}},\ }\bibfield  {title} {\bibinfo {title} {Qubit noise spectroscopy for non-gaussian dephasing environments},\ }\href@noop {} {\bibfield  {journal} {\bibinfo  {journal} {Physical review letters}\ }\textbf {\bibinfo {volume} {116}},\ \bibinfo {pages} {150503} (\bibinfo {year} {2016})}\BibitemShut {NoStop}%
\bibitem [{\citenamefont {Shaw}\ \emph {et~al.}(2024)\citenamefont {Shaw}, \citenamefont {Finkelstein}, \citenamefont {Tsai}, \citenamefont {Scholl}, \citenamefont {Yoon}, \citenamefont {Choi},\ and\ \citenamefont {Endres}}]{shaw2024multi}%
  \BibitemOpen
  \bibfield  {author} {\bibinfo {author} {\bibfnamefont {A.~L.}\ \bibnamefont {Shaw}}, \bibinfo {author} {\bibfnamefont {R.}~\bibnamefont {Finkelstein}}, \bibinfo {author} {\bibfnamefont {R.~B.-S.}\ \bibnamefont {Tsai}}, \bibinfo {author} {\bibfnamefont {P.}~\bibnamefont {Scholl}}, \bibinfo {author} {\bibfnamefont {T.~H.}\ \bibnamefont {Yoon}}, \bibinfo {author} {\bibfnamefont {J.}~\bibnamefont {Choi}},\ \bibnamefont {and}\ \bibinfo {author} {\bibfnamefont {M.}~\bibnamefont {Endres}},\ }\bibfield  {title} {\bibinfo {title} {Multi-ensemble metrology by programming local rotations with atom movements},\ }\href@noop {} {\bibfield  {journal} {\bibinfo  {journal} {Nature Physics}\ }\textbf {\bibinfo {volume} {20}},\ \bibinfo {pages} {195} (\bibinfo {year} {2024})}\BibitemShut {NoStop}%
\bibitem [{\citenamefont {Stilck~Fran{\c{c}}a}\ \emph {et~al.}(2024)\citenamefont {Stilck~Fran{\c{c}}a}, \citenamefont {Markovich}, \citenamefont {Dobrovitski}, \citenamefont {Werner},\ and\ \citenamefont {Borregaard}}]{stilck2024efficient}%
  \BibitemOpen
  \bibfield  {author} {\bibinfo {author} {\bibfnamefont {D.}~\bibnamefont {Stilck~Fran{\c{c}}a}}, \bibinfo {author} {\bibfnamefont {L.~A.}\ \bibnamefont {Markovich}}, \bibinfo {author} {\bibfnamefont {V.}~\bibnamefont {Dobrovitski}}, \bibinfo {author} {\bibfnamefont {A.~H.}\ \bibnamefont {Werner}},\ \bibnamefont {and}\ \bibinfo {author} {\bibfnamefont {J.}~\bibnamefont {Borregaard}},\ }\bibfield  {title} {\bibinfo {title} {Efficient and robust estimation of many-qubit hamiltonians},\ }\href@noop {} {\bibfield  {journal} {\bibinfo  {journal} {Nature Communications}\ }\textbf {\bibinfo {volume} {15}},\ \bibinfo {pages} {311} (\bibinfo {year} {2024})}\BibitemShut {NoStop}%
\bibitem [{\citenamefont {Arshad}\ \emph {et~al.}(2024)\citenamefont {Arshad}, \citenamefont {Bekker}, \citenamefont {Haylock}, \citenamefont {Skrzypczak}, \citenamefont {White}, \citenamefont {Griffiths}, \citenamefont {Gore}, \citenamefont {Morley}, \citenamefont {Salter}, \citenamefont {Smith} \emph {et~al.}}]{arshad2024real}%
  \BibitemOpen
  \bibfield  {author} {\bibinfo {author} {\bibfnamefont {M.~J.}\ \bibnamefont {Arshad}}, \bibinfo {author} {\bibfnamefont {C.}~\bibnamefont {Bekker}}, \bibinfo {author} {\bibfnamefont {B.}~\bibnamefont {Haylock}}, \bibinfo {author} {\bibfnamefont {K.}~\bibnamefont {Skrzypczak}}, \bibinfo {author} {\bibfnamefont {D.}~\bibnamefont {White}}, \bibinfo {author} {\bibfnamefont {B.}~\bibnamefont {Griffiths}}, \bibinfo {author} {\bibfnamefont {J.}~\bibnamefont {Gore}}, \bibinfo {author} {\bibfnamefont {G.~W.}\ \bibnamefont {Morley}}, \bibinfo {author} {\bibfnamefont {P.}~\bibnamefont {Salter}}, \bibinfo {author} {\bibfnamefont {J.}~\bibnamefont {Smith}}, \bibnamefont {et~al.},\ }\bibfield  {title} {\bibinfo {title} {Real-time adaptive estimation of decoherence timescales for a single qubit},\ }\href@noop {} {\bibfield  {journal} {\bibinfo  {journal} {Physical Review Applied}\ }\textbf {\bibinfo {volume} {21}},\ \bibinfo {pages} {024026} (\bibinfo {year} {2024})}\BibitemShut {NoStop}%
\bibitem [{\citenamefont {Tsang}(2023)}]{PhysRevA.107.012611}%
  \BibitemOpen
  \bibfield  {author} {\bibinfo {author} {\bibfnamefont {M.}~\bibnamefont {Tsang}},\ }\bibfield  {title} {\bibinfo {title} {Quantum noise spectroscopy as an incoherent imaging problem},\ }\href {https://doi.org/10.1103/PhysRevA.107.012611} {\bibfield  {journal} {\bibinfo  {journal} {Phys. Rev. A}\ }\textbf {\bibinfo {volume} {107}},\ \bibinfo {pages} {012611} (\bibinfo {year} {2023})}\BibitemShut {NoStop}%
\bibitem [{\citenamefont {Alicki}\ and\ \citenamefont {Lendi}(2007)}]{alicki2007quantum}%
  \BibitemOpen
  \bibfield  {author} {\bibinfo {author} {\bibfnamefont {R.}~\bibnamefont {Alicki}}\ \bibnamefont {and}\ \bibinfo {author} {\bibfnamefont {K.}~\bibnamefont {Lendi}},\ }\href@noop {} {\emph {\bibinfo {title} {Quantum dynamical semigroups and applications}}},\ Vol.\ \bibinfo {volume} {717}\ (\bibinfo  {publisher} {Springer Science \& Business Media},\ \bibinfo {year} {2007})\BibitemShut {NoStop}%
\bibitem [{\citenamefont {Carmichael}(2009)}]{carmichael2009open}%
  \BibitemOpen
  \bibfield  {author} {\bibinfo {author} {\bibfnamefont {H.}~\bibnamefont {Carmichael}},\ }\href@noop {} {\emph {\bibinfo {title} {An open systems approach to quantum optics}}},\ Vol.~\bibinfo {volume} {18}\ (\bibinfo  {publisher} {Springer Science \& Business Media},\ \bibinfo {year} {2009})\BibitemShut {NoStop}%
\bibitem [{\citenamefont {Manzano}(2020)}]{manzano2020short}%
  \BibitemOpen
  \bibfield  {author} {\bibinfo {author} {\bibfnamefont {D.}~\bibnamefont {Manzano}},\ }\bibfield  {title} {\bibinfo {title} {A short introduction to the lindblad master equation},\ }\href@noop {} {\bibfield  {journal} {\bibinfo  {journal} {Aip advances}\ }\textbf {\bibinfo {volume} {10}} (\bibinfo {year} {2020})}\BibitemShut {NoStop}%
\bibitem [{\citenamefont {Campaioli}\ \emph {et~al.}(2024)\citenamefont {Campaioli}, \citenamefont {Cole},\ and\ \citenamefont {Hapuarachchi}}]{campaioli2024quantum}%
  \BibitemOpen
  \bibfield  {author} {\bibinfo {author} {\bibfnamefont {F.}~\bibnamefont {Campaioli}}, \bibinfo {author} {\bibfnamefont {J.~H.}\ \bibnamefont {Cole}},\ \bibnamefont {and}\ \bibinfo {author} {\bibfnamefont {H.}~\bibnamefont {Hapuarachchi}},\ }\bibfield  {title} {\bibinfo {title} {Quantum master equations: Tips and tricks for quantum optics, quantum computing, and beyond},\ }\href@noop {} {\bibfield  {journal} {\bibinfo  {journal} {PRX Quantum}\ }\textbf {\bibinfo {volume} {5}},\ \bibinfo {pages} {020202} (\bibinfo {year} {2024})}\BibitemShut {NoStop}%
\bibitem [{\citenamefont {Albert}(2018)}]{albert2018lindbladians}%
  \BibitemOpen
  \bibfield  {author} {\bibinfo {author} {\bibfnamefont {V.~V.}\ \bibnamefont {Albert}},\ }\bibfield  {title} {\bibinfo {title} {Lindbladians with multiple steady states: theory and applications},\ }\href@noop {} {\bibfield  {journal} {\bibinfo  {journal} {arXiv preprint arXiv:1802.00010}\ } (\bibinfo {year} {2018})}\BibitemShut {NoStop}%
\bibitem [{\citenamefont {Demkowicz-Dobrza{\'n}ski}\ and\ \citenamefont {Maccone}(2014)}]{demkowicz2014using}%
  \BibitemOpen
  \bibfield  {author} {\bibinfo {author} {\bibfnamefont {R.}~\bibnamefont {Demkowicz-Dobrza{\'n}ski}}\ \bibnamefont {and}\ \bibinfo {author} {\bibfnamefont {L.}~\bibnamefont {Maccone}},\ }\bibfield  {title} {\bibinfo {title} {Using entanglement against noise in quantum metrology},\ }\href@noop {} {\bibfield  {journal} {\bibinfo  {journal} {Phys. Rev. Lett.}\ }\textbf {\bibinfo {volume} {113}},\ \bibinfo {pages} {250801} (\bibinfo {year} {2014})}\BibitemShut {NoStop}%
\bibitem [{\citenamefont {Sekatski}\ \emph {et~al.}(2017)\citenamefont {Sekatski}, \citenamefont {Skotiniotis}, \citenamefont {Ko{\l}ody{\'n}ski},\ and\ \citenamefont {D{\"u}r}}]{sekatski2017quantum}%
  \BibitemOpen
  \bibfield  {author} {\bibinfo {author} {\bibfnamefont {P.}~\bibnamefont {Sekatski}}, \bibinfo {author} {\bibfnamefont {M.}~\bibnamefont {Skotiniotis}}, \bibinfo {author} {\bibfnamefont {J.}~\bibnamefont {Ko{\l}ody{\'n}ski}},\ \bibnamefont {and}\ \bibinfo {author} {\bibfnamefont {W.}~\bibnamefont {D{\"u}r}},\ }\bibfield  {title} {\bibinfo {title} {Quantum metrology with full and fast quantum control},\ }\href@noop {} {\bibfield  {journal} {\bibinfo  {journal} {Quantum}\ }\textbf {\bibinfo {volume} {1}},\ \bibinfo {pages} {27} (\bibinfo {year} {2017})}\BibitemShut {NoStop}%
\bibitem [{\citenamefont {Zhou}\ and\ \citenamefont {Jiang}(2021)}]{zhou2021asymptotic}%
  \BibitemOpen
  \bibfield  {author} {\bibinfo {author} {\bibfnamefont {S.}~\bibnamefont {Zhou}}\ \bibnamefont {and}\ \bibinfo {author} {\bibfnamefont {L.}~\bibnamefont {Jiang}},\ }\bibfield  {title} {\bibinfo {title} {Asymptotic theory of quantum channel estimation},\ }\href@noop {} {\bibfield  {journal} {\bibinfo  {journal} {PRX Quantum}\ }\textbf {\bibinfo {volume} {2}},\ \bibinfo {pages} {010343} (\bibinfo {year} {2021})}\BibitemShut {NoStop}%
\bibitem [{\citenamefont {Demkowicz-Dobrza{\'n}ski}\ \emph {et~al.}(2017)\citenamefont {Demkowicz-Dobrza{\'n}ski}, \citenamefont {Czajkowski},\ and\ \citenamefont {Sekatski}}]{demkowicz2017adaptive}%
  \BibitemOpen
  \bibfield  {author} {\bibinfo {author} {\bibfnamefont {R.}~\bibnamefont {Demkowicz-Dobrza{\'n}ski}}, \bibinfo {author} {\bibfnamefont {J.}~\bibnamefont {Czajkowski}},\ \bibnamefont {and}\ \bibinfo {author} {\bibfnamefont {P.}~\bibnamefont {Sekatski}},\ }\bibfield  {title} {\bibinfo {title} {Adaptive quantum metrology under general markovian noise},\ }\href@noop {} {\bibfield  {journal} {\bibinfo  {journal} {Physical Review X}\ }\textbf {\bibinfo {volume} {7}},\ \bibinfo {pages} {041009} (\bibinfo {year} {2017})}\BibitemShut {NoStop}%
\bibitem [{\citenamefont {Zhou}\ \emph {et~al.}(2018)\citenamefont {Zhou}, \citenamefont {Zhang}, \citenamefont {Preskill},\ and\ \citenamefont {Jiang}}]{ZhouNC18AchievingHeisenberg}%
  \BibitemOpen
  \bibfield  {author} {\bibinfo {author} {\bibfnamefont {S.}~\bibnamefont {Zhou}}, \bibinfo {author} {\bibfnamefont {M.}~\bibnamefont {Zhang}}, \bibinfo {author} {\bibfnamefont {J.}~\bibnamefont {Preskill}},\ \bibnamefont {and}\ \bibinfo {author} {\bibfnamefont {L.}~\bibnamefont {Jiang}},\ }\bibfield  {title} {\bibinfo {title} {Achieving the {{Heisenberg}} limit in quantum metrology using quantum error correction},\ }\href {https://doi.org/10.1038/s41467-017-02510-3} {\bibfield  {journal} {\bibinfo  {journal} {Nat. Commun.}\ }\textbf {\bibinfo {volume} {9}},\ \bibinfo {pages} {78} (\bibinfo {year} {2018})}\BibitemShut {NoStop}%
\bibitem [{\citenamefont {Wan}\ and\ \citenamefont {Lasenby}(2022)}]{wan2022bounds}%
  \BibitemOpen
  \bibfield  {author} {\bibinfo {author} {\bibfnamefont {K.}~\bibnamefont {Wan}}\ \bibnamefont {and}\ \bibinfo {author} {\bibfnamefont {R.}~\bibnamefont {Lasenby}},\ }\bibfield  {title} {\bibinfo {title} {Bounds on adaptive quantum metrology under markovian noise},\ }\href@noop {} {\bibfield  {journal} {\bibinfo  {journal} {Phys. Rev. Research}\ }\textbf {\bibinfo {volume} {4}},\ \bibinfo {pages} {033092} (\bibinfo {year} {2022})}\BibitemShut {NoStop}%
\bibitem [{\citenamefont {Sekatski}\ and\ \citenamefont {Perarnau-Llobet}(2022)}]{sekatski2022optimal}%
  \BibitemOpen
  \bibfield  {author} {\bibinfo {author} {\bibfnamefont {P.}~\bibnamefont {Sekatski}}\ \bibnamefont {and}\ \bibinfo {author} {\bibfnamefont {M.}~\bibnamefont {Perarnau-Llobet}},\ }\bibfield  {title} {\bibinfo {title} {Optimal nonequilibrium thermometry in markovian environments},\ }\href@noop {} {\bibfield  {journal} {\bibinfo  {journal} {Quantum}\ }\textbf {\bibinfo {volume} {6}},\ \bibinfo {pages} {869} (\bibinfo {year} {2022})}\BibitemShut {NoStop}%
\bibitem [{\citenamefont {Kurdzia{\l}ek}\ \emph {et~al.}(2023)\citenamefont {Kurdzia{\l}ek}, \citenamefont {G{\'o}recki}, \citenamefont {Albarelli},\ and\ \citenamefont {Demkowicz-Dobrza{\'n}ski}}]{kurdzialek2023using}%
  \BibitemOpen
  \bibfield  {author} {\bibinfo {author} {\bibfnamefont {S.}~\bibnamefont {Kurdzia{\l}ek}}, \bibinfo {author} {\bibfnamefont {W.}~\bibnamefont {G{\'o}recki}}, \bibinfo {author} {\bibfnamefont {F.}~\bibnamefont {Albarelli}},\ \bibnamefont {and}\ \bibinfo {author} {\bibfnamefont {R.}~\bibnamefont {Demkowicz-Dobrza{\'n}ski}},\ }\bibfield  {title} {\bibinfo {title} {Using adaptiveness and causal superpositions against noise in quantum metrology},\ }\href@noop {} {\bibfield  {journal} {\bibinfo  {journal} {Phys. Rev. Letters}\ }\textbf {\bibinfo {volume} {131}},\ \bibinfo {pages} {090801} (\bibinfo {year} {2023})}\BibitemShut {NoStop}%
\bibitem [{\citenamefont {Zhou}\ \emph {et~al.}(2024)\citenamefont {Zhou}, \citenamefont {Manes},\ and\ \citenamefont {Jiang}}]{zhou2024achieving}%
  \BibitemOpen
  \bibfield  {author} {\bibinfo {author} {\bibfnamefont {S.}~\bibnamefont {Zhou}}, \bibinfo {author} {\bibfnamefont {A.~G.}\ \bibnamefont {Manes}},\ \bibnamefont {and}\ \bibinfo {author} {\bibfnamefont {L.}~\bibnamefont {Jiang}},\ }\bibfield  {title} {\bibinfo {title} {Achieving metrological limits using ancilla-free quantum error-correcting codes},\ }\href@noop {} {\bibfield  {journal} {\bibinfo  {journal} {Phys. Rev. A}\ }\textbf {\bibinfo {volume} {109}},\ \bibinfo {pages} {042406} (\bibinfo {year} {2024})}\BibitemShut {NoStop}%
\bibitem [{\citenamefont {Wiseman}\ and\ \citenamefont {Milburn}(2009)}]{Wiseman09QuantumMeasurement}%
  \BibitemOpen
  \bibfield  {author} {\bibinfo {author} {\bibfnamefont {H.~M.}\ \bibnamefont {Wiseman}}\ \bibnamefont {and}\ \bibinfo {author} {\bibfnamefont {G.~J.}\ \bibnamefont {Milburn}},\ }\href {https://doi.org/10.1017/CBO9780511813948} {\emph {\bibinfo {title} {Quantum {{Measurement}} and {{Control}}}}}\ (\bibinfo  {publisher} {Cambridge University Press},\ \bibinfo {address} {Cambridge},\ \bibinfo {year} {2009})\BibitemShut {NoStop}%
\bibitem [{\citenamefont {Pezz{\`e}}\ \emph {et~al.}(2018)\citenamefont {Pezz{\`e}}, \citenamefont {Smerzi}, \citenamefont {Oberthaler}, \citenamefont {Schmied},\ and\ \citenamefont {Treutlein}}]{PezzeRMP18QuantumMetrology}%
  \BibitemOpen
  \bibfield  {author} {\bibinfo {author} {\bibfnamefont {L.}~\bibnamefont {Pezz{\`e}}}, \bibinfo {author} {\bibfnamefont {A.}~\bibnamefont {Smerzi}}, \bibinfo {author} {\bibfnamefont {M.~K.}\ \bibnamefont {Oberthaler}}, \bibinfo {author} {\bibfnamefont {R.}~\bibnamefont {Schmied}},\ \bibnamefont {and}\ \bibinfo {author} {\bibfnamefont {P.}~\bibnamefont {Treutlein}},\ }\bibfield  {title} {\bibinfo {title} {Quantum metrology with nonclassical states of atomic ensembles},\ }\href {https://doi.org/10.1103/RevModPhys.90.035005} {\bibfield  {journal} {\bibinfo  {journal} {Rev. Mod. Phys.}\ }\textbf {\bibinfo {volume} {90}},\ \bibinfo {pages} {035005} (\bibinfo {year} {2018})}\BibitemShut {NoStop}%
\bibitem [{\citenamefont {G{\'o}recki}\ \emph {et~al.}(2022)\citenamefont {G{\'o}recki}, \citenamefont {Riccardi},\ and\ \citenamefont {Maccone}}]{gorecki2022quantum}%
  \BibitemOpen
  \bibfield  {author} {\bibinfo {author} {\bibfnamefont {W.}~\bibnamefont {G{\'o}recki}}, \bibinfo {author} {\bibfnamefont {A.}~\bibnamefont {Riccardi}},\ \bibnamefont {and}\ \bibinfo {author} {\bibfnamefont {L.}~\bibnamefont {Maccone}},\ }\bibfield  {title} {\bibinfo {title} {Quantum metrology of noisy spreading channels},\ }\href {https://doi.org/10.1103/PhysRevLett.129.240503} {\bibfield  {journal} {\bibinfo  {journal} {Phys. Rev. Lett.}\ }\textbf {\bibinfo {volume} {129}},\ \bibinfo {pages} {240503} (\bibinfo {year} {2022})}\BibitemShut {NoStop}%
\bibitem [{\citenamefont {Haine}(2018)}]{haine2018using}%
  \BibitemOpen
  \bibfield  {author} {\bibinfo {author} {\bibfnamefont {S.~A.}\ \bibnamefont {Haine}},\ }\bibfield  {title} {\bibinfo {title} {Using interaction-based readouts to approach the ultimate limit of detection-noise robustness for quantum-enhanced metrology in collective spin systems},\ }\href {https://doi.org/10.1103/PhysRevA.98.030303} {\bibfield  {journal} {\bibinfo  {journal} {Phys. Rev. A}\ }\textbf {\bibinfo {volume} {98}},\ \bibinfo {pages} {030303} (\bibinfo {year} {2018})}\BibitemShut {NoStop}%
\bibitem [{\citenamefont {Demkowicz-Dobrza{\'n}ski}\ \emph {et~al.}(2012)\citenamefont {Demkowicz-Dobrza{\'n}ski}, \citenamefont {Ko{\l}ody{\'n}ski},\ and\ \citenamefont {Gu{\c{t}}{\u{a}}}}]{demkowicz2012elusive}%
  \BibitemOpen
  \bibfield  {author} {\bibinfo {author} {\bibfnamefont {R.}~\bibnamefont {Demkowicz-Dobrza{\'n}ski}}, \bibinfo {author} {\bibfnamefont {J.}~\bibnamefont {Ko{\l}ody{\'n}ski}},\ \bibnamefont {and}\ \bibinfo {author} {\bibfnamefont {M.}~\bibnamefont {Gu{\c{t}}{\u{a}}}},\ }\bibfield  {title} {\bibinfo {title} {The elusive heisenberg limit in quantum-enhanced metrology},\ }\href@noop {} {\bibfield  {journal} {\bibinfo  {journal} {Nature communications}\ }\textbf {\bibinfo {volume} {3}},\ \bibinfo {pages} {1063} (\bibinfo {year} {2012})}\BibitemShut {NoStop}%
\bibitem [{\citenamefont {Ko{\l}ody{\'n}ski}\ and\ \citenamefont {Demkowicz-Dobrza{\'n}ski}(2013)}]{kolodynski2013efficient}%
  \BibitemOpen
  \bibfield  {author} {\bibinfo {author} {\bibfnamefont {J.}~\bibnamefont {Ko{\l}ody{\'n}ski}}\ \bibnamefont {and}\ \bibinfo {author} {\bibfnamefont {R.}~\bibnamefont {Demkowicz-Dobrza{\'n}ski}},\ }\bibfield  {title} {\bibinfo {title} {Efficient tools for quantum metrology with uncorrelated noise},\ }\href@noop {} {\bibfield  {journal} {\bibinfo  {journal} {New J. Phys.}\ }\textbf {\bibinfo {volume} {15}},\ \bibinfo {pages} {073043} (\bibinfo {year} {2013})}\BibitemShut {NoStop}%
\bibitem [{Note1()}]{Note1}%
  \BibitemOpen
  \bibinfo {note} {We use the terms ``Lindblad estimation'' and ``stochastic signal sensing'' interchangeably. If the Born-Markov approximation does not hold, however, then the stochastic signal may not be described by a Lindbladian and a different approach is necessary.}\BibitemShut {Stop}%
\bibitem [{Note2()}]{Note2}%
  \BibitemOpen
  \bibinfo {note} {The operator norm $\|\cdot \|$ is induced by the $\ell _2$ norm of the kets. For a given operator $\protect \hat {A}$, $\|\protect \hat {A}\|$ is also called the spectral norm and equals the largest singular value of $\protect \hat {A}$.}\BibitemShut {Stop}%
\bibitem [{Note3()}]{Note3}%
  \BibitemOpen
  \bibinfo {note} {In the noiseless case, there are no other natural scales to compare to the size of the signal. Instead, we first derive the Kraus operators in Eq.~\ref {eq:Kraus representation} assuming that the short time condition in Eq.~\ref {eq:short time} holds which implies that $\gamma _1t\ll 1$. Then, we take the limit of $\gamma _1t\rightarrow 0$ of the Kraus operators in Eq.~\ref {eq:Kraus representation} to reach Eq.~\ref {eq:K, Kdot}. This is what we mean by the vanishing signal limit in the noiseless case.}\BibitemShut {Stop}%
\bibitem [{\citenamefont {Boulant}\ \emph {et~al.}(2003)\citenamefont {Boulant}, \citenamefont {Havel}, \citenamefont {Pravia},\ and\ \citenamefont {Cory}}]{boulant2003robust}%
  \BibitemOpen
  \bibfield  {author} {\bibinfo {author} {\bibfnamefont {N.}~\bibnamefont {Boulant}}, \bibinfo {author} {\bibfnamefont {T.~F.}\ \bibnamefont {Havel}}, \bibinfo {author} {\bibfnamefont {M.~A.}\ \bibnamefont {Pravia}},\ \bibnamefont {and}\ \bibinfo {author} {\bibfnamefont {D.~G.}\ \bibnamefont {Cory}},\ }\bibfield  {title} {\bibinfo {title} {Robust method for estimating the lindblad operators of a dissipative quantum process from measurements of the density operator at multiple time points},\ }\href@noop {} {\bibfield  {journal} {\bibinfo  {journal} {Phys. Rev. A}\ }\textbf {\bibinfo {volume} {67}},\ \bibinfo {pages} {042322} (\bibinfo {year} {2003})}\BibitemShut {NoStop}%
\bibitem [{\citenamefont {Ben~Av}\ \emph {et~al.}(2020)\citenamefont {Ben~Av}, \citenamefont {Shapira}, \citenamefont {Akerman},\ and\ \citenamefont {Ozeri}}]{ben2020direct}%
  \BibitemOpen
  \bibfield  {author} {\bibinfo {author} {\bibfnamefont {E.}~\bibnamefont {Ben~Av}}, \bibinfo {author} {\bibfnamefont {Y.}~\bibnamefont {Shapira}}, \bibinfo {author} {\bibfnamefont {N.}~\bibnamefont {Akerman}},\ \bibnamefont {and}\ \bibinfo {author} {\bibfnamefont {R.}~\bibnamefont {Ozeri}},\ }\bibfield  {title} {\bibinfo {title} {Direct reconstruction of the quantum-master-equation dynamics of a trapped-ion qubit},\ }\href@noop {} {\bibfield  {journal} {\bibinfo  {journal} {Phys. Rev. A}\ }\textbf {\bibinfo {volume} {101}},\ \bibinfo {pages} {062305} (\bibinfo {year} {2020})}\BibitemShut {NoStop}%
\bibitem [{\citenamefont {Nagourney}\ \emph {et~al.}(1986)\citenamefont {Nagourney}, \citenamefont {Sandberg},\ and\ \citenamefont {Dehmelt}}]{nagourney1986shelved}%
  \BibitemOpen
  \bibfield  {author} {\bibinfo {author} {\bibfnamefont {W.}~\bibnamefont {Nagourney}}, \bibinfo {author} {\bibfnamefont {J.}~\bibnamefont {Sandberg}},\ \bibnamefont {and}\ \bibinfo {author} {\bibfnamefont {H.}~\bibnamefont {Dehmelt}},\ }\bibfield  {title} {\bibinfo {title} {Shelved optical electron amplifier: Observation of quantum jumps},\ }\href@noop {} {\bibfield  {journal} {\bibinfo  {journal} {Phys. Rev. Letters}\ }\textbf {\bibinfo {volume} {56}},\ \bibinfo {pages} {2797} (\bibinfo {year} {1986})}\BibitemShut {NoStop}%
\bibitem [{\citenamefont {Itano}\ \emph {et~al.}(1990)\citenamefont {Itano}, \citenamefont {Heinzen}, \citenamefont {Bollinger},\ and\ \citenamefont {Wineland}}]{itano1990quantum}%
  \BibitemOpen
  \bibfield  {author} {\bibinfo {author} {\bibfnamefont {W.~M.}\ \bibnamefont {Itano}}, \bibinfo {author} {\bibfnamefont {D.~J.}\ \bibnamefont {Heinzen}}, \bibinfo {author} {\bibfnamefont {J.~J.}\ \bibnamefont {Bollinger}},\ \bibnamefont {and}\ \bibinfo {author} {\bibfnamefont {D.~J.}\ \bibnamefont {Wineland}},\ }\bibfield  {title} {\bibinfo {title} {Quantum zeno effect},\ }\href@noop {} {\bibfield  {journal} {\bibinfo  {journal} {Phys. Rev. A}\ }\textbf {\bibinfo {volume} {41}},\ \bibinfo {pages} {2295} (\bibinfo {year} {1990})}\BibitemShut {NoStop}%
\bibitem [{\citenamefont {Tsang}\ \emph {et~al.}(2016)\citenamefont {Tsang}, \citenamefont {Nair},\ and\ \citenamefont {Lu}}]{tsang2016quantum}%
  \BibitemOpen
  \bibfield  {author} {\bibinfo {author} {\bibfnamefont {M.}~\bibnamefont {Tsang}}, \bibinfo {author} {\bibfnamefont {R.}~\bibnamefont {Nair}},\ \bibnamefont {and}\ \bibinfo {author} {\bibfnamefont {X.-M.}\ \bibnamefont {Lu}},\ }\bibfield  {title} {\bibinfo {title} {Quantum theory of superresolution for two incoherent optical point sources},\ }\href {https://doi.org/10.1103/PhysRevX.6.031033} {\bibfield  {journal} {\bibinfo  {journal} {Phys. Rev. X}\ }\textbf {\bibinfo {volume} {6}},\ \bibinfo {pages} {031033} (\bibinfo {year} {2016})}\BibitemShut {NoStop}%
\bibitem [{\citenamefont {Oh}\ \emph {et~al.}(2021)\citenamefont {Oh}, \citenamefont {Zhou}, \citenamefont {Wong},\ and\ \citenamefont {Jiang}}]{oh2021quantum}%
  \BibitemOpen
  \bibfield  {author} {\bibinfo {author} {\bibfnamefont {C.}~\bibnamefont {Oh}}, \bibinfo {author} {\bibfnamefont {S.}~\bibnamefont {Zhou}}, \bibinfo {author} {\bibfnamefont {Y.}~\bibnamefont {Wong}},\ \bibnamefont {and}\ \bibinfo {author} {\bibfnamefont {L.}~\bibnamefont {Jiang}},\ }\bibfield  {title} {\bibinfo {title} {Quantum limits of superresolution in a noisy environment},\ }\href@noop {} {\bibfield  {journal} {\bibinfo  {journal} {Physical Review Letters}\ }\textbf {\bibinfo {volume} {126}},\ \bibinfo {pages} {120502} (\bibinfo {year} {2021})}\BibitemShut {NoStop}%
\bibitem [{\citenamefont {Shi}\ \emph {et~al.}(2024)\citenamefont {Shi}, \citenamefont {Brady}, \citenamefont {G{\'o}recki}, \citenamefont {Maccone}, \citenamefont {Di~Candia},\ and\ \citenamefont {Zhuang}}]{shi2024quantum}%
  \BibitemOpen
  \bibfield  {author} {\bibinfo {author} {\bibfnamefont {H.}~\bibnamefont {Shi}}, \bibinfo {author} {\bibfnamefont {A.~J.}\ \bibnamefont {Brady}}, \bibinfo {author} {\bibfnamefont {W.}~\bibnamefont {G{\'o}recki}}, \bibinfo {author} {\bibfnamefont {L.}~\bibnamefont {Maccone}}, \bibinfo {author} {\bibfnamefont {R.}~\bibnamefont {Di~Candia}},\ \bibnamefont {and}\ \bibinfo {author} {\bibfnamefont {Q.}~\bibnamefont {Zhuang}},\ }\bibfield  {title} {\bibinfo {title} {Quantum-enhanced dark matter detection with in-cavity control: mitigating the rayleigh curse},\ }\href@noop {} {\bibfield  {journal} {\bibinfo  {journal} {arXiv preprint arXiv:2409.04656}\ } (\bibinfo {year} {2024})}\BibitemShut {NoStop}%
\bibitem [{\citenamefont {Haine}\ and\ \citenamefont {Szigeti}(2015)}]{haine2015quantum}%
  \BibitemOpen
  \bibfield  {author} {\bibinfo {author} {\bibfnamefont {S.~A.}\ \bibnamefont {Haine}}\ \bibnamefont {and}\ \bibinfo {author} {\bibfnamefont {S.~S.}\ \bibnamefont {Szigeti}},\ }\bibfield  {title} {\bibinfo {title} {Quantum metrology with mixed states: When recovering lost information is better than never losing it},\ }\href@noop {} {\bibfield  {journal} {\bibinfo  {journal} {Physical Review A}\ }\textbf {\bibinfo {volume} {92}},\ \bibinfo {pages} {032317} (\bibinfo {year} {2015})}\BibitemShut {NoStop}%
\bibitem [{rep()}]{repo}%
  \BibitemOpen
  \href@noop {} {\bibinfo {title} {slipperyslider}},\ \bibinfo {howpublished} {J.~W.~Gardner. \emph{slipperySlider}. 2024. \url{https://git.ligo.org/jameswalter.gardner/slipperyslider}}\BibitemShut {NoStop}%
\bibitem [{Note4()}]{Note4}%
  \BibitemOpen
  \bibinfo {note} {Here, the complex phase $\varphi _1$ in $\protect \hat {L}_{1,(s)}$ and $\protect \hat {L}_{2,(n)}$ has been compensated by setting $\phi \DOTSB \mapstochar \rightarrow \phi + \varphi _1 / 2$ such that we may assume that $\varphi _1=0$ without loss of generality.}\BibitemShut {Stop}%
\bibitem [{\citenamefont {Macieszczak}\ \emph {et~al.}(2014)\citenamefont {Macieszczak}, \citenamefont {Fraas},\ and\ \citenamefont {Demkowicz-Dobrza{\'n}ski}}]{macieszczak2014bayesian}%
  \BibitemOpen
  \bibfield  {author} {\bibinfo {author} {\bibfnamefont {K.}~\bibnamefont {Macieszczak}}, \bibinfo {author} {\bibfnamefont {M.}~\bibnamefont {Fraas}},\ \bibnamefont {and}\ \bibinfo {author} {\bibfnamefont {R.}~\bibnamefont {Demkowicz-Dobrza{\'n}ski}},\ }\bibfield  {title} {\bibinfo {title} {Bayesian quantum frequency estimation in presence of collective dephasing},\ }\href@noop {} {\bibfield  {journal} {\bibinfo  {journal} {New Journal of Physics}\ }\textbf {\bibinfo {volume} {16}},\ \bibinfo {pages} {113002} (\bibinfo {year} {2014})}\BibitemShut {NoStop}%
\bibitem [{\citenamefont {Direkci}\ \emph {et~al.}(2024)\citenamefont {Direkci}, \citenamefont {Finkelstein}, \citenamefont {Endres},\ and\ \citenamefont {Gefen}}]{direkci2024heisenberg}%
  \BibitemOpen
  \bibfield  {author} {\bibinfo {author} {\bibfnamefont {S.}~\bibnamefont {Direkci}}, \bibinfo {author} {\bibfnamefont {R.}~\bibnamefont {Finkelstein}}, \bibinfo {author} {\bibfnamefont {M.}~\bibnamefont {Endres}},\ \bibnamefont {and}\ \bibinfo {author} {\bibfnamefont {T.}~\bibnamefont {Gefen}},\ }\bibfield  {title} {\bibinfo {title} {Heisenberg-limited bayesian phase estimation with low-depth digital quantum circuits},\ }\href@noop {} {\bibfield  {journal} {\bibinfo  {journal} {arXiv preprint arXiv:2407.06006}\ } (\bibinfo {year} {2024})}\BibitemShut {NoStop}%
\bibitem [{\citenamefont {Mok}\ \emph {et~al.}(2024)\citenamefont {Mok}, \citenamefont {Poddar}, \citenamefont {Sierra}, \citenamefont {Rusconi}, \citenamefont {Preskill},\ and\ \citenamefont {Asenjo-Garcia}}]{mok2024universal}%
  \BibitemOpen
  \bibfield  {author} {\bibinfo {author} {\bibfnamefont {W.-K.}\ \bibnamefont {Mok}}, \bibinfo {author} {\bibfnamefont {A.}~\bibnamefont {Poddar}}, \bibinfo {author} {\bibfnamefont {E.}~\bibnamefont {Sierra}}, \bibinfo {author} {\bibfnamefont {C.~C.}\ \bibnamefont {Rusconi}}, \bibinfo {author} {\bibfnamefont {J.}~\bibnamefont {Preskill}},\ \bibnamefont {and}\ \bibinfo {author} {\bibfnamefont {A.}~\bibnamefont {Asenjo-Garcia}},\ }\bibfield  {title} {\bibinfo {title} {Universal scaling laws for correlated decay of many-body quantum systems},\ }\href@noop {} {\bibfield  {journal} {\bibinfo  {journal} {arXiv preprint arXiv:2406.00722}\ } (\bibinfo {year} {2024})}\BibitemShut {NoStop}%
\bibitem [{\citenamefont {Bohnet}\ \emph {et~al.}(2012)\citenamefont {Bohnet}, \citenamefont {Chen}, \citenamefont {Weiner}, \citenamefont {Meiser}, \citenamefont {Holland},\ and\ \citenamefont {Thompson}}]{bohnet2012steady}%
  \BibitemOpen
  \bibfield  {author} {\bibinfo {author} {\bibfnamefont {J.~G.}\ \bibnamefont {Bohnet}}, \bibinfo {author} {\bibfnamefont {Z.}~\bibnamefont {Chen}}, \bibinfo {author} {\bibfnamefont {J.~M.}\ \bibnamefont {Weiner}}, \bibinfo {author} {\bibfnamefont {D.}~\bibnamefont {Meiser}}, \bibinfo {author} {\bibfnamefont {M.~J.}\ \bibnamefont {Holland}},\ \bibnamefont {and}\ \bibinfo {author} {\bibfnamefont {J.~K.}\ \bibnamefont {Thompson}},\ }\bibfield  {title} {\bibinfo {title} {A steady-state superradiant laser with less than one intracavity photon},\ }\href@noop {} {\bibfield  {journal} {\bibinfo  {journal} {Nature}\ }\textbf {\bibinfo {volume} {484}},\ \bibinfo {pages} {78} (\bibinfo {year} {2012})}\BibitemShut {NoStop}%
\bibitem [{\citenamefont {Ferioli}\ \emph {et~al.}(2023)\citenamefont {Ferioli}, \citenamefont {Glicenstein}, \citenamefont {Ferrier-Barbut},\ and\ \citenamefont {Browaeys}}]{ferioli2023non}%
  \BibitemOpen
  \bibfield  {author} {\bibinfo {author} {\bibfnamefont {G.}~\bibnamefont {Ferioli}}, \bibinfo {author} {\bibfnamefont {A.}~\bibnamefont {Glicenstein}}, \bibinfo {author} {\bibfnamefont {I.}~\bibnamefont {Ferrier-Barbut}},\ \bibnamefont {and}\ \bibinfo {author} {\bibfnamefont {A.}~\bibnamefont {Browaeys}},\ }\bibfield  {title} {\bibinfo {title} {A non-equilibrium superradiant phase transition in free space},\ }\href@noop {} {\bibfield  {journal} {\bibinfo  {journal} {Nature Physics}\ }\textbf {\bibinfo {volume} {19}},\ \bibinfo {pages} {1345} (\bibinfo {year} {2023})}\BibitemShut {NoStop}%
\bibitem [{\citenamefont {Pichler}\ \emph {et~al.}(2015)\citenamefont {Pichler}, \citenamefont {Ramos}, \citenamefont {Daley},\ and\ \citenamefont {Zoller}}]{pichler2015quantum}%
  \BibitemOpen
  \bibfield  {author} {\bibinfo {author} {\bibfnamefont {H.}~\bibnamefont {Pichler}}, \bibinfo {author} {\bibfnamefont {T.}~\bibnamefont {Ramos}}, \bibinfo {author} {\bibfnamefont {A.~J.}\ \bibnamefont {Daley}},\ \bibnamefont {and}\ \bibinfo {author} {\bibfnamefont {P.}~\bibnamefont {Zoller}},\ }\bibfield  {title} {\bibinfo {title} {Quantum optics of chiral spin networks},\ }\href@noop {} {\bibfield  {journal} {\bibinfo  {journal} {Physical Review A}\ }\textbf {\bibinfo {volume} {91}},\ \bibinfo {pages} {042116} (\bibinfo {year} {2015})}\BibitemShut {NoStop}%
\bibitem [{\citenamefont {Gottesman}\ \emph {et~al.}(2001)\citenamefont {Gottesman}, \citenamefont {Kitaev},\ and\ \citenamefont {Preskill}}]{GottesmanPRA01EncodingQubit}%
  \BibitemOpen
  \bibfield  {author} {\bibinfo {author} {\bibfnamefont {D.}~\bibnamefont {Gottesman}}, \bibinfo {author} {\bibfnamefont {A.}~\bibnamefont {Kitaev}},\ \bibnamefont {and}\ \bibinfo {author} {\bibfnamefont {J.}~\bibnamefont {Preskill}},\ }\bibfield  {title} {\bibinfo {title} {Encoding a qubit in an oscillator},\ }\href {https://doi.org/10.1103/PhysRevA.64.012310} {\bibfield  {journal} {\bibinfo  {journal} {Phys. Rev. A}\ }\textbf {\bibinfo {volume} {64}},\ \bibinfo {pages} {012310} (\bibinfo {year} {2001})}\BibitemShut {NoStop}%
\bibitem [{\citenamefont {Michael}\ \emph {et~al.}(2016)\citenamefont {Michael}, \citenamefont {Silveri}, \citenamefont {Brierley}, \citenamefont {Albert}, \citenamefont {Salmilehto}, \citenamefont {Jiang},\ and\ \citenamefont {Girvin}}]{michael2016new}%
  \BibitemOpen
  \bibfield  {author} {\bibinfo {author} {\bibfnamefont {M.~H.}\ \bibnamefont {Michael}}, \bibinfo {author} {\bibfnamefont {M.}~\bibnamefont {Silveri}}, \bibinfo {author} {\bibfnamefont {R.}~\bibnamefont {Brierley}}, \bibinfo {author} {\bibfnamefont {V.~V.}\ \bibnamefont {Albert}}, \bibinfo {author} {\bibfnamefont {J.}~\bibnamefont {Salmilehto}}, \bibinfo {author} {\bibfnamefont {L.}~\bibnamefont {Jiang}},\ \bibnamefont {and}\ \bibinfo {author} {\bibfnamefont {S.~M.}\ \bibnamefont {Girvin}},\ }\bibfield  {title} {\bibinfo {title} {New class of quantum error-correcting codes for a bosonic mode},\ }\href@noop {} {\bibfield  {journal} {\bibinfo  {journal} {Phys. Rev. X}\ }\textbf {\bibinfo {volume} {6}},\ \bibinfo {pages} {031006} (\bibinfo {year} {2016})}\BibitemShut {NoStop}%
\bibitem [{\citenamefont {Kim}\ and\ \citenamefont {Carosi}(2010)}]{kim2010axions}%
  \BibitemOpen
  \bibfield  {author} {\bibinfo {author} {\bibfnamefont {J.~E.}\ \bibnamefont {Kim}}\ \bibnamefont {and}\ \bibinfo {author} {\bibfnamefont {G.}~\bibnamefont {Carosi}},\ }\bibfield  {title} {\bibinfo {title} {Axions and the strong ${{C P}}$ problem},\ }\href {https://doi.org/10.1103/RevModPhys.82.557} {\bibfield  {journal} {\bibinfo  {journal} {Rev. Mod. Phys.}\ }\textbf {\bibinfo {volume} {82}},\ \bibinfo {pages} {557} (\bibinfo {year} {2010})}\BibitemShut {NoStop}%
\bibitem [{\citenamefont {Choi}\ \emph {et~al.}(2021)\citenamefont {Choi}, \citenamefont {Im},\ and\ \citenamefont {Shin}}]{choi2021recent}%
  \BibitemOpen
  \bibfield  {author} {\bibinfo {author} {\bibfnamefont {K.}~\bibnamefont {Choi}}, \bibinfo {author} {\bibfnamefont {S.~H.}\ \bibnamefont {Im}},\ \bibnamefont {and}\ \bibinfo {author} {\bibfnamefont {C.~S.}\ \bibnamefont {Shin}},\ }\bibfield  {title} {\bibinfo {title} {Recent progress in the physics of axions and axion-like particles},\ }\href {https://doi.org/10.1146/annurev-nucl-120720-031147} {\bibfield  {journal} {\bibinfo  {journal} {Annu. Rev. Nucl. Part. Sci.}\ }\textbf {\bibinfo {volume} {71}},\ \bibinfo {pages} {225} (\bibinfo {year} {2021})}\BibitemShut {NoStop}%
\bibitem [{\citenamefont {Cameron}\ \emph {et~al.}(1993)\citenamefont {Cameron}, \citenamefont {Cantatore}, \citenamefont {Melissinos}, \citenamefont {Ruoso}, \citenamefont {Semertzidis}, \citenamefont {Halama}, \citenamefont {Lazarus}, \citenamefont {Prodell}, \citenamefont {Nezrick}, \citenamefont {Rizzo} \emph {et~al.}}]{cameron1993search}%
  \BibitemOpen
  \bibfield  {author} {\bibinfo {author} {\bibfnamefont {R.}~\bibnamefont {Cameron}}, \bibinfo {author} {\bibfnamefont {G.}~\bibnamefont {Cantatore}}, \bibinfo {author} {\bibfnamefont {A.}~\bibnamefont {Melissinos}}, \bibinfo {author} {\bibfnamefont {G.}~\bibnamefont {Ruoso}}, \bibinfo {author} {\bibfnamefont {Y.}~\bibnamefont {Semertzidis}}, \bibinfo {author} {\bibfnamefont {H.}~\bibnamefont {Halama}}, \bibinfo {author} {\bibfnamefont {D.}~\bibnamefont {Lazarus}}, \bibinfo {author} {\bibfnamefont {A.}~\bibnamefont {Prodell}}, \bibinfo {author} {\bibfnamefont {F.}~\bibnamefont {Nezrick}}, \bibinfo {author} {\bibfnamefont {C.}~\bibnamefont {Rizzo}}, \bibnamefont {et~al.},\ }\bibfield  {title} {\bibinfo {title} {Search for nearly massless, weakly coupled particles by optical techniques},\ }\href {https://doi.org/10.1103/PhysRevD.47.3707} {\bibfield  {journal} {\bibinfo  {journal} {Phys. Rev. D}\ }\textbf {\bibinfo {volume} {47}},\ \bibinfo {pages} {3707} (\bibinfo {year} {1993})}\BibitemShut {NoStop}%
\bibitem [{\citenamefont {Du}\ \emph {et~al.}(2018)\citenamefont {Du}, \citenamefont {Force}, \citenamefont {Khatiwada}, \citenamefont {Lentz}, \citenamefont {Ottens}, \citenamefont {Rosenberg}, \citenamefont {Rybka}, \citenamefont {Carosi}, \citenamefont {Woollett}, \citenamefont {Bowring} \emph {et~al.}}]{du2018search}%
  \BibitemOpen
  \bibfield  {author} {\bibinfo {author} {\bibfnamefont {N.}~\bibnamefont {Du}}, \bibinfo {author} {\bibfnamefont {N.}~\bibnamefont {Force}}, \bibinfo {author} {\bibfnamefont {R.}~\bibnamefont {Khatiwada}}, \bibinfo {author} {\bibfnamefont {E.}~\bibnamefont {Lentz}}, \bibinfo {author} {\bibfnamefont {R.}~\bibnamefont {Ottens}}, \bibinfo {author} {\bibfnamefont {L.}~\bibnamefont {Rosenberg}}, \bibinfo {author} {\bibfnamefont {G.}~\bibnamefont {Rybka}}, \bibinfo {author} {\bibfnamefont {G.}~\bibnamefont {Carosi}}, \bibinfo {author} {\bibfnamefont {N.}~\bibnamefont {Woollett}}, \bibinfo {author} {\bibfnamefont {D.}~\bibnamefont {Bowring}}, \bibnamefont {et~al.},\ }\bibfield  {title} {\bibinfo {title} {Search for invisible axion dark matter with the axion dark matter experiment},\ }\href {https://doi.org/10.1103/PhysRevLett.120.151301} {\bibfield  {journal} {\bibinfo  {journal} {Phys. Rev. Lett.}\ }\textbf {\bibinfo {volume} {120}},\ \bibinfo {pages} {151301} (\bibinfo {year} {2018})}\BibitemShut {NoStop}%
\bibitem [{\citenamefont {Dixit}\ \emph {et~al.}(2021)\citenamefont {Dixit}, \citenamefont {Chakram}, \citenamefont {He}, \citenamefont {Agrawal}, \citenamefont {Naik}, \citenamefont {Schuster},\ and\ \citenamefont {Chou}}]{DixitPRL21SearchingDark}%
  \BibitemOpen
  \bibfield  {author} {\bibinfo {author} {\bibfnamefont {A.~V.}\ \bibnamefont {Dixit}}, \bibinfo {author} {\bibfnamefont {S.}~\bibnamefont {Chakram}}, \bibinfo {author} {\bibfnamefont {K.}~\bibnamefont {He}}, \bibinfo {author} {\bibfnamefont {A.}~\bibnamefont {Agrawal}}, \bibinfo {author} {\bibfnamefont {R.~K.}\ \bibnamefont {Naik}}, \bibinfo {author} {\bibfnamefont {D.~I.}\ \bibnamefont {Schuster}},\ \bibnamefont {and}\ \bibinfo {author} {\bibfnamefont {A.}~\bibnamefont {Chou}},\ }\bibfield  {title} {\bibinfo {title} {Searching for {{Dark Matter}} with a {{Superconducting Qubit}}},\ }\href {https://doi.org/10.1103/PhysRevLett.126.141302} {\bibfield  {journal} {\bibinfo  {journal} {Phys. Rev. Lett.}\ }\textbf {\bibinfo {volume} {126}},\ \bibinfo {pages} {141302} (\bibinfo {year} {2021})}\BibitemShut {NoStop}%
\bibitem [{Note5()}]{Note5}%
  \BibitemOpen
  \bibinfo {note} {On these short timescales, whether the signal is entirely stochastic or instead needs to be modelled as a combination of a coherent part and an incoherent part is a topic of ongoing research~\cite {gardner2024stochastic,shi2024quantum}.}\BibitemShut {Stop}%
\bibitem [{\citenamefont {Deng}\ \emph {et~al.}(2024)\citenamefont {Deng}, \citenamefont {Li}, \citenamefont {Chen}, \citenamefont {Ni}, \citenamefont {Cai}, \citenamefont {Mai}, \citenamefont {Zhang}, \citenamefont {Zheng}, \citenamefont {Yu}, \citenamefont {Zou} \emph {et~al.}}]{deng2024quantum}%
  \BibitemOpen
  \bibfield  {author} {\bibinfo {author} {\bibfnamefont {X.}~\bibnamefont {Deng}}, \bibinfo {author} {\bibfnamefont {S.}~\bibnamefont {Li}}, \bibinfo {author} {\bibfnamefont {Z.-J.}\ \bibnamefont {Chen}}, \bibinfo {author} {\bibfnamefont {Z.}~\bibnamefont {Ni}}, \bibinfo {author} {\bibfnamefont {Y.}~\bibnamefont {Cai}}, \bibinfo {author} {\bibfnamefont {J.}~\bibnamefont {Mai}}, \bibinfo {author} {\bibfnamefont {L.}~\bibnamefont {Zhang}}, \bibinfo {author} {\bibfnamefont {P.}~\bibnamefont {Zheng}}, \bibinfo {author} {\bibfnamefont {H.}~\bibnamefont {Yu}}, \bibinfo {author} {\bibfnamefont {C.-L.}\ \bibnamefont {Zou}}, \bibnamefont {et~al.},\ }\bibfield  {title} {\bibinfo {title} {Quantum-enhanced metrology with large fock states},\ }\href {https://doi.org/10.1038/s41567-024-02619-5} {\bibfield  {journal} {\bibinfo  {journal} {Nat. Phys.}\ ,\ \bibinfo {pages} {1}} (\bibinfo {year} {2024})}\BibitemShut {NoStop}%
\bibitem [{\citenamefont {Eickbusch}\ \emph {et~al.}(2022)\citenamefont {Eickbusch}, \citenamefont {Sivak}, \citenamefont {Ding}, \citenamefont {Elder}, \citenamefont {Jha}, \citenamefont {Venkatraman}, \citenamefont {Royer}, \citenamefont {Girvin}, \citenamefont {Schoelkopf},\ and\ \citenamefont {Devoret}}]{eickbusch2022fast}%
  \BibitemOpen
  \bibfield  {author} {\bibinfo {author} {\bibfnamefont {A.}~\bibnamefont {Eickbusch}}, \bibinfo {author} {\bibfnamefont {V.}~\bibnamefont {Sivak}}, \bibinfo {author} {\bibfnamefont {A.~Z.}\ \bibnamefont {Ding}}, \bibinfo {author} {\bibfnamefont {S.~S.}\ \bibnamefont {Elder}}, \bibinfo {author} {\bibfnamefont {S.~R.}\ \bibnamefont {Jha}}, \bibinfo {author} {\bibfnamefont {J.}~\bibnamefont {Venkatraman}}, \bibinfo {author} {\bibfnamefont {B.}~\bibnamefont {Royer}}, \bibinfo {author} {\bibfnamefont {S.~M.}\ \bibnamefont {Girvin}}, \bibinfo {author} {\bibfnamefont {R.~J.}\ \bibnamefont {Schoelkopf}},\ \bibnamefont {and}\ \bibinfo {author} {\bibfnamefont {M.~H.}\ \bibnamefont {Devoret}},\ }\bibfield  {title} {\bibinfo {title} {Fast universal control of an oscillator with weak dispersive coupling to a qubit},\ }\href {https://doi.org/10.1038/s41567-022-01776-9} {\bibfield  {journal} {\bibinfo  {journal} {Nature Physics}\ }\textbf {\bibinfo {volume} {18}},\ \bibinfo {pages} {1464} (\bibinfo {year}
  {2022})}\BibitemShut {NoStop}%
\bibitem [{\citenamefont {v.~Neumann}(1928)}]{v1928theorie}%
  \BibitemOpen
  \bibfield  {author} {\bibinfo {author} {\bibfnamefont {J.}~\bibnamefont {v.~Neumann}},\ }\bibfield  {title} {\bibinfo {title} {Zur theorie der gesellschaftsspiele},\ }\href@noop {} {\bibfield  {journal} {\bibinfo  {journal} {Mathematische annalen}\ }\textbf {\bibinfo {volume} {100}},\ \bibinfo {pages} {295} (\bibinfo {year} {1928})}\BibitemShut {NoStop}%
\end{thebibliography}%

\end{document}